\documentclass[12pt,letterpaper]{article}
\usepackage[usenames, dvipsnames]{xcolor}
\usepackage{jcapmod}

\usepackage{tocloft}
\usepackage[]{todonotes}
\usepackage{verbatim}
\usepackage{mathrsfs}
\usepackage{cleveref}
\usepackage[shortlabels]{enumitem}

\usepackage{pgf}
\usepackage{tikz-cd}
\usetikzlibrary{shapes,arrows}
\usetikzlibrary{calc}
\usepackage{pgfplots}
\pgfplotsset{compat=1.12}
\usepackage{tikz-3dplot}

\usepackage{tikz}
\usepackage{setspace,caption}
\usepackage{lipsum}
\usetikzlibrary{matrix,arrows,calc}
\makeatletter
\newsavebox\myboxA
\newsavebox\myboxB
\newlength\mylenA

\definecolor{cornellRed}{HTML}{B31B1B}
\definecolor{cornellBlue}{HTML}{0068AC}
\definecolor{cornellGreen}{HTML}{6EB43F}

\newtheorem{theorem}{Theorem}
\newtheorem{lemma}[theorem]{Lemma}
\newtheorem{definition}[theorem]{Definition}
\newtheorem{proposition}[theorem]{Proposition}
\newtheorem{conjecture}[theorem]{Conjecture}
\newtheorem{corollary}[theorem]{Corollary}
\newtheorem{example}[theorem]{Example}

\usepackage{bbm} 					
\usepackage{slashed} 			
\usepackage{graphicx}				
\usepackage{subcaption}			
\usepackage{psfrag}				
\usepackage{tensor}				
\usepackage{fouridx}				
\usepackage{bm}					
\usepackage{mdframed}				
\usepackage{multirow}				
\usepackage{soul}					
\usepackage{bbold}				
\usepackage{multicol}				
\usepackage{tikz-cd}
\usepackage{rotating}

\usetikzlibrary{arrows}

\tikzset{
commutative diagrams/.cd,
arrow style=tikz,
diagrams={>=latex}}

\usepackage{amsmath}
\usepackage{amssymb}
\usepackage{amsfonts}
\usepackage{mathtools}

\usepackage{feynmf}
\usepackage{marvosym}

\usepackage{import}

\newcommand*\xoverline[2][0.75]{
    \sbox{\myboxA}{$\m@th#2$}
    \setbox\myboxB\null
    \ht\myboxB=\ht\myboxA
    \dp\myboxB=\dp\myboxA
    \wd\myboxB=#1\wd\myboxA
    \sbox\myboxB{$\m@th\overline{\copy\myboxB}$}
    \setlength\mylenA{\the\wd\myboxA}
    \addtolength\mylenA{-\the\wd\myboxB}
    \ifdim\wd\myboxB<\wd\myboxA
       \rlap{\hskip 0.5\mylenA\usebox\myboxB}{\usebox\myboxA}
    \else
        \hskip -0.5\mylenA\rlap{\usebox\myboxA}{\hskip 0.5\mylenA\usebox\myboxB}
    \fi}
\makeatother

\newcommand{\cO}{\mathcal{O}}

\newcommand{\PP}{\mathbb{P}}

\newcommand{\ZZ}{\mathbb{Z}}

\newcommand{\im}{\,\mathrm{Im}\,}

\definecolor{cobalt}{RGB}{44, 98, 120}
\definecolor{celadon}{rgb}{0.67, 0.88, 0.69}
\definecolor{dm}{cmyk}{.20, 0, .30, 0}
\definecolor{burgundy}{rgb}{0.5, 0.0, 0.13}
\definecolor{plotBlue}{RGB}{94, 130, 181}

\newcommand{\RV}{\mathscr{R}}
\newcommand{\Pf}{\mathscr{P}}
\newcommand{\C}{\mathbb{C}}
\newcommand{\R}{\mathbb{R}}
\newcommand{\Z}{\mathbb{Z}}

\DeclareMathOperator{\Ima}{Im}
\DeclareMathOperator{\minface}{minface}

\DeclareMathOperator{\coker}{\rm coker}
\hypersetup{
  colorlinks,
  citecolor=Violet,
  linkcolor=cobalt,
  urlcolor=Blue}

\DeclareSymbolFontAlphabet{\mathbb}{AMSb}

\newif\iffastcompile

\fastcompilefalse

\iffastcompile
\newcommand{\cl}[1]{}
\newcommand{\lm}[1]{}
\newcommand{\ms}[1]{}
\newcommand{\ab}[1]{}
\newcommand{\bs}[1]{}
\else
\newcommand{\cl}[1]{\todo[color=burgundy!30, size=\scriptsize, bordercolor=burgundy!30]{CL: #1}}
\newcommand{\lm}[1]{\todo[color=dm!90, size=\scriptsize, bordercolor=dm!90]{LM: #1}}
\newcommand{\ms}[1]{\todo[color=dm!90, size=\scriptsize, bordercolor=dm!90]{MS: #1}}
\newcommand{\ab}[1]{\todo[color=blue!30, size=\scriptsize, bordercolor=blue!50]{AB: #1}}
\newcommand{\bs}[1]{\todo[color=blue!30, size=\scriptsize, bordercolor=blue!50]{BS: #1}}

\fi

\newcommand{\email}[1]{\href{mailto:#1}{#1}}

\ProvideTextCommandDefault{\Dbar}{
\leavevmode\lower.5ex\rlap{\hskip-.07em\accent"16}D
}

\begin{document}
	\newcommand{\main}{.}
\begin{titlepage}

\setcounter{page}{1} \baselineskip=15.5pt \thispagestyle{empty}
\setcounter{tocdepth}{2}

\bigskip\

\vspace{1cm}
\begin{center}
{\fontsize{22}{28} \bfseries The Hodge Numbers of Divisors \\ \vspace{0.25cm}
of Calabi-Yau Threefold Hypersurfaces}
\end{center}

\vspace{0.45cm}

\begin{center}
\scalebox{0.95}[0.95]{{\fontsize{14}{30}\selectfont Andreas P. Braun,$^{a}$ Cody Long,$^{b}$ Liam McAllister,$^{c}$}} \vspace{0.25cm}
\scalebox{0.95}[0.95]{{\fontsize{14}{30}\selectfont Michael Stillman,$^{d}$ and Benjamin Sung$^{b}$}}

\end{center}

\begin{center}

\textsl{$^{a}$Mathematical Institute, University of Oxford, Oxford OX2 6GG, UK}\\
\textsl{$^{b}$Department of Physics, Northeastern University, Boston, MA 02115, USA}\\
\textsl{$^{c}$Department of Physics, Cornell University, Ithaca, NY 14853, USA}\\
\textsl{$^{d}$Department of Mathematics, Cornell University, Ithaca, NY 14853, USA}\\

\vspace{0.25cm}

\vskip .3cm
\email{\tt andreas.braun@maths.ox.ac.uk, co.long@northeastern.edu, mcallister@cornell.edu, mike@math.cornell.edu, b.sung@northeastern.edu}
\end{center}

\vspace{0.6cm}
\noindent
We prove a formula for the Hodge numbers of square-free divisors of Calabi-Yau threefold hypersurfaces in toric varieties. Euclidean branes wrapping divisors affect the vacuum structure of Calabi-Yau compactifications of type IIB string theory, M-theory, and F-theory. Determining the nonperturbative couplings due to Euclidean branes on a
divisor $D$
requires counting fermion zero modes, which depend on the Hodge numbers $h^i({\cal{O}}_D)$. Suppose that $X$ is a smooth Calabi-Yau threefold hypersurface in a toric variety $V$, and let $D$ be the restriction to $X$ of a square-free divisor of $V$. We give a formula for $h^i({\cal{O}}_D)$ in terms of combinatorial data. Moreover, we construct a CW complex $\Pf_D$ such that $h^i({\cal{O}}_D)=h_i(\Pf_D)$.
We describe an efficient algorithm that makes possible for the first time the computation of sheaf cohomology for such divisors at large $h^{1,1}$.
As an illustration we compute the Hodge numbers of a class of divisors in a threefold with $h^{1,1}=491$.
Our results are a step toward a systematic computation of Euclidean brane superpotentials in Calabi-Yau hypersurfaces.

\noindent
\vspace{2.1cm}

\noindent\today

\end{titlepage}
\tableofcontents

\newpage

\section{Introduction}

Compactifications of type IIB string theory on orientifolds of Calabi-Yau threefolds, and of F-theory on Calabi-Yau fourfolds, provide important classes of four-dimensional effective theories with $\mathcal{N}=1$ supersymmetry.  The vacuum structure of these theories
depends on
the potential for the K\"ahler moduli, which parameterize the sizes of holomorphic submanifolds in the Calabi-Yau manifold.  In particular, in lieu of a potential for the K\"ahler moduli,
positive vacuum energy in four dimensions induces an instability of the overall volume of the compactification,
and therefore realistic particle physics and cosmology require a computation of the K\"ahler moduli potential.

In principle a minimum of the K\"ahler moduli potential could be created by competition among purely perturbative corrections to the
leading-order K\"ahler potential.  However, at present
the best-understood constructions of metastable vacua require
contributions to the superpotential
for the K\"ahler moduli \cite{KKLT,LVS1,LVS2}.
Because of the non-renormalization theorem,
such terms are necessarily nonperturbative, resulting from Euclidean branes wrapping cycles --- in fact, divisors --- in the compact space.\footnote{Strong gauge dynamics in four dimensions, arising on seven-branes wrapping four-cycles in the compact space, is an alternative.  The geometric requirements on such cycles closely parallel those in the Euclidean brane case, and we will refer only to the latter in this work.}

An important goal is to determine which divisors in a Calabi-Yau threefold or fourfold
support nonperturbative superpotential terms from Euclidean branes.
Witten has shown
\cite{Witten:1996bn} that Euclidean M5-branes on a smooth effective vertical divisor $D$ of a smooth Calabi-Yau fourfold $Y$ give a nonvanishing contribution to the superpotential whenever $D$ is \emph{rigid}, meaning that the Hodge numbers $h^{0,i}(D)=h^i(D,{\cal{O}}_D)=h^i({\cal{O}}_D)$ obey
\begin{equation} \label{rig}
h^{0,0}(D)=1\,,\,\, h^{0,1}(D)=0\,,\,\,  h^{0,2}(D)=0\,,\,\,  h^{0,3}(D)=0\,,
\end{equation}
which we abbreviate as $h^{\bullet}({\cal{O}}_D)=(1,0,0,0)$.
Rigidity corresponds to the absence of massless bosonic deformations, and implies that the only zero modes of the Dirac operator on the M5-brane are the two universal Goldstino modes that result from the supersymmetries broken by the M5-brane.
In more general circumstances --- when either $D$ or $Y$ is singular, when fluxes are included on $D$ or in $Y$, or when $D$ is a divisor of a Calabi-Yau threefold --- the conditions for a nonperturbative superpotential are more subtle, but the
Hodge numbers are still essential data.

For this reason, a long-term aim
is to compute the Hodge numbers of effective divisors $D$ of Calabi-Yau manifolds $X$.
This is most manageable in the case of
smooth Calabi-Yau hypersurfaces in toric varieties:
for many years there have been computational
algorithms and implementations for
computing sheaf cohomology of coherent sheaves on toric varieties
\cite{EMS2000,M2,GGSmith}, and there are faster
implementations for computing sheaf cohomology of line bundles on
toric varieties \cite{Blumenhagen:2010pv, Blumenhagen:2010ed}.  Unfortunately, all
of these implementations fail to finish for even very modest sizes of
$h^{1,1}(X)$, say for $h^{1,1}(X) \gtrsim 10-20$.  A key open
problem in computational algebraic geometry is to find algorithms that
work for effective divisors or coherent sheaves on all Calabi-Yau threefold hypersurfaces
arising from the Kreuzer-Skarke database \cite{KSdatabase} of four-dimensional
reflexive polytopes, as well as for Calabi-Yau fourfold hypersurfaces in toric varieties.

We will not arrive at a fully general answer.  However, in the
important special case in which $X$ is a smooth Calabi-Yau threefold
hypersurface in a toric variety $V$, and $D$ is the restriction to $X$
of a square-free effective divisor $\widehat{D}$ on $V$, we will
establish a formula for $h^i({\cal{O}}_D)$ in terms of the
combinatorial data of $V$.  This formula, given in
Theorem~\ref{MasterTheorem}, is our main result.   As we will see,
Theorem~\ref{MasterTheorem} allows one to read off the Hodge numbers of many divisors by inspection, and moreover it is
straightforward to turn this formula into an algorithm that computes
the Hodge numbers of any square-free divisor of any
$X$ arising from the Kreuzer-Skarke database \cite{StringTorics}.

The principal tools in our proof are stratification; the hypercohomology spectral sequence of the Mayor-Vietoris complex (\ref{eq:mvcy}); and a correspondence
that we establish
between square-free divisors of a Calabi-Yau hypersurface and CW complexes constructed on a triangulation of the associated reflexive polytope $\Delta^{\circ}$.  Let us briefly summarize these results.  Stratification is the decomposition of an $n$-dimensional toric variety $V$ into
tori,
and leads to the extremely simple expressions (\ref{hodgesubv}) for the Hodge numbers $h^{0,i}$ of particular subvarieties of a Calabi-Yau hypersurface $X \subset V$.
Among these subvarieties are prime toric divisors $D_i$, $i=1,\ldots h^{1,1}(X)+4$, each of which corresponds to a lattice point of a reflexive polytope $\Delta^{\circ}$, as we review in \S\ref{sec:preliminaries}.
The Hodge numbers $h^{\bullet}({\cal{O}}_{D_i})$ and the intersections of the prime toric divisors $D_i$ are fully specified by elementary combinatorial data: namely,
by the simplicial complex $\cal{T}$ corresponding to a
triangulation
of $\Delta^{\circ}$, together with the number of lattice points interior to each face of $\Delta$.

A \emph{square-free divisor} $D$ is the union of a collection of distinct prime toric divisors, $D=\sum D_i$.
In order to compute $h^i({\cal{O}}_D)$, one can appropriately combine the data characterizing the constituent prime toric divisors $D_i \subset D$.
To achieve this, we establish (in Appendix \ref{sec:mvc}) that the Mayer-Vietoris sheaf sequence associated to $D$ is exact, and we then examine the corresponding hypercohomology spectral sequence.  Formally, these methods are already sufficient to derive an expression for $h^i({\cal{O}}_D)$, but we find it valuable to carry out the computation, and to express the result, in terms of a particular CW complex $\mathscr{P}$ that encodes the data of $\cal{T}$ and $\Delta$.
The construction of $\mathscr{P}$ amounts to
attaching cells to each $j$-face $\Theta^\circ$ of $\Delta^\circ$, in a manner determined by the
number of lattice points in the relative interior of the dual face $\Theta \subset \Delta$.
A square-free divisor $D$ naturally determines a subcomplex $\mathscr{P}_D \subset \mathscr{P}$, and via the hypercohomology spectral sequence we are able to relate
the sheaf cohomology of ${\cal{O}}_D$ to the
cellular homology
of the CW complex $\mathscr{P}_D$.  In particular, we show
that $h^i({\cal{O}}_D)=h_i(\Pf_D)$, proving the theorem.

A subtlety in applying our results to compute Euclidean brane superpotentials is that any
nontrivial sum $D=\sum D_i$ of prime toric divisors $D_i$ that is rigid is necessarily reducible, and
involves normal crossing singularities where the $D_i$ intersect.
Normal crossing singularities present no obstacle to defining and computing $h^i({\cal{O}}_D)$, but they do complicate the connection between $h^i({\cal{O}}_D)$ and the number of fermion zero modes: new zero modes can appear at the intersection loci.
Thus, the Hodge numbers $h^i({\cal{O}}_D)$  that we compute here mark a significant step toward computing the superpotential, but do not provide a final answer.
Systematically counting the fermion zero modes associated to normal crossing singularities, along the lines of \cite{Donagi:2010pd,Donagi:2012ts,Clingher:2012rg}, is an important problem for the future \cite{wip}.

The organization of this paper is as follows.  In \S\ref{sec:preliminaries} we set notation and recall elementary properties of Calabi-Yau hypersurfaces in toric varieties.
Then, given any square-free divisor $D$ in a Calabi-Yau threefold hypersurface, we define a corresponding $\Delta$-complex $\mathscr{R}_D$ and a CW complex $\mathscr{P}_D$, which are constructed so that their homology encodes the sheaf cohomology of ${\cal{O}}_D$.  In \S\ref{sec:proofsec} we prove our main result, Theorem~\ref{MasterTheorem}, which asserts that $h^i({\cal{O}}_D)=h_i(\mathscr{P}_D)$.
In \S\ref{sec:example} we illustrate our findings in the example  of a Calabi-Yau threefold with $h^{1,1}=491$.
We conclude in \S\ref{sec:conclusions}.  Appendix \ref{sec:mvc} defines Mayer-Vietoris complexes and proves some relevant properties.
In Appendix \ref{sec:stratification} we review key results from stratification.
In Appendix \ref{s:theproof} we directly compute $h^2({\cal{O}}_D)$ by counting lattice points in dual faces, for the special case where $\mathcal{T}_D$ is restricted to a single 2-face of $\Delta^{\circ}$.

\section{Notation and Preliminaries} \label{sec:preliminaries}

\subsection{Polytopes and toric varieties}

In this paper we consider Calabi-Yau threefold hypersurfaces $X$ in
simplicial
toric varieties $V$,
as studied by Batyrev in \cite{Batyrev} (see also Appendix \ref{sec:stratification} for a review). Such Calabi-Yau threefolds are constructed from pairs $(\Delta, \Delta^\circ)$  of four-dimensional polytopes with vertices in $\ZZ^4$, obeying
\begin{equation}\label{dualdef}
\langle \Delta, \Delta^\circ \rangle \geq -1 \, .
\end{equation}
A pair of $d$-dimensional lattice polytopes obeying (\ref{dualdef}) is called a reflexive pair. In each dimension, there are only a finite number of reflexive pairs, up to equivalence, and those in dimension $d \leq 4$ have all been enumerated \cite{Kreuzer:1995cd,Kreuzer:1998vb,Kreuzer:2000xy,KSdatabase}.

Given a four-dimensional reflexive polytope $\Delta^\circ$, we choose a fine star regular\footnote{Fine means that the  triangulation uses all the lattice points of $\Delta^\circ$.
A triangulation is star if the origin, which is the unique interior point of a reflexive polytope $\Delta^\circ$, is contained in every four-dimensional simplex of $\widehat{{\cal T}}$. Regularity is a condition that ensures that $V$, and hence $X$, is projective.} triangulation $\widehat{{\cal T}}$ of $\Delta^\circ$.  Since each simplex in $\widehat{{\cal T}}$ contains the origin, this triangulation determines a fan $\Sigma$, and
$V := \mathbb{P}_\Sigma$ is the corresponding simplicial toric variety.
If $F$ is a generic linear combination of the monomials in the Cox ring of $V$ that correspond to the lattice points of the polytope $\Delta$, then $X := \{F=0\} \subset V$ is a smooth Calabi-Yau threefold hypersurface in $V$ \cite{Batyrev}.
The toric variety $V$ in general has pointlike orbifold singularities, but for generic $F$, $X$ does not intersect these points, and is smooth.

We denote the set of faces of $\Delta^\circ$ by $\cal F$, the set of faces of dimension at most $j$ by ${\cal F}(\le j)$, and
the sets of vertices, edges, and 2-faces by ${\cal F}(0)$, ${\cal F}(1)$, and ${\cal F}(2)$, respectively.
For each face $\Theta^\circ$ in $\cal F$, there is a unique face $\Theta$ of $\Delta$ defined by
\begin{equation}
\langle \Theta^\circ,\Theta \rangle = -1 \, .
\end{equation}
Given any face $\Theta \subset \Delta$, we denote by $\ell^*(\Theta)$ the number of lattice points in the relative interior of $\Theta$.
Given a face $\Theta^\circ$ of $\Delta^\circ$, we define its genus  $g(\Theta^\circ)$ by
\begin{equation}
g(\Theta^\circ):= \ell^*(\Theta)  \,,
\end{equation}
i.e.~we define the genus of $\Theta^\circ$ to be the number of interior lattice points of its dual face.

If $\sigma \in \widehat{{\cal T}}$ is a simplex, we define the corresponding minimal face, $\minface(\sigma)$, to be the lowest-dimensional face of $\Delta^\circ$ containing all of the lattice points $\{ p_I \mid I \in \sigma \}$.
We define
\begin{equation}
\mu(\sigma) := \dim \minface(\sigma) \,
\end{equation} and
\begin{equation}
g(\sigma) := g(\minface(\sigma)) \,.
\end{equation}

\subsection{The Picard group of $X$}\label{secpic}

For each nonzero lattice point $p_I$ on $\Delta^\circ$ there is an associated ray of the fan $\Sigma$ and a corresponding homogeneous coordinate $z_I$ of the toric variety $V$. We may hence associate the toric divisor $\widehat{D}_I$ given by $\{z_I = 0\}$ with the point $p_I$. This notion can be extended to each $d$-simplex $\sigma$ of the triangulation $\widehat{{\cal T}}$
by associating $\sigma$ with the subvariety $V_\sigma := \{z_I = 0 \, \forall\, p_I \in \sigma \}$. Let $X_\sigma:=V_\sigma\cap X$ be the intersection  of $V_\sigma$ with the Calabi-Yau hypersurface $X$.  This intersection is nonzero if and only if the simplex $\sigma$ is contained in a 2-face of $\Delta^\circ$.
It is therefore useful to define
\begin{equation}
{\cal T} := \{ \sigma \in \widehat{{\cal T}} \mid \mu(\sigma)\le 2\} \,,
\end{equation} which omits simplices of $\widehat{{\cal T}}$ that pass through facets (3-faces) of $\Delta^\circ$, and so correspond to varieties that do not intersect a generic Calabi-Yau hypersurface $X$.  In a slight abuse of language, we refer to $\mathcal{T}$ as a triangulation.

The Hodge numbers of subvarieties $X_\sigma \subset X$ are given by rather simple formulae \cite{Klemm:1996ts,Braun:2016igl,Greiner}.
For the divisor $D_I$ of $X$ associated with a lattice point $\sigma = p_I \in {\cal T}$ we find (see Appendix \ref{sec:stratification})
\begin{equation}\label{redprimetoric}
h^{\bullet}(\mathcal{O}_{D_I}) =
\left\{\begin{array}{ccc}
1,&0,&g(\sigma) \\
1,&g(\sigma),&0 \\
1+g(\sigma),&0,&0 \\
\end{array}\right\}
~\mbox{for}~\mu(\sigma) =
\left\{\begin{array}{c}
0 \\
1 \\
2
\end{array}\right. \, .
\end{equation}
In particular, divisors associated with points interior to 2-faces $\Theta^\circ$ with $g(\Theta^\circ)>0$ are reducible. A 1-simplex $\sigma$ of ${\cal T}$ connecting a pair of lattice points $p_I$ and $p_J$ in ${\cal T}$
corresponds to the intersection $C_{IJ} := D_I \cap D_J$, with Hodge numbers
\begin{equation}
h^{\bullet}(\mathcal{O}_C) =
\left\{\begin{array}{cc}
1,&g(\sigma) \\
1+g(\sigma),&0 \\
\end{array}\right\}
~\mbox{for}~\mu(\sigma) =
\left\{\begin{array}{c}
1 \\
2
\end{array}\right.\, .
\end{equation}
That is, a curve $C$ associated with a 1-simplex $\sigma$ interior to a 1-face $\Theta^\circ$ is irreducible, of genus $g(\Theta^{\circ})$.  A curve $C$ associated with a 1-simplex $\sigma$ interior to a 2-face $\Theta^\circ$ is a union of $g(\Theta^\circ)+1$ disjoint $\mathbb{P}^1$s.  A 2-simplex $\sigma$ of ${\cal T}$ containing three lattice points $p_I, p_J, p_K$ corresponds to the intersection of three divisors,
and the corresponding $X_\sigma$ consists of $1+g(\sigma)$ points.

The above results can be summarized as
\begin{equation} \label{hodgesubv}
h^{0,i}(X_\sigma) = \delta_{i,0}  + \delta_{i,2-\mu(\sigma)}g(\sigma) \,.
\end{equation}
This result easily generalizes to arbitrary dimension: see Appendix \ref{sec:anydim}.
Moreover, in Appendix \ref{sect:topcy3divs} we give similar, albeit slightly more complicated and triangulation-dependent, formulae for the Hodge numbers $h^{1,1}(X_{\sigma})$.

As shown in \cite{Batyrev}, the Hodge numbers of a Calabi-Yau hypersurface $X$ obey simple combinatorial relations as well. In particular, the rank of the Picard group of
$X$ satisfies
\begin{equation}
h^{1,1}(X) = \sum_{\Theta^\circ \in {\cal F}(\le 2)}\ell^*(\Theta^\circ) - 4  + \sum_{\Theta^\circ \in {\cal F}(2)} \ell^*(\Theta^\circ)\ell^*(\Theta)
\end{equation}
We can identify $N:= h^{1,1}(X)+4$ divisors that obey four linear relations, and that generate the Picard group of $X$.
First of all, the $K \le N$ divisors of $X$ associated with lattice points in ${\cal F}(\le 1)$ are irreducible.
However, the divisor associated with a lattice point $p_r$ interior to a 2-face $\Theta^\circ$ has $1+g(\Theta^\circ)$ irreducible, connected components, which we denote $D_r^{\alpha}$, $\alpha \in \{0,\ldots, g(\Theta^\circ)\}$.
The
$N$  irreducible divisors
\begin{equation}
\{D_1,\ldots D_K, D_{K+1}^{0},\ldots, D_{K+1}^{g(p_{K+1})},D_{K+2}^{0},\ldots\}
\end{equation}
then generate the Picard group of $X$.
These can be written collectively as $\{D_I^{\alpha}\}$ with $\alpha \in \{0,\ldots,\delta_{\mu(p_I),2}\,g(p_I)\}$, but we will reindex them as
as $$ D_i \in \{D_1, \ldots, D_N\}.$$

We refer to the $D_i$ as {\bf{prime toric divisors}}, even though the $D_r^{\alpha}$  do not all descend from prime toric divisors on $V$ unless $X$ is favorable,
i.e.~unless
$g(\Theta) g(\Theta^\circ)= 0~\forall~\Theta^\circ \in {\cal{F}}(2)$.  By (\ref{redprimetoric}), a prime toric divisor $D_i$ associated with a lattice point $\sigma$ has
\begin{equation} \label{irredprimetoric}
h^{\bullet}(\mathcal{O}_{D_i}) =
\left\{\begin{array}{ccc}
1,&0,&g(\sigma) \\
1,&g(\sigma),&0 \\
1,&0,&0 \\
\end{array}\right\}
~\mbox{for}~\mu(\sigma) =
\left\{\begin{array}{c}
0 \\
1 \\
2
\end{array}\right. \, .
\end{equation}
For any subset $G \subseteq \{1,\ldots,N \}$, there is an associated divisor $D$ (possibly reducible) defined by
\begin{equation}\label{eq:decompD_G}
D = \sum_{i \in G} D_i \, .
\end{equation}
We call such a $D$ a {\bf{square-free divisor}}.
The main purpose of this work is to analyze square-free divisors.

\subsection{The ravioli complex $\RV$}

Every $d$-simplex of $\mathcal{T}$ contained in the relative interior of a face $\Theta^\circ \in {\cal{F}}(j)$ for $j \leq 2$ gives rise to a closed subvariety $X_\sigma \subset X$, of complex dimension $2-d$.
When $j \le 1$, the simplices in $\mathcal{T}$ correspond to irreducible subvarieties of $X$.
However, for $j=2$, i.e.~when $\sigma \in \mathcal{T}$ is in the interior of a 2-face $\Theta^\circ$ of $\Delta^\circ$, the simplex $\sigma$ corresponds to a subvariety of $X$ that has $1 + g(\Theta^\circ)$ connected (and irreducible) components, which we denote by $X_\sigma^\alpha$, for $\alpha = 0, \ldots, g(\sigma)$.

The intersection structure of the $X_\sigma$ is determined by $\Delta$ and by the triangulation ${\cal T}$ of $\Delta^\circ$.
For each 2-face $\Theta^{\circ}$ of $\Delta^\circ$, we can choose an ordering of $\alpha = 0, \ldots, g(\sigma)$
such that for any $\sigma, \lambda \in \mathcal{T}$ with $\minface(\sigma)= \minface(\lambda) = \Theta^{\circ}$, we have
\begin{align}
\,\,\,  \, \, \, \, \, \,X^\alpha_{\sigma} \cap X^\beta_{\lambda}  = \begin{cases}
   \delta^{\alpha \beta} X^\alpha_{\sigma \cup \lambda}\, \quad &\text{if } \sigma\cup \lambda \in \mathcal{T}\,,\\
   \emptyset \, \quad& \text{if }  \sigma \cup \lambda \notin \mathcal{T}\,.
   \end{cases}\, \label{eqn:starint1}
  \end{align}

Next, for each $\tau \in \mathcal{T}$ with $\mu(\tau) \leq 1$, the intersection structure with the $X^\alpha_{\sigma}$ can be written as
\begin{align}
  X_\tau \cap X^\alpha_{\sigma}  = \begin{cases}
  X^\alpha_{\tau \cup \sigma}\, \quad &\text{if } \tau \cup \sigma \in \mathcal{T}\,, \\
   \emptyset\, \quad &\text{if } \tau \cup \sigma \notin \mathcal{T}\,.
  \end{cases} \label{eqn:starint2}
  \end{align}
Finally, for $\tau, \omega \in \mathcal{T}$, with $\mu(\tau) \leq 1$ and $\mu(\omega) \leq 1$, we have:
  \begin{align}
  X_{\tau} \cap X_{\omega}   = \begin{cases}
  X_{\tau \cup \omega}\, \quad &\text{if } \tau \cup \omega \in \mathcal{T}\,, \\
   \emptyset\ \quad &\text{if }  \tau \cup \omega \notin \mathcal{T}\,.
  \end{cases}\label{eqn:starint3}
  \end{align}
In the special case of \eqref{eqn:starint3} in which
$\tau \cup \omega \in \mathcal{T}$ with $\minface(\tau \cup \omega) = \Theta^{\circ} \in {\cal{F}}(2)$,
we can write \eqref{eqn:starint3} as
\begin{eqnarray}
  X_{\tau} \cap X_{\omega}  & = &\bigcup\limits_{\alpha = 0}^{g(\Theta^{\circ})} X^\alpha_{\tau\cup\omega}\, .  \label{eqn:starint4}
\end{eqnarray}

We will now define a complex, called the {\bf{ravioli complex}} $\RV$, that accounts for the intersection structure \eqref{eqn:starint1}-\eqref{eqn:starint4}.
Recall that the simplices in $\mathcal{T}$ correspond to subvarieties in $X$, which are possibly disconnected and reducible.  The cells in $\RV$ will correspond to the connected, irreducible components of the subvarieties in $X$ encoded by $\mathcal{T}$.
The ravioli complex is defined by
\begin{equation}\label{eq:ravdef}
 \RV = \Bigl\{ X_\sigma^{\alpha} \big| \sigma \in \mathcal{T}, \mu(\sigma) = 2, 0 \le \alpha \le g(\sigma) \Bigr\}\cup \Bigl\{X_\tau \big| \tau \in \mathcal{T}, \mu(\tau) \le 1 \Bigr\}\,,
\end{equation} which as a set consists of the irreducible connected components of intersections of the $D_i$.
The elements of $\RV$ that have dimension $2-i$ in $X$ are called $i$-cells.
For those $X$ such that\footnote{Notice that even in the case where $X$ is favorable, i.e.~obeying $g(\Theta) g(\Theta^\circ)= 0~\forall~\Theta^\circ \in {\cal{F}}(2)$,  $\RV$ is not necessarily equal to $\mathcal{T}$.} $g(\Theta^\circ)=0~\forall~\Theta^\circ \in {\cal F}(2)$, $\RV$ is the simplicial complex ${\mathcal{T}}$.  In the general case, the $i$-cells of $\RV$ are the same as the $i$-simplices of $\mathcal{T}$, except that each simplex $\sigma$ in the interior of any two-dimensional face $\Theta^\circ \in {\cal F}(2)$ of $\Delta^\circ$ is replaced by $1 + g(\Theta^\circ)$ disjoint copies of itself, after which the ($1$-cell) boundaries of the $1 + g(\Theta^\circ)$ disjoint copies $\mathcal{T}\mid_{\Theta^\circ}$ of are identified with each other.

\begin{figure}
 \begin{center}
  \includegraphics[height=3cm]{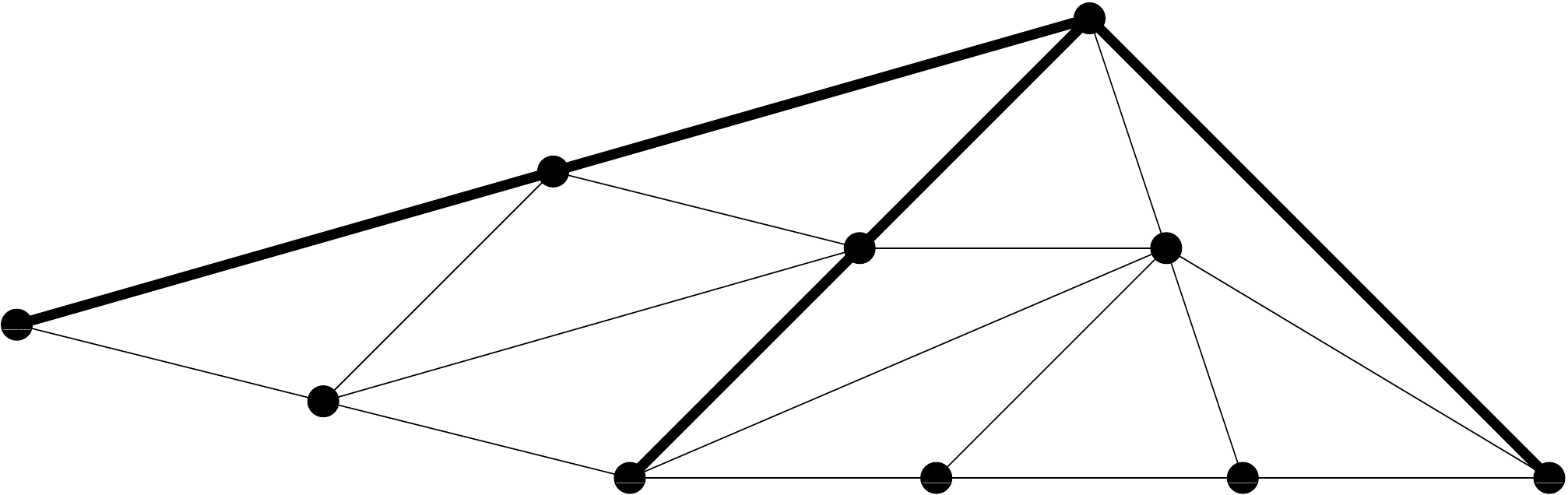} \\
  \vspace{.5cm}
  \includegraphics[height=3cm]{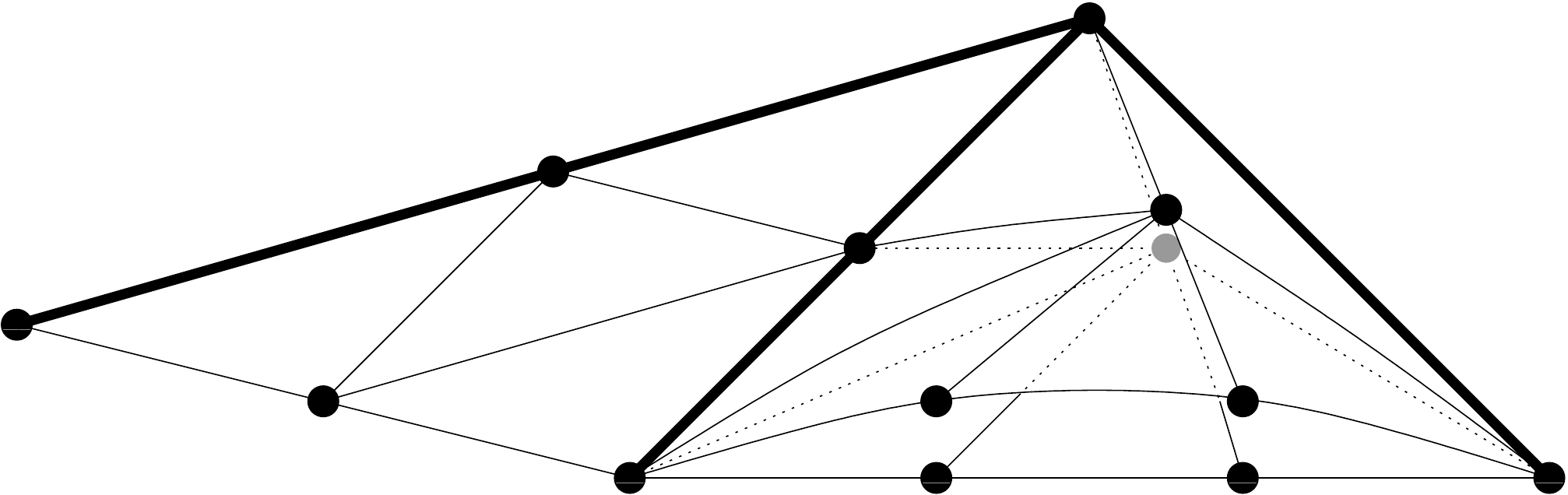} \\
  \vspace{.5cm}
  \includegraphics[height=3cm]{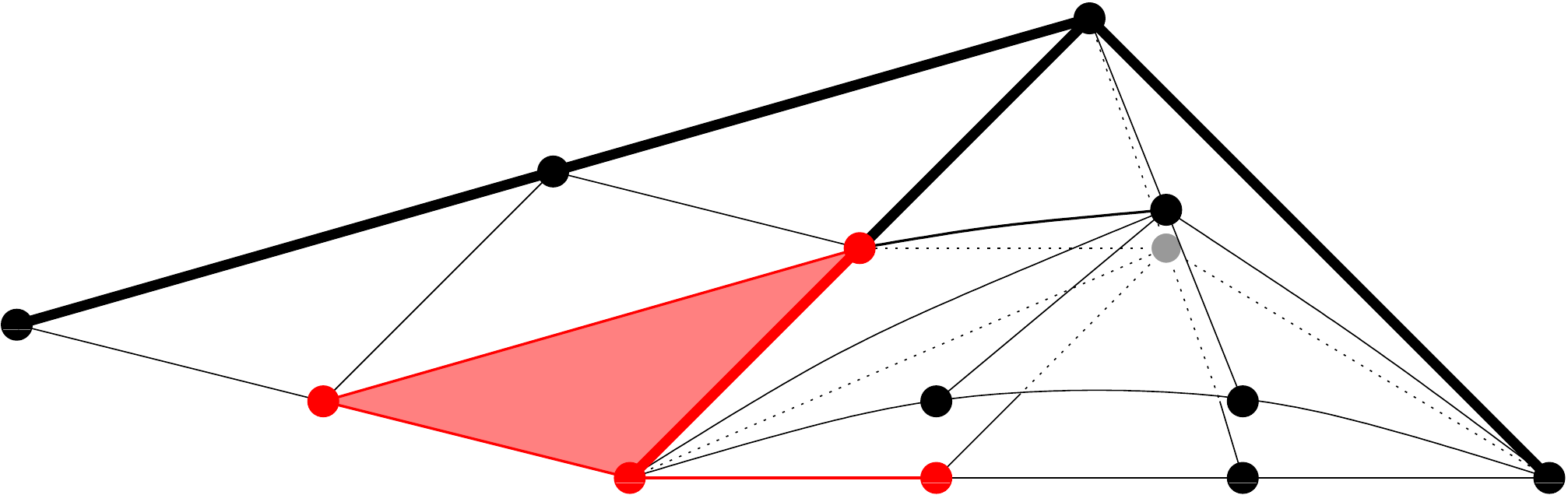}
  \caption{\label{raviolifig1} The simplicial complex $\mathcal{T}$ defined by a triangulation, the corresponding ravioli complex $\RV$, and
  the subcomplex $\RV_D \subset \RV$ associated to a divisor $D$.  The upper figure shows two adjacent two-dimensional faces, separated by a thick line, and a triangulation. The middle figure shows the associated ravioli complex in case the genus of the face on the left is zero and the genus of the face on the right is one, so that $\RV$ has two sheets over the right face.  The lower figure shows $\RV_D$ for $D$ the union of the four irreducible divisors $D_i$ associated with the points $p_i$ colored in red.}
 \end{center}
\end{figure}

\begin{figure}
\captionsetup{singlelinecheck = false, justification=justified}
\begin{center}
		\tdplotsetmaincoords{75}{30}
\begin{tikzpicture}[tdplot_main_coords]
	\coordinate (A) at (0, 0, 1.);
	\draw[fill=cornellBlue, fill opacity=0.1] (2, 2, 0)--(2, -2, 0)--(-2, -2, 0)--(-2, 2, 0) --(2, 2, 0);
 	\foreach \x in {-1.75, -1.5, ..., 1.75}
 	{
 		\draw[smooth, domain=-2:2, variable=\y, opacity=0.75] plot(\x, \y, {cos(45*\x)*cos(45*\y)});
 		\draw[smooth, domain=-2:2, variable=\y, opacity=0.5, densely dashed] plot(\x, \y, {-cos(45*\x)*cos(45*\y)});
 	}
	\begin{scope}[]
	\draw[smooth, domain=0:2, variable=\x] plot(\x, \x, {cos(45*\x)*cos(45*\x)});
	\draw[smooth, domain=0:2, variable=\x] plot(\x, -\x, {cos(45*\x)*cos(45*\x)});
	\draw[smooth, domain=0:2, variable=\x] plot(-\x, \x, {cos(45*\x)*cos(45*\x)});
	\draw[smooth, domain=0:2, variable=\x] plot(-\x, -\x, {cos(45*\x)*cos(45*\x)});
	\end{scope}
	\begin{scope}[dashed, opacity=0.75]
		\draw (2, 2, 0)--(0, 0, 0);
		\draw (2, -2, 0)--(0, 0, 0);
		\draw (-2, 2, 0)--(0, 0, 0);
		\draw (-2, -2, 0)--(0, 0, 0);
		\draw (0, 0, 0)--(A);
	\end{scope}
	\draw (A) node[shape=circle, fill=cornellRed, scale=0.25] {};
 	\draw[opacity=0.75] (0, 0, 0) node[shape=circle, fill=cornellRed, scale=0.25] {};
\end{tikzpicture}
	\end{center}
\caption{\label{raviolifig2} The ravioli complex over a 2-face.}
\end{figure}
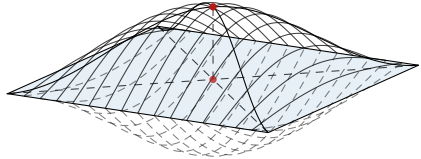

In general $\RV$ is not a simplicial complex, but instead a $\Delta$-complex:\footnote{The term $\Delta$-complex is standard in topology, and the symbol $\Delta$ appearing in the name should not be confused with the polytope $\Delta$. See \cite{Hatcher} for background on $\Delta$-complexes and CW complexes.} the $1$-cells and $2$-cells are not necessarily uniquely specified by the $0$-cells that contain them (in $X$).  The homology and cohomology complexes of $\RV$ are readily obtained, and the fact that $\RV$ is not always a simplicial complex does not present difficulties for computation or visualization.
The origin of the name should be clear from Figures \ref{raviolifig1} and \ref{raviolifig2}.

We can now associate a $\Delta$-complex ${\RV}_D \subseteq \RV$ to any square-free divisor $D$ as follows. The points of ${\RV}_D$ are the points (0-cells) $p_i$ in $\RV$ corresponding to the divisors appearing in \eqref{eq:decompD_G}. Some pairs (triples) of the $p_i$ may in general be connected by 1-cells (2-cells) of the complex $\RV$.  The  $p_i$, together with the set of all 1-cells and 2-cells connecting them, therefore define a unique subcomplex $\RV_D \subseteq \RV$. An example is shown in the bottom image in Figure \ref{raviolifig1}. The points in $\RV_D$ correspond to codimension-one subvarieties in $X$, while the higher-dimensional cells in $\RV_D$ encode intersections among the prime toric divisors. One-cells in $\RV_D$ correspond to the intersections of pairs of divisors in $X$, which are irreducible curves in $X$, and
2-cells in $\RV_D$ correspond to triple intersections of divisors in $X$. Note that a 2-cell in $\RV_D$ always corresponds to a single point on $X$.

\subsection{The puff complex $\mathscr{P}$}\label{generalized}

Let us briefly recapitulate.  By (\ref{hodgesubv}), the sheaf cohomology of prime toric divisors and their intersections is fully specified once one knows the simplicial complex $\mathcal{T}$ determined by the triangulation $\cal{T}$, together with the genera $g(\Theta^\circ)$ of the faces $\Theta^\circ \in {\cal{F}}(\le 2)$.
For $\Theta^\circ$ a 2-face, the number $g(\Theta^\circ)$ records the extent to which subvarieties corresponding to simplices contained in $\Theta^\circ$ are reducible.
By promoting the simplicial complex $\mathcal{T}$ to the $\Delta$-complex $\RV$, we have encoded the information about reducibility directly in the complex $\RV$.
Heuristically, viewed as  sets of data about subvarieties,
\begin{equation}
\RV \leftrightarrow \Bigl\{\mathcal{T},\bigl\{g(\Theta^\circ)|\Theta^\circ \in {\cal{F}}(2)\bigr\}\Bigr\}\,.
\end{equation}
The next step is to account for the data of $\{g(\Theta^\circ)|\Theta^\circ \in {\cal{F}}(\le 1)\}$.  In close analogy to the promotion $\mathcal{T} \to \RV$, we now define a CW complex\footnote{The property of CW complexes that is relevant here is that a $2$-cell or $1$-cell $\sigma$ can be attached to the complex by a map that identifies the boundary $\partial \sigma$ with a $0$-cell in the complex.} $\mathscr{P}$ that encodes this data.  Heuristically,
\begin{equation}
\mathscr{P} \leftrightarrow \Bigl\{\mathcal{T},\bigl\{g(\Theta^\circ)|\Theta^\circ \in {\cal{F}}(\le 2)\bigr\}\Bigr\}\,,
\end{equation} which is made precise by the following:

\medskip
\begin{definition} \label{puffdef}
The {\bf{puff complex}} $\mathscr{P}$ is a CW complex constructed from the ravioli complex $\mathscr{R}$ as follows. For each vertex $v$ and edge $e$ in $\mathscr{R}$, we have
natural inclusions $v \xhookrightarrow{} S^{2}$ and $e \xhookrightarrow{} S^{2}$, where the latter inclusion induces a cellular structure on $S^{2}$ with an $S^{1}$ attached to each interior 0-cell. These induce the inclusions $i_{v} \colon v \xhookrightarrow{} \bigvee\limits_{i=1}^{g(v)} S^{2}$ and $i_{e} \colon e \xhookrightarrow{} \mathscr{X}$ where $\mathscr{X}$ is defined by the following pushout diagram:
\begin{center}
\begin{tikzpicture}[baseline= (a).base]
\node[scale=1.2] (a) at (0,0){
\begin{tikzcd}
    \amalg_{i=1}^{g(e)}\,e\arrow{r}\arrow[hookrightarrow]{d}&e\arrow{d}{\varphi}\\
    \amalg_{i}\, S^{2} \ar{r}&\mathscr{X}
\end{tikzcd}
};
\end{tikzpicture}
\end{center}
Then $\mathscr{P}$ is defined by the following pushout diagram:
\begin{center}
\begin{tikzpicture}[baseline= (a).base]
\node[scale=1.2] (a) at (0,0){
\begin{tikzcd}
    \amalg_{i}\,v_{i}\amalg_{j}e_{j}\arrow{r}\arrow[hookrightarrow]{d}{\amalg i_{v}\, \amalg i_{e}}&\mathscr{R}\arrow{d}{\varphi}\\
    \amalg_{i}\bigvee S^{2}\amalg_{j}\mathscr{X} \ar{r}&\mathscr{P}
\end{tikzcd}
};
\end{tikzpicture}
\end{center}
The CW complex $\mathscr{P}_{D}$ associated to a divisor $D$ is the sub-complex of $\mathscr{P}$ corresponding to the image of $\mathscr{R}_{D}$ under the morphism $\varphi$.
\end{definition}
In other words, to construct $\mathscr{P}_{D}$, we attach a bouquet of $g(v)$ two-spheres to each vertex $v \in G$, a bouquet of $g(p)$ circles to each point $p \in G$ interior to an edge, and a bouquet of cylinders $S^1 \times I$ to each connected component of $\mathscr{R}_D$ restricted to $e$.  These spaces are glued together along their common points: in particular, the $g(e)$ cylinders over a complete edge $e \subset \mathscr{R}_D$ are pinched down into the two vertices bounding $e$, forming a collection of $g(e)$ two-dimensional voids.  

\medskip\noindent We will find it useful to divide the puff complex into layers:
\begin{definition}
For $0\le j \le 2$, the $j$th {\bf{layer}} $\mathscr{P}^{(j)}$ of the puff complex $\mathscr{P}$ is the subset of $\mathscr{P}$ resulting from replacing points interior to each $(2-j)$-face $\Theta^\circ$ of $\Delta^\circ$ with $g(\Theta^\circ)$ $j$-spheres, replacing 1-simplices interior to $\Theta^\circ$ with $g(\Theta^\circ)$ cylinders $S^j \times I$, etc., as in Definition \ref{puffdef}.  In particular, $\mathscr{P}^{(0)}=\mathscr{R}$.
\end{definition}
\begin{figure}
  \begin{center}
   \scalebox{.5}{ 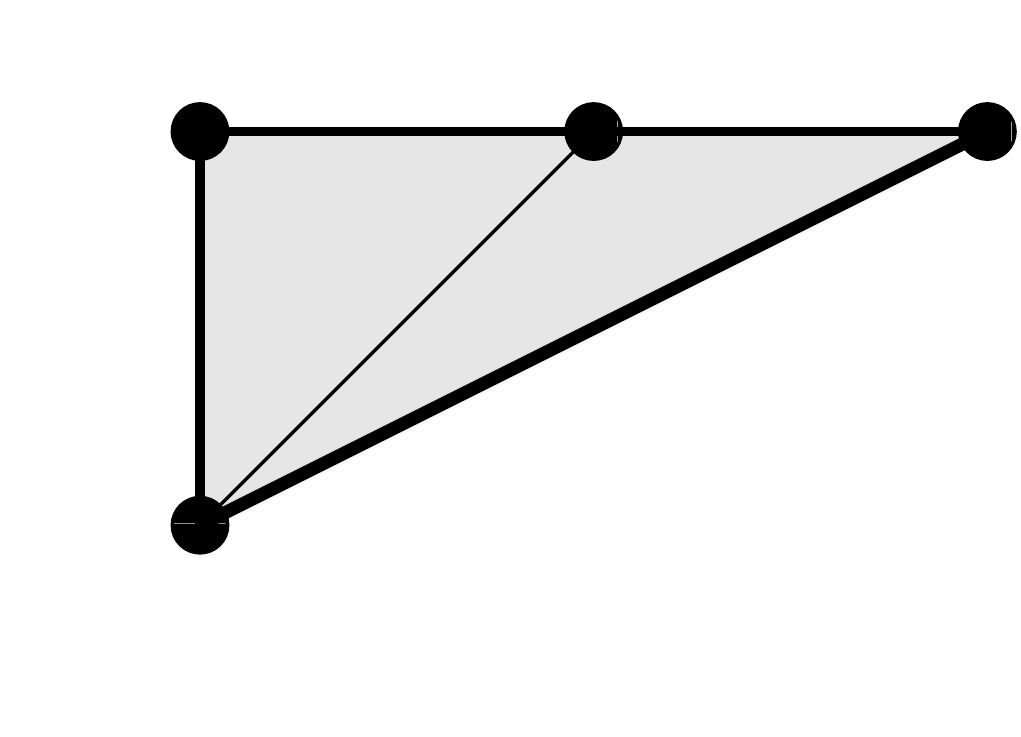 } \\
  \includegraphics[height=4cm]{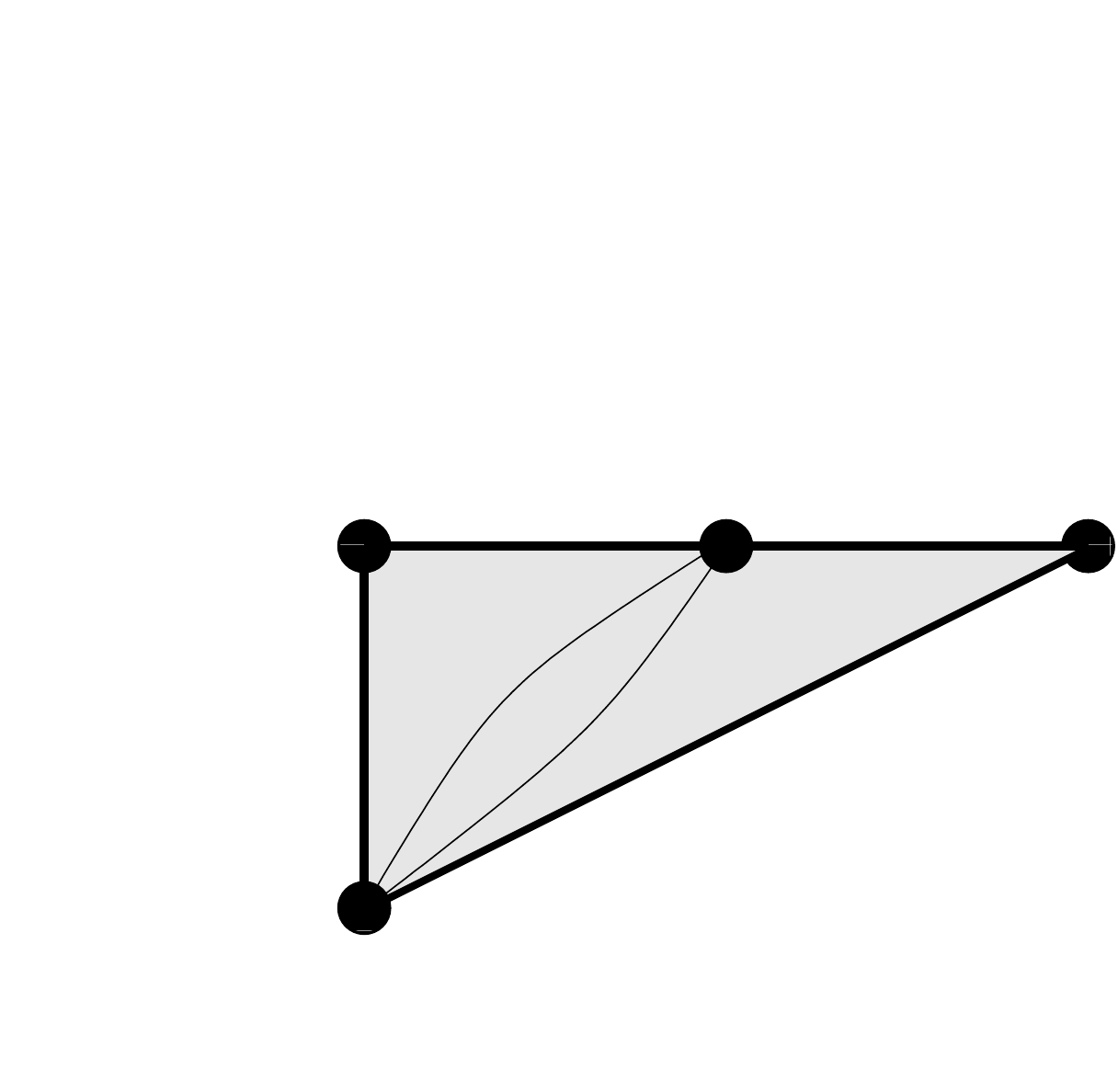}
  \includegraphics[height=4cm]{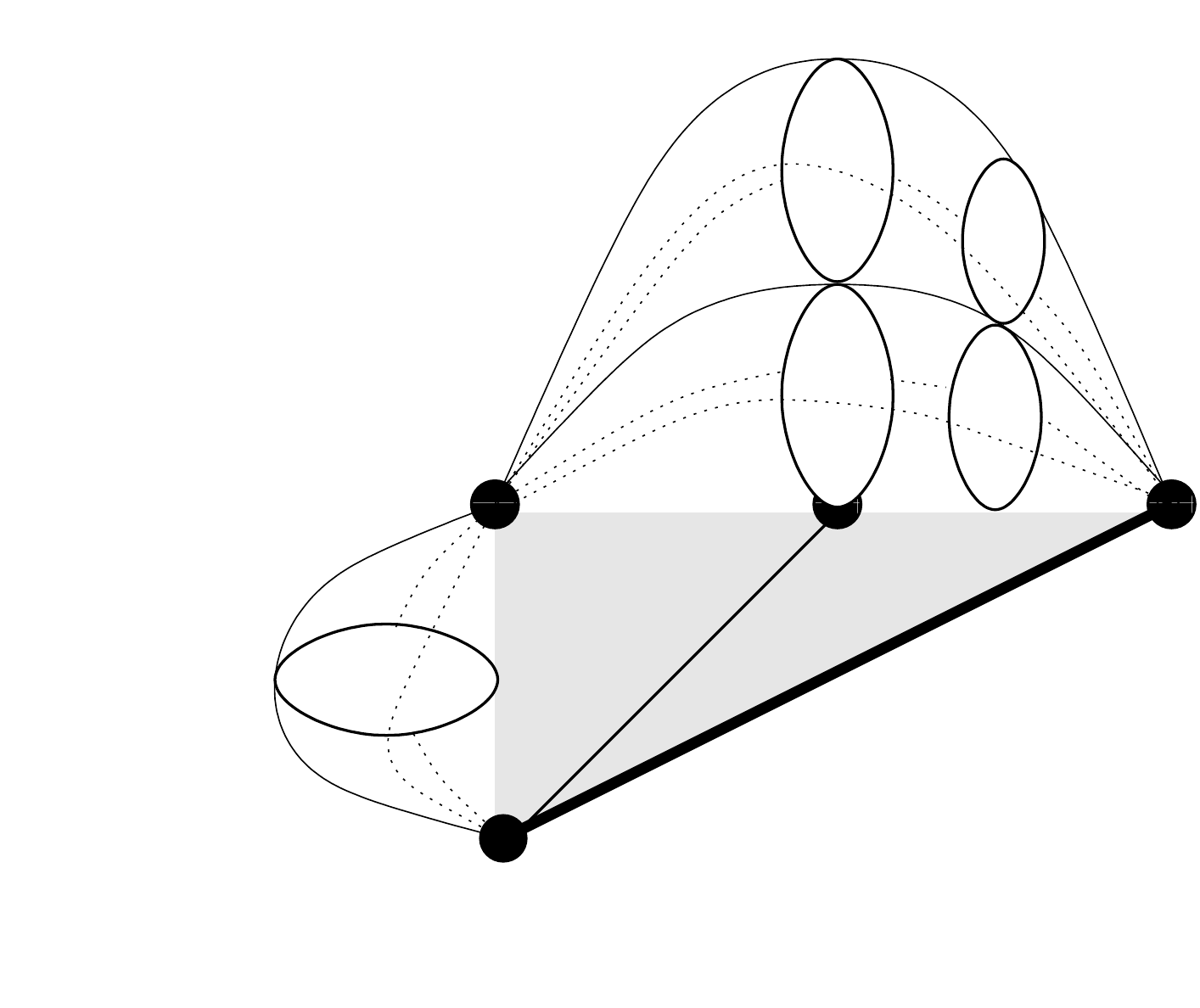}
  \includegraphics[height=4cm]{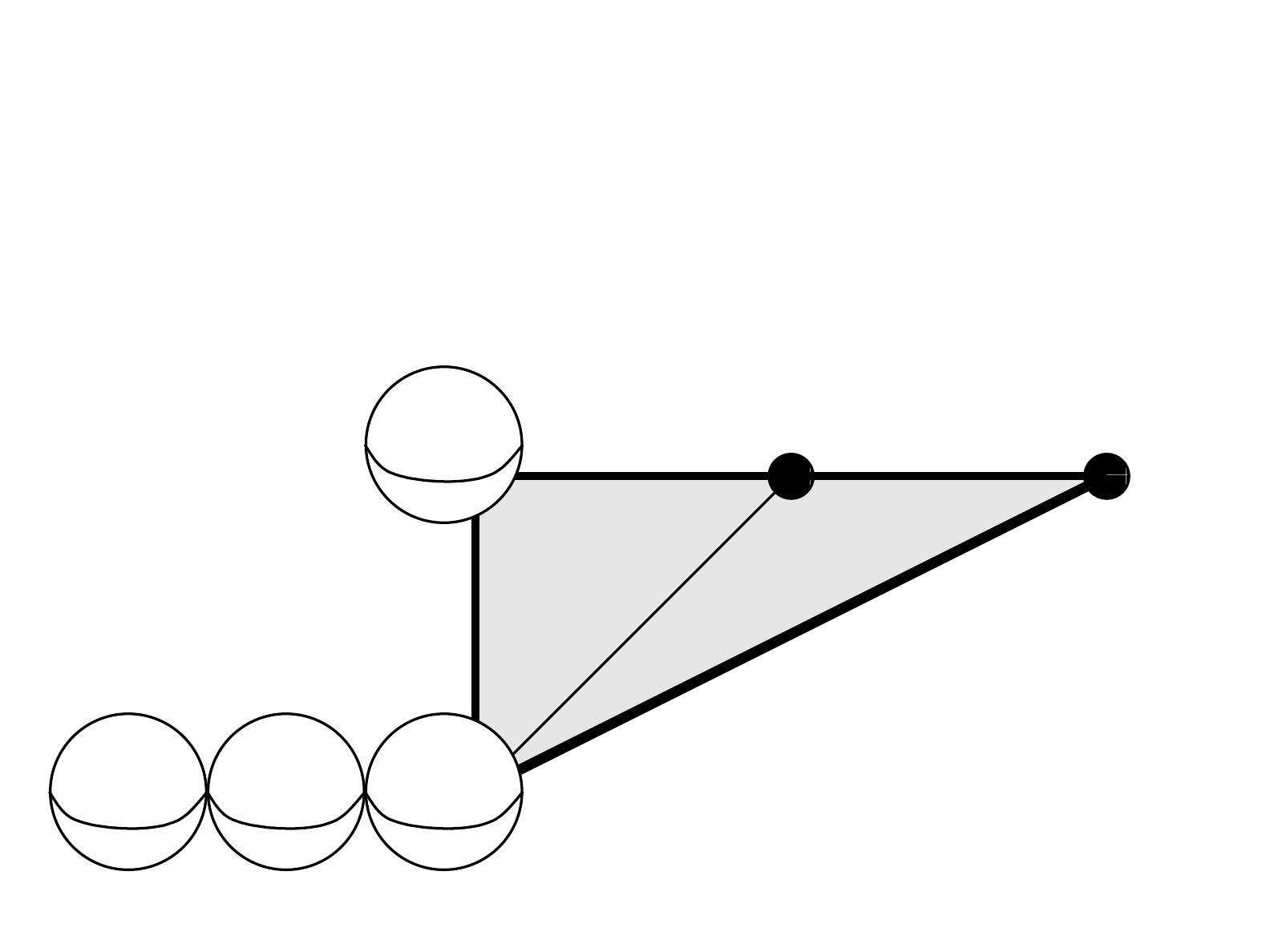}
  \caption{\label{fig:pufflayers}The layers of the puff complex for a single face.
  The lower row shows the three layers of $\mathscr{P}$ for the triangulated face at the top of the figure.  In this example, the genera of the various faces are such that $g(\Theta^{\circ [2]})=1$, $g(\Theta^{\circ [1]}_i)=(2,1,0)$ and  $g(\Theta^{\circ [0]}_i)=(1,3,0)$.}
 \end{center}
\end{figure}
Figure \ref{fig:pufflayers} contains a sketch of the different layers of the puff complex for a simple example.

The homology of $\mathscr{P}_{D}$ is readily obtained.  Distinct connected components of $\mathscr{P}_{D}$ can be examined separately, so we may take $h_0(\mathscr{P}_{D})=1$ without loss of generality.
Contributions to $h_1(\mathscr{P}_{D})$ come from
\begin{enumerate}[(a)]
\item{}One-cycles in $\mathscr{R}_{D}$.
\item{}For each edge $e \subset \Delta^{\circ}$, a bouquet of $g(p)$ cylinders over each connected component of ${\mathscr{R}_D}\bigl|_e$ that is strictly interior to $e$.
\end{enumerate}
Contributions to $h_2(\mathscr{P}_{D})$ come from
\begin{enumerate}[resume*]
\item{}Two-cycles in $\mathscr{R}_{D}$.
\item{}The bouquet of $g(e)$ pinched cylinders over each edge $e$ such that ${\mathscr{R}_D}\bigl|_e=e$.
\item{}The bouquet of $g(v)$ two-spheres over each vertex $v$, i.e.~the layer $\mathscr{P}^{(2)}$.
\end{enumerate}
We will prove in \S\ref{sec:spectral} that these classes of contributions are in one-to-one correspondence with classes of contributions in the hypercohomology spectral sequence that computes the cohomology of ${\cal O}_D$.  In other words, the homology of the CW complex $\mathscr{P}_{D}$ encodes the cohomology of ${\cal O}_D$.  As we will see, this correspondence has both computational and heuristic utility.

\section{Spectral Sequence Computation of Hodge Numbers}\label{sec:proofsec}
In this section we will prove our principal result:
\begin{theorem} \label{MasterTheorem}
Let $D=\sum_{i\in G} D_i$, with $G \subseteq \{1,\ldots,N\}$, be a square-free divisor, as defined above.  Denote by $\Pf$ and
$\Pf_D$
the puff complex
and the associated subcomplex determined by $D$, respectively.
Then the Hodge numbers $h^i({\cal{O}}_D)$ are given by
\begin{equation} \label{MasterFormulaGenRav}
h^i({\cal{O}}_D) = h_i(\Pf_D) \,.
\end{equation}
 \end{theorem}

In the rest of this section, we prove Theorem~\ref{MasterTheorem}. Our method is based on examining a hypercohomology spectral sequence.
We use stratification to identify the $E^2$ page of that sequence, and then we use a somewhat indirect argument to show that the differentials
on the $E^2$ page are all zero.

\subsection{Hypercohomology spectral sequence}\label{sec:spectral}

To start, let $\{D_1, \ldots, D_N\}$ be the prime toric divisors, as defined in \S2.
Given
a square-free toric divisor in $X$,
\[ D := \sum_{i\in G} D_i,\]
we will compute the cohomology $h^\bullet(\cO_D)$.  Suppose that we have indexed the $D_i$ so that $D = \sum_{i=1}^{r} D_i$, for some $r \le N$.

The set of divisors $\{D_1, \ldots, D_r\}$
is dimensionally transverse on a smooth variety, and so by Proposition~\ref{MVprop}
of Appendix \ref{sec:mvc}, the
generalized Mayer-Vietoris sequence
  \begin{equation}\label{eq:mvcy}
0 \longrightarrow \cO_{D}
	\longrightarrow \bigoplus_{i=1}^r \cO_{D_i}
	\longrightarrow \bigoplus_{i<j} \cO_{D_{ij}}
	\longrightarrow \bigoplus_{i<j<k} \cO_{D_{ijk}}
	\longrightarrow
	0
\end{equation}
is an exact sequence of sheaves.  The hypercohomology spectral sequence of this complex will allow us to
compute the cohomology of $\cO_D$.

Define
\begin{align}
&F_0 := \bigoplus_{i=1}^r \cO_{D_i}\, , \nonumber \\
&F_1 :=  \bigoplus_{i<j} \cO_{D_{ij}}\, , \nonumber \\
&F_2 := \bigoplus_{i<j<k} \cO_{D_{ijk}}\, .
\end{align}
The complex of sheaves (\ref{eq:mvcy})
gives rise to a hypercohomology spectral sequence $E^r_{p,q}(D)$ with differentials denoted by
  \[ \delta^r_{p,q \rightarrow p',q'} : E^r_{p,q}(D) \longrightarrow E^r_{p',q'}(D) \]
  where $p' = p+r$ and $q' = q - r + 1$.
  The first page of this spectral sequence is
  \begin{center}
\begin{tikzpicture}[descr/.style={fill=white,inner sep=1.5pt}]
        \matrix (m) [
            matrix of math nodes,
            row sep=1em,
            column sep=2.5em,
            text height=1.5ex, text depth=0.25ex
        ]
        {
           H^2(F_0)  & 0& 0\\
           H^1(F_0) & H^1(F_1)  &0\\
           H^0(F_0) &H^0(F_1) &H^0(F_2)\\
                 };

        \path[overlay,->, font=\scriptsize,>=latex]
        (m-1-1) edge (m-1-2)
        (m-1-2) edge (m-1-3)
         (m-2-1) edge (m-2-2)
        (m-2-2) edge (m-2-3)
          (m-3-1) edge (m-3-2)
        (m-3-2) edge (m-3-3);
        \node at (-1.3,.3) {$\alpha$};
        \node at (-1.3,-.7) {$\beta$};
        \node at (1.3,-.7) {$\gamma$};
\end{tikzpicture}
\end{center}
These are indexed such that $E^1_{p,q}(D) = H^q(F_p)$, and e.g. $\delta^1_{0,1 \rightarrow 1,1} = \alpha$.

The second page, $E^2(D)$  reads

\begin{center}
\begin{tikzpicture}[descr/.style={fill=white,inner sep=1.5pt}]
        \matrix (m) [
            matrix of math nodes,
            row sep=1em,
            column sep=2.5em,
            text height=1.5ex, text depth=0.25ex
        ]
        {
           0 & H^2(F_0)  & 0& 0&0 \\
           0 & \ker \alpha & \coker \alpha  &0&0 \\
           0 & \ker \beta & \ker \gamma/\im\beta &\coker \gamma & 0\\
                 };

        \path[overlay,->, font=\scriptsize,>=latex]
        (m-2-2) edge  (m-3-4);
       \node at (1.3,-.4) {$\zeta$};
\end{tikzpicture}
\end{center}
where $\zeta := \delta^2_{0,1 \rightarrow 2,0}$ is the only nonzero differential on this page.

The third and final page reads
\begin{center}
\begin{tikzpicture}[descr/.style={fill=white,inner sep=1.5pt}]
        \matrix (m) [
            matrix of math nodes,
            row sep=1em,
            column sep=2.5em,
            text height=1.5ex, text depth=0.25ex
        ]
        {
           0 & H^2(F_0)  & 0& 0&0 \\
           0 & \ker \zeta & \coker \alpha  &0&0 \\
           0 & \ker \beta & \ker \gamma/\im\beta &\coker \zeta & 0\\
                 };
\end{tikzpicture}
\end{center}
On this page, the differential $\delta^3$ is zero.

The hypercohomology spectral sequence converges to the cohomology of $\cO_D$, meaning that
\begin{eqnarray}
\dim H^0(\mathcal{O}_D) &=& \dim\ker\beta \\
\dim H^1(\mathcal{O}_D) &=& \dim\ker\zeta +  \dim (\ker \gamma/\im\beta) \\
\dim H^2(\mathcal{O}_D) &=& \dim H^2(F_0) + \dim \coker \alpha + \dim \coker \zeta
\end{eqnarray}
In the special case that $\zeta$ is the zero map, we have the simplifications
\begin{eqnarray}
\dim H^1(\mathcal{O}_D) &=& \dim \ker\alpha +  \dim (\ker \gamma/\im\beta) \\
\dim H^2(\mathcal{O}_D) &=& \dim H^2(F_0) + \dim \coker \alpha + \dim \coker \gamma
  \end{eqnarray}
If the dependence on a divisor is understood, we denote the maps $\alpha$, $\beta$, $\gamma$, $\zeta$, while if a divisor $D$ is specified, we write
$\alpha(D)$, etc.

\subsection{Cohomology from stratification}

We now use stratification to compute the cohomology organized by the spectral sequence given above.

\begin{proposition} \label{SSprop1}
  Row 0
  of page 1 of the hypercohomology spectral sequence has cohomology equal to that of the ravioli complex ${\mathscr{R}}_D = \Pf^{(0)}_D$, that is:
  \begin{eqnarray}
    \dim \ker \beta &=& \dim H_0({\mathscr{R}}_D) \\
    \dim (\ker \gamma/\im\beta) &=& \dim H_1({\mathscr{R}}_D) \\
    \dim \coker \gamma &=& \dim H_2({\mathscr{R}}_D)
    \end{eqnarray}
\end{proposition}
Note: this can be written more succinctly in the form
\begin{equation}
 \dim E^2_{p,0}(D) = \dim H_p(\Pf^{(0)}_D)\,,
\end{equation}
for all $p$, even though $\dim E^2_{p,0}(D)$ can only be nonzero for $p=0,1,$ or $2$.
\medskip

\noindent{\bf Proof.}
We will identify row 0
of page 1 of the spectral sequence with the cohomology complex of the complex $\RV_D$.
In the following, we will take $\RV_D(k)$ to be the set of k-faces of $\RV_D$. In particular,
$\RV_D(0)$ is the set of 0-dimensional faces (i.e. the $D_i$), $\RV_D(1)$ is the set of 1-dimensional faces,
i.e.~one for each connected component of a $D_i \cap D_j$.
The same connected component $C$ cannot occur for two different such
intersections, as otherwise the component $C$ would be contained in the intersection of three or four of the $D_i$, which cannot happen since
the $D_i$ are dimensionally transverse.  Finally, $\RV_D(2)$ corresponds to the points in each of the triple intersections of three divisors,
$D_i \cap D_j \cap D_k$.  Again, each point can only appear in one such intersection. Notice that $\RV_D$
is a Delta complex whose cohomology sequence is
  \begin{equation}\label{eq:ravcpx}
0 \longrightarrow \bigoplus_{i=1}^r \C e_{D_i}
	\stackrel{\delta_0}{\longrightarrow} \bigoplus_{C \in \RV_D(1)} \C e_C
	\stackrel{\delta_1}{\longrightarrow} \bigoplus_{P \in \RV_D(2)} \C e_P
	\longrightarrow
	0
\end{equation}
  where the maps $\delta_a$ are as follows: if $C$ is a connected
  component of $D_i \cap D_j$, for $i < j$, $(\delta_0(e_{D_i}))_C =
  -1$ and $(\delta_0(e_{D_j}))_C = 1$, and all other components are
  zero.  If $C$ is an irreducible component of $D_i \cap D_j$, and if $P$ is a point of the intersection $D_i \cap D_j \cap
  D_k$, (for $i < j$), then
  \[ (\delta_1(e_C))_P = {\rm{sign}}(i,j,k). \]

Now consider the complex of row 0 in page 1 of the
spectral sequence.  We see that $H^0(F_0) = \bigoplus_{i=1}^r \C
  e_{D_i}$, $H^0(F_1) = \bigoplus_{C \in \RV_D(1)} \C e_C$, and
  $H^0(F_2) = \bigoplus_{P \in \RV_D(2)} \C e_P$, i.e.~the
  complex
  \[
\begin{tikzcd}
0 \arrow{r}& H^{0}(F_{0}) \arrow{r}{\beta}& H^{0}(F_{1}) \arrow{r}{\gamma}& H^{0}(F_{2}) \arrow{r}& 0
\end{tikzcd}
\]
and the complex (\ref{eq:ravcpx})
have the same terms.  Under this identification, it is easy to see
  that the maps are identical too, as the morphisms $\beta$ and $\gamma$ are canonically induced by the generalized Mayer-Vietoris complex.
  $\square$
\bigskip

Proposition \ref{SSprop1} generalizes to all rows of the second page:

\begin{proposition} \label{e2pqprop}
  For all $p$ and $q$,
  \[ \dim E^2_{p,q}(D) = \dim H_{p+q}(\Pf^{(q)}_D). \]
\end{proposition}

\noindent{\bf Proof.}
We have already established the case $q=0$.  For $q=2$, we note that stratification gives $h^2(\cO_{D_i}) = 0$, unless $D_i$ corresponds to a vertex of $\Delta^{\circ}$ having positive genus $g(v)$. Summing these gives
\begin{equation}
    \dim H^2(F_0) = \sum\limits_{\substack{v \in {\cal F}(0)\\ v \in D}} g(v)\,,
    \end{equation}
proving Proposition \ref{e2pqprop} for $q=2$.

The case $q=1$ can be established by noting that row 1 of page 1 of
the spectral sequence, because of stratification, breaks up into a
direct sum of complexes, summed over all edges $e$ in $\Delta^\circ$
intersecting $D$:
\begin{equation} \bigoplus_{e \in \mathcal{F}(1)} \C^{g(e)} \otimes \left( 0 \longrightarrow \bigoplus_{\tau \in \mathcal{T}_{D \cap e}(0), \tau \notin \partial e} \C \longrightarrow
  \bigoplus_{\tau \in \mathcal{T}_{D \cap e}(1)} \C \longrightarrow 0
  \right)  \label{q1directsum}
\end{equation}
The complex in parentheses is simply the relative cochain complex
for the pair $(\mathcal{T}_{D \cap e}, \mathcal{T}_{D \cap \partial
  e})$.  Note that because $e$ is an edge, $(\mathcal{T}_{D \cap e}, \mathcal{T}_{D \cap \partial
  e}) = (\RV_{D \cap e}, \RV_{D \cap \partial e})$.

Therefore the $p$-th cohomology (for $p=0,1$) of the direct sum of
complexes \eqref{q1directsum} is the relative cohomology $\bigoplus_e \C^{g(e)} \otimes
H^p(\RV_{D \cap e}, \RV_{D \cap \partial e})$, which is the same as
$\bigoplus_e \C^{g(e)} \otimes H_p(\RV_{D \cap e}, \RV_{D \cap
  \partial e})$, which by the definition of $\mathscr{P}$ equals
$H_{p+1}(\Pf^{(1)}_D)$.  $\square$

\begin{corollary}~\label{cor:SSpuff}
  Given a square-free divisor $D$, if $\zeta(D) = \delta^2_{0,1 \rightarrow 2,0}(D)$ then
  \begin{align}
    \zeta(D) = 0 \iff& \dim H^i(\cO_D) = \dim H_i(\Pf_D) \mbox{ for all $i$} \\
    \iff& \dim H^1(\cO_D) = \dim H_1(\Pf_D)\\
    \iff& \dim H^2(\cO_D) = \dim H_2(\Pf_D)\,.
    \end{align}
  \end{corollary}

\subsection{Proof that $\zeta=0$}\label{sec:delta}

In this section we will prove the following, which also, by Corollary~\ref{cor:SSpuff}, proves Theorem~\ref{MasterTheorem}.

\begin{lemma} \label{zetazero}
  If $D$ corresponds to a square-free divisor on $X$, then the map
  \[ \zeta(D) = \delta^2_{0,1 \rightarrow 2,0}(D)\]
is the zero map.
\end{lemma}

We prove this result by first splitting $D = A+B_1 +\ldots + B_m$, where $A$ and the $B_{i}$ are all square-free
divisors having disjoint support.  Consider the subgraph 
$\mathcal{T}_D(\le 1)$.  Suppose that this graph has $m$ components each of
which is contained in the strict interior of an edge of $\Delta^\circ$.
Let $B_{i}$ be the sum of the divisors corresponding to the lattice points in
the $i$th component.  Let $A$ be the sum of the rest of the divisors.
Define $A^{j} := A + \sum_{i=1}^{j} B_i$, with $A^0 = A$ and $A^{m}=D$.

\begin{lemma}\label{zeta1}
  The map $\zeta(A)$ is zero, and so $\dim H^i(\cO_{A}) = \dim H_i(\Pf_{A})$.
  \end{lemma}

\noindent{\bf Proof.}
The map is zero, because its domain, $E^2_{0,1}(A)$, which has dimension $\dim H_1(\Pf^{(1)}_{A})$, is zero by construction.
$\square$

\begin{lemma}\label{zeta2}
  $\dim H^2(\cO_{B_{i}}) = 0$.
  \end{lemma}

\noindent{\bf Proof.}
If the given component $B_{i}$ consists of divisors associated to lattice points interior to
an edge of genus $g$, then $h^{\bullet}(\cO_{B_{i}})=(1, g, 0)$.
$\square$

\begin{lemma}\label{zeta3}
  Fix an integer $1 \le i \le m$, and
  let $C := A^{i-1} \cap B_i$.  Then $C$ is a curve whose irreducible components are all $\PP^1$'s,
  two intersect in at most one point, and $H^1(\cO_C) = 0$.
  \end{lemma}

\noindent{\bf Proof.}
$C := A^{i-1} \cap B_{i}$ is a collection of rational curves
$C = \cup_M C_M$.  As a
complex, $C$ corresponds to a collection of 1-simplices $\sigma_{M}$,
each with one end in $\mathscr{R}_{B_i}$, and the other end in
an edge not containing $\RV_{B_i}$, in a 2-face, or in a vertex.
Two curves $C_{M}$ and $C_{N}$ intersect if and only if $\sigma_{M}$ and
$\sigma_{N}$ share a 2-simplex $\sigma_{M,N}$.

Now notice that $h^1({\cal{O}}_C) \neq 0$ is possible only if the
$C_{M}$ intersect in such a way as to form a nontrivial loop.  But
this would require a closed loop in which three or more 1-simplices
$\sigma_1,\ldots, \sigma_K$ are connected by 2-simplices $\sigma_{1,2},\ldots,
\sigma_{K-1,K}, \sigma_{K,1}$.  However, because $\mathscr{R}_{B_{i}}$ is in
the strict interior of an edge, no such loop can be formed.  Thus,
$h^1({\cal{O}}_C)=0$.
$\square$

\begin{lemma}\label{zeta4}
  $\dim H^2(\cO_{A}) = \dim H^2(\cO_D)$.
  \end{lemma}

\noindent{\bf Proof.}
We prove this by induction.  Consider $A^{i} = A^{i-1} + B_{i}$, where $A^{0} = A$.
We use induction to show that $\dim H^2(\cO_{A}) = \dim H^2(\cO_{A^{i}})$, for $i = 0, \ldots, m$.
The statement is trivial for $i = 0$, and for $i > 0$, we use the long exact sequence
associated to the Mayer-Vietoris sequence
\[ 0 \longrightarrow \cO_{A^{i}} \longrightarrow \cO_{A^{i-1}} \oplus \cO_{B_{i}} \longrightarrow \cO_C \longrightarrow 0.\]
combined with Lemmas~\ref{zeta2} and \ref{zeta3}.
One obtains that $H^2(\cO_{A^{i}}) = H^2(\cO_{A^{i-1}})$.
$\square$

\begin{lemma}\label{zeta5}
  $\dim H^2(\Pf_{A}) = \dim H^2(\Pf_D)$.
\end{lemma}

\noindent{\bf Proof.}
There are four cases where an element of $H^2$ can appear: $D$ contains a vertex of genus $>0$,
$D$ contains an entire edge of genus $>0$, $D$ contains at least two full sheets over a 2-face, and finally,
$D$ contains a collection of full 2-faces that contain a void.
Since by definition, not all of the lattice points of the edge containing $B_{i}$ are in $D$, and $B_{i}$ contains no vertices, $\Pf_{A}$ and $\Pf_{D}$ must have the same $H^2$.
$\square$

\bigskip

\noindent{\bf Proof of Lemma~\ref{zetazero}, and therefore also the proof of Theorem~\ref{MasterTheorem}.}
Lemmas~\ref{zeta1}, \ref{zeta4}, and \ref{zeta5} imply that
\[ \dim H^2(\Pf_D) = \dim H^2(\Pf_{A}) = \dim H^2(\cO_{A}) = \dim H^2(\cO_D), \]
and so by Corollary~\ref{cor:SSpuff}, we have $\zeta(D) = 0$, proving the theorem.
$\square$

\subsection{Generalization to Calabi-Yau $n$-folds}\label{sec:fourfoldnfold}

Most aspects of the computation above generalize immediately to Calabi-Yau $n$-folds.
\begin{definition} \label{GeneralizationofMasterDef}
Let $V$ be a simplicial toric variety of dimension $n+1$, and let $X$ be a smooth Calabi-Yau $n$-fold hypersurface in $V$, with $N_n:=h^{1,1}(X)+n+1$.
Let $D=\sum_{i\in G} D_i$, with $G \subseteq \{1,\ldots,N_n\}$, be a square-free divisor.
The construction of Definition \ref{puffdef} generalizes to any $n \ge 3$, and we denote by $\Pf$ and $\Pf_D$ the puff complex and the associated subcomplex determined by $D$, respectively.
\end{definition}
We note that the constructions of $\Pf$ and $\Pf_D$ are immediate because our results on the stratification of subvarieties $X_{\sigma} \subset X$ apply for any $n$.

Next, the Mayer-Vietoris sequence generalizing (\ref{eq:mvcy}) contains $n+1$ nonzero terms.
The resulting hypercohomology spectral sequence again converges to the cohomology of $\cO_D$, and it is easy to see that
row zero of page one of the hypercohomology spectral sequence again has cohomology equal to that of
$\RV_D$.  Moreover, Proposition 5 holds for arbitrary $n$ \cite{wip}.
However, for $n>3$ there are more diagonal maps, generalizing $\zeta$.

If for every $n$ the diagonal maps were shown to be identically zero, as we have proved in the case $n=3$, then we would have $h^i({\cal{O}}_D) = h_i(\Pf_D)$ for all $n$.
However, examining the diagonal maps directly in the same manner as done for $\zeta$ in \S\ref{sec:delta} would be somewhat involved.  We defer a proper analysis of these maps to \cite{wip}, and state here only the following:
\begin{conjecture} \label{GeneralizationofMasterTheorem}
Let $V$, $X$, $D$, and $\Pf_D$ be as in Definition \ref{GeneralizationofMasterDef}, and set $n=4$.
For each such $D$ there exists $k(D) \in \mathbb{Z}_{\ge 0}$ such that
\begin{equation} \label{GeneralizationofMasterFormulaGenRav}
h^\bullet({\cal{O}}_D) = h_\bullet(\Pf_D) - (0,k(D),k(D),0)\,.
\end{equation}
\end{conjecture}
\medskip
We show in \cite{wip} that $k(D)=0$ for all $D$ except those obeying a rather restrictive condition.

\section{Interpretation and an Example}  \label{sec:example}

We will now briefly discuss the interpretation of our result, and then illustrate the utility of our formula in an example.

\subsection{Contractible graphs} \label{ss:contraction}

Equation~\ref{MasterFormulaGenRav} depends on only certain topological properties of the complex $\Pf_D$,
and so two complexes that correspond to divisors that are related to one another by deformations that preserve these topological properties will have identical Hodge numbers. This leads to a useful tool in enumerating certain divisors, as we now explain.

Given a triangulation ${\cal T}$ and the associated simplicial complex ${\mathcal{T}}$, a square-free divisor $D$ determines a unique simplicial complex ${\mathcal{T}}_D \subset {\mathcal{T}}$, as well as a corresponding CW complex $\Pf_D \subset \Pf$.
The number ${\cal{N}}$ of distinct square-free divisors of a given puff complex $\Pf$ that corresponds to a Calabi-Yau hypersurface $X$ is $2^{h^{1,1}(X) + 4}$.  Thus, the task of working out the Hodge numbers of all possible square-free divisors appears formidable for large $h^{1,1}(X)$.
Fortunately, as we will now explain, Theorem~\ref{MasterTheorem} implies that square-free divisors fall into equivalence classes.

Suppose that we begin with a set of lattice points $G_1 \subset \{1,\ldots,N\}$, which define a square-free divisor $D_1$ and a CW complex $\Pf_{G_1}$.  Now let us add or remove one lattice point from $G_1$, so that $G_1$ is changed to some $G_2 \subset \{1,\ldots,N\}$.  This operation uniquely defines a new square-free divisor $D_2$ and a new CW complex $\Pf_{G_2}$.
By Equation~(\ref{MasterFormulaGenRav}), if $h_i(\Pf_{G_1})=h_i(\Pf_{G_2})$ then
$h^i({\cal{O}}_{D_1})=h^i({\cal{O}}_{D_2})$.

We define a {\bf{single contraction}} to be the operation of changing $D_1\to D_{2}$ by adding or removing a lattice point, as specified above, in such a way that $h^i(\Pf_{D_1})=h^i(\Pf_{D_{2}})$.
We define a {\bf{contraction}} to be an arbitrary composition of single contractions. Contraction is an equivalence relation on square-free divisors, and all members of a contraction equivalence class have the same Hodge numbers.  However, two divisors with the same Hodge numbers do not necessarily belong to the same contraction equivalence class.

\begin{figure}[ht]\label{fig:cont1}
  \centering
  \begin{tikzpicture}[scale=.52]
    \draw[line width=0.3mm] (-2.1,-2.1) grid[step=2cm] (4,4);
    \draw[line width=0.3mm] (-2,4) -- (0,2);
    \draw[line width=0.3mm] (0,4) -- (2,2);
    \draw[line width=0.3mm] (2,4) -- (4,2);
    \draw[line width=0.3mm] (-2,2) -- (0,0);
    \draw[line width=0.3mm] (-2,0) -- (0,-2);
    \draw[line width=0.3mm] (0,0) -- (2,-2);
    \draw[line width=0.3mm] (0,2) -- (2,0);
    \draw[line width=0.3mm] (2,2) -- (4,0);
     \draw[line width=0.3mm] (2,0) -- (4,-2);
   \foreach \x in {-1,...,2}{
      \foreach \y in {-1,...,2}{
        \node[draw,circle,inner sep=2pt,fill] at (2*\x,2*\y) {};
      }
    }
    \node[draw,circle,inner sep=2pt,fill,cornellRed] at (0,0) {};
    \node [label={[xshift=-.3cm, yshift=-.75cm]\footnotesize{$D_1$}}] at (0,0) {};
     \node [label={[xshift=-.3cm, yshift=-.75cm]\footnotesize{$D_2$}}] at (0,2) {};
       \node [label={[xshift=-.3cm, yshift=-.75cm]\footnotesize{$D_3$}}]  at (2,0) {};
     \node[draw,circle,inner sep=2pt,fill,cornellRed] at (2,0) {};
      \node[draw,circle,inner sep=2pt,fill,cornellRed] at (0,2) {};
     \draw[->,line width=1.3pt] (4.5,1) -- (7.8,1);
      \draw [line width=0.4mm, draw=cornellRed, fill=cornellRed, fill opacity=0.2]
       (0,0) -- (0,2) -- (2,0) -- cycle;
	 \path[->]
           (4.5,1)    edge  node[sloped, anchor=center, above, text width=2.0cm] { \footnotesize{$\, \, \, D_2 \rightarrow D_1$}}      (7.8,1);
  \end{tikzpicture}
   \begin{tikzpicture}[scale=.52]
          \draw[line width=0.3mm] (-2.1,-2.1) grid[step=2cm] (4,4);
    \draw[line width=0.3mm] (-2,4) -- (0,2);
    \draw[line width=0.3mm] (0,4) -- (2,2);
    \draw[line width=0.3mm] (2,4) -- (4,2);
    \draw[line width=0.3mm] (-2,2) -- (0,0);
    \draw[line width=0.3mm] (-2,0) -- (0,-2);
    \draw[line width=0.3mm] (0,0) -- (2,-2);
    \draw[line width=0.3mm] (0,2) -- (2,0);
    \draw[line width=0.3mm] (2,2) -- (4,0);
     \draw[line width=0.3mm] (2,0) -- (4,-2);
    \foreach \x in {-1,...,2}{
      \foreach \y in {-1,...,2}{
        \node[draw,circle,inner sep=2pt,fill] at (2*\x,2*\y) {};
      }
    }
    \node[draw,circle,inner sep=2pt,fill,cornellRed] at (0,0) {};
    \node [label={[xshift=-.3cm, yshift=-.75cm]\footnotesize{$D_1$}}] at (0,0) {};
       \node [label={[xshift=-.3cm, yshift=-.75cm]\footnotesize{$D_3$}}]  at (2,0) {};
     \node[draw,circle,inner sep=2pt,fill,cornellRed] at (2,0) {};
     \draw[->,line width=1.3pt] (4.5,1) -- (7.8,1);
         \draw[->,line width=1.3pt] (4.5,1) -- (7.8,1);
          \draw [line width=0.4mm, draw=cornellRed]
       (0,0) -- (2,0) -- cycle;

          \path[->]
           (4.5,1)    edge  node[sloped, anchor=center, above, text width=2.0cm] { \footnotesize{$\, \, \, D_3 \rightarrow D_1$}}      (7.8,1);
  \end{tikzpicture}
     \begin{tikzpicture}[scale=.52]
        \draw[line width=0.3mm] (-2.1,-2.1) grid[step=2cm] (4,4);
      \draw[line width=0.3mm] (-2,4) -- (0,2);
    \draw[line width=0.3mm] (0,4) -- (2,2);
    \draw[line width=0.3mm] (2,4) -- (4,2);
    \draw[line width=0.3mm] (-2,2) -- (0,0);
    \draw[line width=0.3mm] (-2,0) -- (0,-2);
    \draw[line width=0.3mm] (0,0) -- (2,-2);
    \draw[line width=0.3mm] (0,2) -- (2,0);
    \draw[line width=0.3mm] (2,2) -- (4,0);
     \draw[line width=0.3mm] (2,0) -- (4,-2);
      \foreach \x in {-1,...,2}{
      \foreach \y in {-1,...,2}{
        \node[draw,circle,inner sep=2pt,fill] at (2*\x,2*\y) {};
      }
    }
    \node[draw,circle,inner sep=2pt,fill,cornellRed] at (0,0) {};
    \node [label={[xshift=-.3cm, yshift=-.75cm]\footnotesize{$D_1$}}] at (0,0) {};
  \end{tikzpicture}

  \caption{Example of contraction for a graph $D = D_1 + D_2 + D_3$, where the $D_i$ correspond to points interior to a 2-face. First $D_2$ is removed, and then $D_3$ is removed.}
    \label{figure:contraction}
\end{figure}
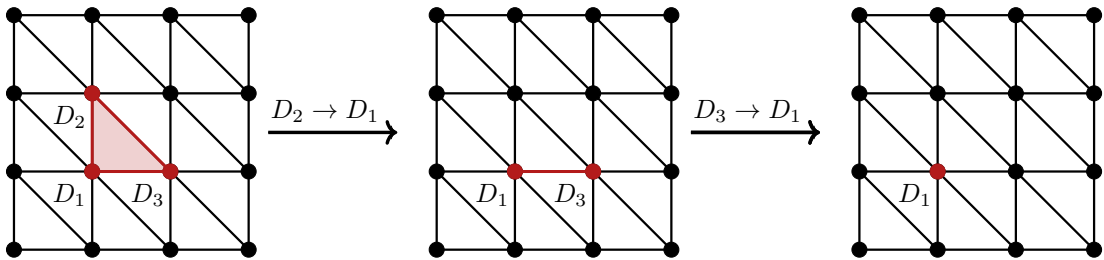

As an
example consider the complex in Figure~\ref{figure:contraction}, with associated divisor $D = D_1 + D_2 + D_3$, where the $D_i$ correspond to the points $p_i$. Equation~\ref{MasterFormulaGenRav} gives
$h^\bullet (\mathcal{O}_D) = (1 + g(f),0,0)$, where $f$ is the 2-face containing $G$. We can perform a contraction by, for example, first deleting $p_2$, and then deleting $p_3$, as in Figure~\ref{figure:contraction}. We will present a more involved example in $\S$\ref{sec:largeone}.

\subsection{Example with $h^{1,1}=491$}\label{sec:largeone}
As a demonstration of the utility of Theorem~\ref{MasterTheorem} we will now calculate the Hodge numbers of some nontrivial divisors in a Calabi-Yau hypersurface $X$ in a toric variety corresponding to a fine, star, regular triangulation of the largest polytope $\Delta_L^\circ$ in the Kreuzer-Skarke database~\cite{KSdatabase}.   In this example, $\Delta_L^\circ$ is the convex hull of the columns of
\begin{equation}\label{eqn:mat}
  B = \begin{pmatrix}
    -1 & -1 & -1 & -1 & 1 \\
    -1 & -1 & -1 & 2 & -1 \\
    -1 & -1 & 6 & -1 & -1 \\
    -1 & 83 & -1 & -1 & -1
  \end{pmatrix}
  \end{equation}
There are 679 nonzero lattice points in  $\Delta^\circ$, but 184 of these are interior to 3-faces, and therefore do not intersect $X$.
$X$ has $h^{1,1}(X) = 491$, with 495 toric divisors. The face structure is very simple: $\Delta_L^{\circ}$ is a $4$-simplex, with vertices indexed by \{0,1,2,3,4\}, corresponding to the columns of $B$. There are 5 vertices, 10 edges, and 10 triangles in  $\Delta_L^\circ$, before triangulation. The genera of all the faces are given in Table~\ref{tab:largegenus}.
   \begin{table}
 \begin{center}
  \begin{tabular}{|l||r|r|r|}
    \hline
    dim & face & \# int pts & genus \\
    \hline
0 & 0 & 1 & 0 \\
0 & 1 & 1 & 0 \\
0 & 2 & 1 & 1 \\
0 & 3 & 1 & 3 \\
0 & 4 & 1 & 6 \\ \hline
1 & 0,1 & 83 & 0 \\
1 & 0,2 & 6 & 0 \\
1 & 0,3 & 2 & 0 \\
1 & 0,4 & 1 & 0 \\
1 & 1,2 & 6 & 0 \\
1 & 1,3 & 2 & 0 \\
1 & 1,4 & 1 & 0 \\
1 & 2,3 & 0 & 1 \\
1 & 2,4 & 0 & 2 \\
1 & 3,4 & 0 & 6 \\ \hline
2 & 0,1,2 & 246 & 0 \\
2 & 0,1,3 & 82 & 0 \\
2 & 0,1,4 & 41 & 0 \\
2 & 0,2,3 & 6 & 0 \\
2 & 0,2,4 & 3 & 0 \\
2 & 0,3,4 & 1 & 0 \\
2 & 1,2,3 & 6 & 0 \\
2 & 1,2,4 & 3 & 0 \\
2 & 1,3,4 & 1 & 0 \\
2 & 2,3,4 & 0 & 1 \\
    \hline
  \end{tabular}
   \end{center}
   \caption{The genera of all vertices, edges, and faces in $\Delta_L^\circ$.  }
   \label{tab:largegenus}
 \end{table}

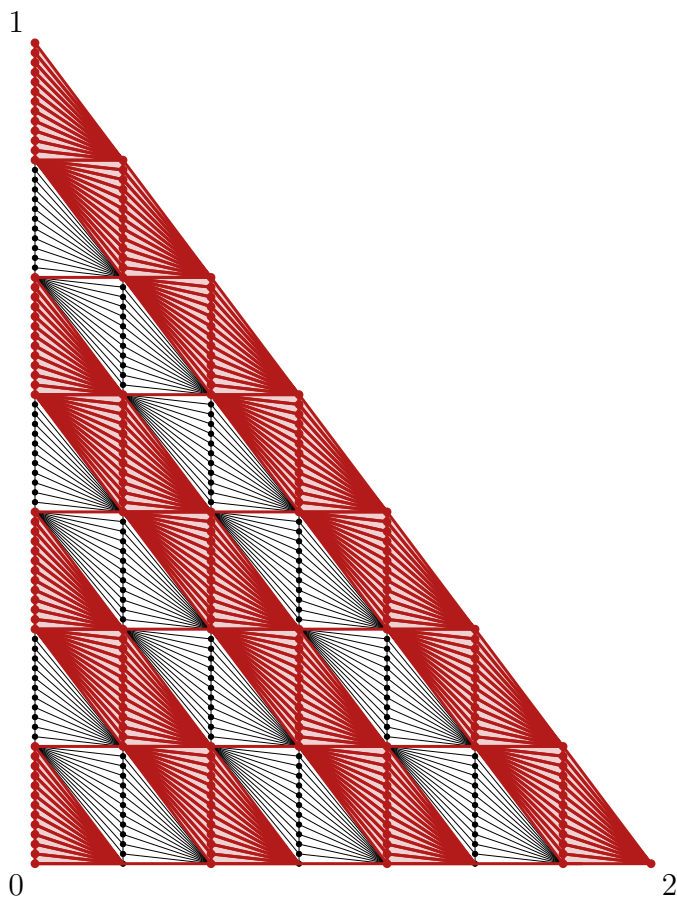
\begin{figure}
    \begin{center}
   \begin{tikzpicture}[scale=.13]
    \node [below left] at (-1,-1) {$0$};
    \node [above left] at (-1,83) {$1$};
    \node [below right] at (62,-1) {$2$};
                  \draw[line width=0.1mm] (-1,-1) -- (-1,83);
                   \draw[line width=0.1mm] (-1,-1) -- (62,-1);
                    \draw[line width=0.1mm] (-1,83) -- (62,-1);
      \foreach \y in {-1,...,83}{
        \node[draw,circle,inner sep=.7pt,fill] at (-1,\y) {};
    }
      \foreach \y in {-1,...,71}{
        \node[draw,circle,inner sep=.7pt,fill] at (8,\y) {};
    }
     \draw[line width=0.1mm] (8,-1) -- (8,71);

      \foreach \y in {-1,...,59}{
        \node[draw,circle,inner sep=.7pt,fill] at (17,\y) {};
    }
  \draw[line width=0.1mm] (17,-1) -- (17,59);
 \foreach \y in {-1,...,47}{
        \node[draw,circle,inner sep=.7pt,fill] at (26,\y) {};
    }
      \draw[line width=0.1mm] (26,-1) -- (26,47);
     \foreach \y in {-1,...,35}{
        \node[draw,circle,inner sep=.7pt,fill] at (35,\y) {};
    }
      \draw[line width=0.1mm] (35,-1) -- (35,35);
     \foreach \y in {-1,...,23}{
        \node[draw,circle,inner sep=.7pt,fill] at (44,\y) {};
    }
      \draw[line width=0.1mm] (44,-1) -- (44,23);
     \foreach \y in {-1,...,11}{
        \node[draw,circle,inner sep=.7pt,fill] at (53,\y) {};
    }
     \foreach \y in {-1,...,-1}{
        \node[draw,circle,inner sep=.7pt,fill] at (62,\y) {};
    }
\foreach \x in {1,...,7}{
 \foreach \q in {-1,...,11}{
        \draw[line width=0.1mm] (9*\x-1,-1) -- (9*\x-9-1,\q);;
    }
 }

  \foreach \x in {0,...,5}{
 \foreach \q in {-1,...,11}{
        \draw[line width=0.1mm] (9*\x-1,11) -- (9*\x+9-1,12-\q-2);;
    }
 }

 \foreach \x in {1,...,6}{
 \foreach \q in {-1,...,11}{
        \draw[line width=0.1mm] (9*\x-1,-1+12) -- (9*\x-9-1,\q+12);;
    }
 }
 \foreach \x in {0,...,4}{
 \foreach \q in {-1,...,11}{
        \draw[line width=0.1mm] (9*\x-1,11+12) -- (9*\x+9-1,12-\q-2+12);;
    }
 }
  \foreach \x in {1,...,5}{
 \foreach \q in {-1,...,11}{
        \draw[line width=0.1mm] (9*\x-1,-1+24) -- (9*\x-9-1,\q+24);;
    }
 }
 \foreach \x in {0,...,3}{
 \foreach \q in {-1,...,11}{
        \draw[line width=0.1mm] (9*\x-1,11+24) -- (9*\x+9-1,12-\q-2+24);;
    }
 }

   \foreach \x in {1,...,4}{
 \foreach \q in {-1,...,11}{
        \draw[line width=0.1mm] (9*\x-1,-1+36) -- (9*\x-9-1,\q+36);;
    }
 }
 \foreach \x in {0,...,2}{
 \foreach \q in {-1,...,11}{
        \draw[line width=0.1mm] (9*\x-1,11+36) -- (9*\x+9-1,12-\q-2+36);;
    }
 }

   \foreach \x in {1,...,3}{
 \foreach \q in {-1,...,11}{
        \draw[line width=0.1mm] (9*\x-1,-1+48) -- (9*\x-9-1,\q+48);;
    }
 }
 \foreach \x in {0,...,1}{
 \foreach \q in {-1,...,11}{
        \draw[line width=0.1mm] (9*\x-1,11+48) -- (9*\x+9-1,12-\q-2+48);;
    }
 }
   \foreach \x in {1,...,2}{
 \foreach \q in {-1,...,11}{
        \draw[line width=0.1mm] (9*\x-1,-1+60) -- (9*\x-9-1,\q+60);;
    }
 }
 \foreach \x in {0,...,0}{
 \foreach \q in {-1,...,11}{
        \draw[line width=0.1mm] (9*\x-1,11+60) -- (9*\x+9-1,12-\q-2+60);;
    }
 }
  \foreach \x in {1,...,1}{
 \foreach \q in {-1,...,11}{
        \draw[line width=0.1mm] (9*\x-1,-1+72) -- (9*\x-9-1,\q+72);
    }
 }
   \foreach \p in {0,2,4,6}{
       \foreach \q in {-1,...,11}{
        \draw (-1,\p*12+\q) node[shape=circle, fill=cornellRed, scale=.3] {};
}
    \foreach \q in {-1,...,10}{
      \draw [line width=0.4mm, draw=cornellRed, fill=cornellRed, fill opacity=0.2]
       (-1,\p*12+\q) -- (-1,\p*12+\q + 1) -- (8,\p*12-1) -- cycle;
       }
       }
       \foreach \p in {1,3,5}{
        \foreach \q in {-1,...,10}{
      \draw [line width=0.4mm, draw=cornellRed, fill=cornellRed, fill opacity=0.2]
       (-1,11+12*\p) -- (8,\q+12*\p ) -- (8,\q+1+12*\p) -- cycle;
       }
       }
        \foreach \p in {2,4,6}{
          \foreach \q in {-1,...,11}{
        \draw  (8,\q+12*\p-12 )  node[shape=circle, fill=cornellRed, scale=.3] {};
}
}

   \foreach \p in {1,3,5}{
   \foreach \q in {-1,...,10}{
      \draw [line width=0.4mm, draw=cornellRed, fill=cornellRed, fill opacity=0.2]
       (-1+9,\p*12+\q) -- (-1+9,\p*12+\q + 1) -- (8+9,\p*12-1) -- cycle;
       }
       }

       \foreach \p in {0,2,4}{
        \foreach \q in {-1,...,10}{
      \draw [line width=0.4mm, draw=cornellRed, fill=cornellRed, fill opacity=0.2]
       (-1+9,11+12*\p) -- (8+9,\q+12*\p ) -- (8+9,\q+1+12*\p) -- cycle;
       }
       }
        \foreach \p in {1,3,5}{
          \foreach \q in {-1,...,11}{
        \draw  (8+9,\q+12*\p-12 )  node[shape=circle, fill=cornellRed, scale=.3] {};
}
}
   \foreach \p in {0,2,4}{
     \foreach \q in {-1,...,10}{
      \draw [line width=0.4mm, draw=cornellRed, fill=cornellRed, fill opacity=0.2]
       (-1+18,\p*12+\q) -- (-1+18,\p*12+\q + 1) -- (8+18,\p*12-1) -- cycle;
       }
       }

       \foreach \p in {1,3}{
        \foreach \q in {-1,...,10}{
      \draw [line width=0.4mm, draw=cornellRed, fill=cornellRed, fill opacity=0.2]
       (-1+18,11+12*\p) -- (8+18,\q+12*\p ) -- (8+18,\q+1+12*\p) -- cycle;
       }
       }
        \foreach \p in {2,4}{
          \foreach \q in {-1,...,11}{
        \draw  (8+18,\q+12*\p-12 )  node[shape=circle, fill=cornellRed, scale=.3] {};
}
}
   \foreach \p in {1,3}{
     \foreach \q in {-1,...,10}{
      \draw [line width=0.4mm, draw=cornellRed, fill=cornellRed, fill opacity=0.2]
       (-1+27,\p*12+\q) -- (-1+27,\p*12+\q + 1) -- (8+27,\p*12-1) -- cycle;
       }
       }

       \foreach \p in {0,2}{
        \foreach \q in {-1,...,10}{
      \draw [line width=0.4mm, draw=cornellRed, fill=cornellRed, fill opacity=0.2]
       (-1+27,11+12*\p) -- (8+27,\q+12*\p ) -- (8+27,\q+1+12*\p) -- cycle;
       }
       }
        \foreach \p in {1,3}{
          \foreach \q in {-1,...,11}{
        \draw  (8+27,\q+12*\p-12 )  node[shape=circle, fill=cornellRed, scale=.3] {};
}
}
   \foreach \p in {0,2}{
    \foreach \q in {-1,...,10}{
      \draw [line width=0.4mm, draw=cornellRed, fill=cornellRed, fill opacity=0.2]
       (-1+36,\p*12+\q) -- (-1+36,\p*12+\q + 1) -- (8+36,\p*12-1) -- cycle;
       }
       }

       \foreach \p in {1}{
        \foreach \q in {-1,...,10}{
      \draw [line width=0.4mm, draw=cornellRed, fill=cornellRed, fill opacity=0.2]
       (-1+36,11+12*\p) -- (8+36,\q+12*\p ) -- (8+36,\q+1+12*\p) -- cycle;
       }
       }
        \foreach \p in {2}{
          \foreach \q in {-1,...,11}{
        \draw  (8+36,\q+12*\p-12 )  node[shape=circle, fill=cornellRed, scale=.3] {};
}
}
   \foreach \p in {1}{
     \foreach \q in {-1,...,10}{
      \draw [line width=0.4mm, draw=cornellRed, fill=cornellRed, fill opacity=0.2]
       (-1+45,\p*12+\q) -- (-1+45,\p*12+\q + 1) -- (8+45,\p*12-1) -- cycle;
       }
       }

       \foreach \p in {0}{
        \foreach \q in {-1,...,10}{
      \draw [line width=0.4mm, draw=cornellRed, fill=cornellRed, fill opacity=0.2]
       (-1+45,11+12*\p) -- (8+45,\q+12*\p ) -- (8+45,\q+1+12*\p) -- cycle;
       }
       }
        \foreach \p in {1}{
          \foreach \q in {-1,...,11}{
        \draw  (8+45,\q+12*\p-12 )  node[shape=circle, fill=cornellRed, scale=.3] {};
}
}
   \foreach \p in {0}{
     \foreach \q in {-1,...,10}{
      \draw [line width=0.4mm, draw=cornellRed, fill=cornellRed, fill opacity=0.2]
       (-1+54,\p*12+\q) -- (-1+54,\p*12+\q + 1) -- (8+54,\p*12-1) -- cycle;
       }
       }

        \draw  (8+54,-1 )  node[shape=circle, fill=cornellRed, scale=.3] {};
            \draw[line width=0.4mm,draw=cornellRed] (-1,-1) -- (55,-1);
             \draw[line width=0.4mm,draw=cornellRed] (-1,11) -- (44,11);
             \draw[line width=0.4mm,draw=cornellRed] (-1,23) -- (35,23);
              \draw[line width=0.4mm,draw=cornellRed] (-1,35) -- (26,35);
              \draw[line width=0.4mm,draw=cornellRed] (-1,47) -- (17,47);
                \draw[line width=0.4mm,draw=cornellRed] (-1,59) -- (8,59);
    \end{tikzpicture}
     \caption{The largest 2-face $f_{\ell}$ in the largest four-dimensional reflexive polytope $\Delta_L^\circ$, with a `banded' complex show in red. The Hodge numbers of the corresponding divisor $D$ are easily read off as $h^\bullet({\cal{O}}_D) = (1,9,1)$.}
     \label{fig:biggie}
         \end{center}
    \end{figure}

Consider the largest 2-face $f_\ell$ in $\Delta_L^\circ$, which is a triangle with vertices labeled by $\{0,1,2\}$.  This face has $344$ lattice points, as shown in Figure~\ref{fig:biggie}. All (sub)faces except vertex 2 have genus 0, and so the puff complex $\mathscr{P}$ restricted to $f_{\ell}$ is simply $\mathcal{T}$ restricted to $f_{\ell}$, with a single $S^2$ cell attached to vertex 2.
If we choose, excluding point $2$, a set of points on $f_{\ell}$ whose corresponding complex is connected and has no cycles, then the corresponding divisor will be rigid. The divisor corresponding to the entire face has Hodge numbers $(1,0,1)$. The divisor corresponding to the line connecting 2 to 4 has Hodge numbers $(1,0,9)$.  As a more nontrivial example we can form simplicial complexes with cycles on $f_\ell$ and $\Delta_L^\circ$. The complex defined by taking all the boundary points of $f_\ell$, and none of the interior points, defines a divisor with Hodge numbers $(1,1,1)$.
In Figure~\ref{fig:biggie} we show a triangulation\footnote{This triangulation of $f_\ell$ is for illustrative purposes; the triangulation we have obtained from a regular triangulation of $\Delta_L^\circ$ is difficult to render.}
of $f_\ell$, and a choice of a more complicated subcomplex containing cycles.  Theorem~\ref{MasterTheorem} allows us to easily read off the Hodge numbers of the corresponding divisor $D$ as $h^\bullet({\cal{O}}_D) = (1,9,1)$.

\newpage

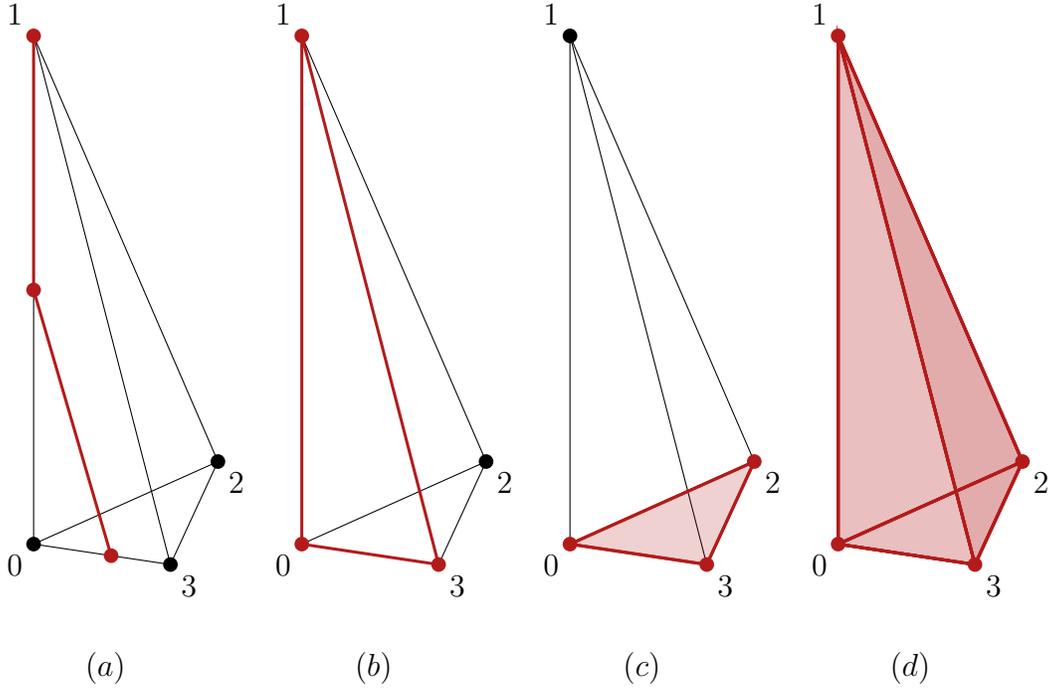
\begin{figure}
\captionsetup{singlelinecheck = false, justification=justified}
\begin{center}
		\tdplotsetmaincoords{75}{30}
\begin{tikzpicture}[tdplot_main_coords,scale = .7]
	
	\coordinate (A) at (-1, -1, -1);
	\coordinate (B) at (-1, -1, 9);
	\coordinate (C) at (-1, 6, -1);
	\coordinate (D) at (2, -1, -1);
	
	\coordinate (E) at (-1,-1, 4);
	\coordinate (F) at (.7,-1, -1);
	
	\begin{scope}
		\draw (A)--(B);
		\draw (A)--(C);
		\draw (A)--(D);
		\draw (B)--(C);
		\draw (B)--(D);
		\draw (C)--(D);
	\end{scope}
	 \draw[line width=0.4mm, color=cornellRed] (B) -- (E);
	  \draw[line width=0.4mm, color=cornellRed] (E) -- (F);
	\draw (A) node[shape=circle, fill=black, scale=.5] {};
 	\draw (B) node[shape=circle, fill=cornellRed, scale=.5] {};
	\draw (C) node[shape=circle, fill=black, scale=.5] {};
	\draw (D) node[shape=circle, fill=black, scale=.5] {};
	\draw (E) node[shape=circle, fill=cornellRed, scale=.5] {};
	\draw (F) node[shape=circle, fill=cornellRed, scale=.5] {};
	 \node [below left] at (A) {$0$};
	 \node [above left] at (B) {$1$};
	 \node [below right] at (C) {$2$};
	 \node [below right] at (D) {$3$};
	 \node [below] at (0,0,-3) {$(a)$};

\end{tikzpicture}
\begin{tikzpicture}[tdplot_main_coords,scale = .7]
	
	\coordinate (A) at (-1, -1, -1);
	\coordinate (B) at (-1, -1, 9);
	\coordinate (C) at (-1, 6, -1);
	\coordinate (D) at (2, -1, -1);
	
	\begin{scope}
		\draw (A)--(B)[fill=cornellRed];
		\draw (A)--(C);
		\draw (A)--(D);
		\draw (B)--(C);
		\draw (B)--(D);
		\draw (C)--(D);
	\end{scope}
	  \draw[line width=0.4mm, color=cornellRed] (A) -- (B);
	   \draw[line width=0.4mm, color=cornellRed] (A) -- (D);
	   \draw[line width=0.4mm, color=cornellRed] (B) -- (D);
	
	\draw (A) node[shape=circle, fill=cornellRed, scale=.5] {};
 	\draw (B) node[shape=circle, fill=cornellRed, scale=.5] {};
	\draw (C) node[shape=circle, fill=black, scale=.5] {};
	\draw (D) node[shape=circle, fill=cornellRed, scale=.5] {};
	 \node [below left] at (A) {$0$};
	 \node [above left] at (B) {$1$};
	 \node [below right] at (C) {$2$};
	 \node [below right] at (D) {$3$};
 \node [below] at (0,0,-3) {$(b)$};
\end{tikzpicture}
\begin{tikzpicture}[tdplot_main_coords,scale = .7]
	
	\coordinate (A) at (-1, -1, -1);
	\coordinate (B) at (-1, -1, 9);
	\coordinate (C) at (-1, 6, -1);
	\coordinate (D) at (2, -1, -1);
	
	\begin{scope}
		\draw (A)--(B);
		\draw (A)--(C);
		\draw (A)--(D);
		\draw (B)--(C);
		\draw (B)--(D);
		\draw (C)--(D);
	\end{scope}
	
	 \draw[line width=0.4mm, color=cornellRed] (A) -- (C);
	   \draw[line width=0.4mm, color=cornellRed] (A) -- (D);
	   \draw[line width=0.4mm, color=cornellRed] (C) -- (D);
	    \draw [line width=0.4mm, draw=cornellRed, fill=cornellRed, fill opacity=0.2]
       (A) -- (C) -- (D) -- cycle;
	
	\draw (A) node[shape=circle, fill=cornellRed, scale=.5] {};
 	\draw (B) node[shape=circle, fill=black, scale=.5] {};
	\draw (C) node[shape=circle, fill=cornellRed, scale=.5] {};
	\draw (D) node[shape=circle, fill=cornellRed, scale=.5] {};
	 \node [below left] at (A) {$0$};
	 \node [above left] at (B) {$1$};
	 \node [below right] at (C) {$2$};
	 \node [below right] at (D) {$3$};
	  \node [below] at (0,0,-3) {$(c)$};

\end{tikzpicture}
\begin{tikzpicture}[tdplot_main_coords,scale = .7]
	
	\coordinate (A) at (-1, -1, -1);
	\coordinate (B) at (-1, -1, 9);
	\coordinate (C) at (-1, 6, -1);
	\coordinate (D) at (2, -1, -1);
	
	\begin{scope}
		\draw (A)--(B);
		\draw (A)--(C);
		\draw (A)--(D);
		\draw (B)--(C);
		\draw (B)--(D);
		\draw (C)--(D);
	\end{scope}
	
	 \draw[line width=0.4mm, color=cornellRed] (A) -- (C);
	   \draw[line width=0.4mm, color=cornellRed] (A) -- (D);
	   \draw[line width=0.4mm, color=cornellRed] (C) -- (D);
	    \draw [line width=0.4mm, draw=cornellRed, fill=cornellRed, fill opacity=0.2]
       (A) -- (C) -- (D) -- cycle;

        \draw[line width=0.4mm, color=cornellRed] (B) -- (C);
	   \draw[line width=0.4mm, color=cornellRed] (B) -- (D);
	   \draw[line width=0.4mm, color=cornellRed] (C) -- (D);
	    \draw [line width=0.4mm, draw=cornellRed, fill=cornellRed, fill opacity=0.2]
       (B) -- (C) -- (D) -- cycle;

        \draw[line width=0.4mm, color=cornellRed] (B) -- (A);
	   \draw[line width=0.4mm, color=cornellRed] (B) -- (D);
	   \draw[line width=0.4mm, color=cornellRed] (A) -- (D);
	    \draw [line width=0.4mm, draw=cornellRed, fill=cornellRed, fill opacity=0.1]
       (B) -- (A) -- (D) -- cycle;

            \draw[line width=0.4mm, color=cornellRed] (B) -- (A);
	   \draw[line width=0.4mm, color=cornellRed] (B) -- (C);
	   \draw[line width=0.4mm, color=cornellRed] (C) -- (A);
	    \draw [line width=0.4mm, draw=cornellRed, fill=cornellRed, fill opacity=0.2]
       (B) -- (A) -- (C) -- cycle;

	\draw (A) node[shape=circle, fill=cornellRed, scale=.5] {};
 	\draw (B) node[shape=circle, fill=cornellRed, scale=.5] {};
	\draw (C) node[shape=circle, fill=cornellRed, scale=.5] {};
	\draw (D) node[shape=circle, fill=cornellRed, scale=.5] {};
	 \node [below left] at (A) {$0$};
	 \node [above left] at (B) {$1$};
	 \node [below right] at (C) {$2$};
	 \node [below right] at (D) {$3$};
	  \node [below] at (0,0,-3) {$(d)$};

\end{tikzpicture}

	\end{center}
\caption{ A facet of $\Delta_L^\circ$ defined by the points $\{0,1,2,3\}$, with several  complexes drawn on it. The red points indicate lattice points included in the complex, and the red lines cycle-less paths that may include additional lattice points. The red triangles indicate entire faces that are included in the complex. Note that the point $1$, corresponding to $(-1,-1,-1,83)$, has been scaled in, for visualization purposes. In complex $(d)$ we have included the entire boundary of the facet. From Theorem~\ref{MasterTheorem}
the Hodge numbers of the corresponding divisors are $h^\bullet({\cal{O}}_{D_a}) = (1, 0, 0) $, $h^\bullet({\cal{O}}_{D_b}) =(1,1,3) $, $h^\bullet({\cal{O}}_{D_c}) = (1,0,5)$, and $h^\bullet({\cal{O}}_{D_d}) = (1,0,6)$.}
\label{fig:bigfaces}
\end{figure}

As a further example of contraction, one can consider complexes that are not just restricted to a single 2-face, as depicted in Figure~\ref{fig:bigfaces}. Here we show a facet of $\Delta_L^\circ$ defined by the points $\{0,1,2,3\}$, with several  complexes drawn on it. The red points indicate lattice points included in the complex, and the red lines cycle-free paths that may include additional lattice points. The red triangles indicate entire faces that are included in the complex. Note that the point $1$, corresponding to $(-1,-1,-1,83)$, has been scaled in, for visualization purposes. In complex $(d)$ we have included the entire boundary of the facet. From Theorem~\ref{MasterTheorem} we can immediately read off the Hodge numbers of the corresponding divisors as $h^\bullet({\cal{O}}_{D_a}) = (1, 0, 0) $, $h^\bullet({\cal{O}}_{D_b}) =(1,1,3) $, $h^\bullet({\cal{O}}_{D_c}) = (1,0,5)$, and $h^\bullet({\cal{O}}_{D_d}) = (1,0,6)$.

\bigbreak

\section{Conclusions}  \label{sec:conclusions}

In this work we have computed the Hodge numbers $h^{i}({\cal{O}}_D)$ of square-free divisors $D$ of Calabi-Yau threefold
hypersurfaces in toric varieties.  Given the data of a simplicial complex $\mathcal{T}$ corresponding to a
triangulation of a four-dimensional reflexive polytope $\Delta^{\circ}$, we constructed a CW complex $\mathscr{P}$ that simultaneously encodes data about $\Delta^{\circ}$ and its dual $\Delta$.  The construction of $\mathscr{P}$ involves attaching certain cells to $\mathcal{T}$ in a manner determined by $\Delta$.
The specification of a square-free divisor $D$ determines a subcomplex $\mathscr{P}_D$, as well as an exact Mayer-Vietoris sheaf sequence.
By examining the corresponding hypercohomology spectral sequence, we proved that $h^{i}({\cal{O}}_D)=h_i(\mathscr{P}_D)$.  Here the left-hand side is manifestly the dimension of a sheaf cohomology group, but the right-hand side is the dimension of a simplicial (or cellular) homology group.

Our results are a step forward in the study of divisors in Calabi-Yau hypersurfaces.
Theorem~\ref{MasterTheorem} permits extremely efficient computation of the Hodge numbers of square-free divisors in threefolds,
even for $h^{1,1} \gg 1$.  Conjecture~\ref{GeneralizationofMasterTheorem} extends these methods to fourfolds, enlarging the range of divisors with computable Hodge numbers \cite{wip}.

The ultimate goal of this work was to determine which effective divisors $D$ of a Calabi-Yau hypersurface
support Euclidean brane superpotential terms.  Theorem~\ref{MasterTheorem} represents significant progress toward this goal, but further advances will be necessary to give a completely general answer.  First of all, although the Hodge numbers $h^{i}({\cal{O}}_D)$ provide essential information about  the number of fermion zero modes of a Euclidean brane on $D$,  when $D$ is not smooth there can be additional zero modes associated to singular loci.
The only smooth, \emph{rigid}, square-free divisors are the prime toric divisors $D_i$ themselves; a nontrivial square-free divisor that is rigid necessarily involves normal crossing singularities where the components $D_i$ intersect.  One should therefore ask how to count fermion zero modes on a divisor with normal crossings.
In some cases, normal crossings can be shown to yield no new fermion zero modes, so that rigidity is a sufficient condition for a superpotential \cite{wip}.  Moreover, there exist nontrivial smooth square-free divisors $D$ with Hodge numbers $h^{\bullet}({\cal{O}}_D)=(1,0,n)$, with $n>0$.  In a suitably magnetized Euclidean D3-brane wrapping such a divisor, worldvolume flux can lift the zero modes counted by $h^2({\cal{O}}_D)$, and so lead to a superpotential term, even though $D$ is not rigid.  Finally, Donagi and Wijnholt have proposed that in general
the fermion zero modes can be counted by the logarithmic cohomology of $D$, $h^{\bullet}_{\rm{log}}({\cal{O}}_D)$ \cite{Donagi:2010pd}.  Verifying and applying this idea is a natural task for the future.

There are many possible extensions of this work.
The effects of worldvolume flux could plausibly be incorporated along the lines of \cite{Bianchi:2011qh}, but including the effects of bulk fluxes appears more challenging.
A further step would be to compute the Hodge numbers of effective divisors that are not square-free.
Advances in these directions would allow for a truly systematic computation of the nonperturbative superpotential for the K\"ahler moduli in compactifications on Calabi-Yau hypersurfaces.

\section*{Acknowledgments}

We are grateful to Yuri Berest, Ralph Blumenhagen, Thomas Grimm, Jim Halverson, Tristan H\"ubsch, John Stout,
Wati Taylor, and Timo Weigand for discussions, and we thank John Stout for creating Figure 2.  We thank the organizers of String Phenomenology 2017 at Virginia Tech for providing a stimulating environment for discussing this work.  L.M.~thanks the theory groups at MIT, Northeastern, and Stanford for their hospitality while portions of this work were carried out. C.L.~and B.S.~thank the theory group at MIT for their hospitality while portions of this work were carried out.
The work of A.P.B.~was supported by the SCFT grant ST/L000474/1 and the
ERC grant 682608 (HIGGSBNDL).  The work of C.L.~was supported by NSF grant PHY-1620526. The work of L.M.~was supported in part by NSF grant PHY-1719877.  The work of M.S.~was supported in part by NSF grant DMS-1502294.  The work of B.S.~was supported by NSF RTG grant DMS-1645877.

\appendix

\newpage

\section{Mayer-Vietoris Complexes} \label{sec:mvc}

\noindent In this section we establish the exactness of
the generalized Mayer-Vietoris sequence, and we give related background on spectral sequences.

Given closed subvarieties (or subschemes) $D$ and $E$ of a variety (or scheme) $X$, the Mayer-Vietoris complex is
\begin{equation}
0 \longrightarrow \cO_{D \cup E}  \longrightarrow \cO_D \oplus \cO_E \longrightarrow \cO_{D \cap E} \longrightarrow 0
\end{equation}
This is always an exact sequence, and one can view it as an inclusion-exclusion sequence.  If $I$ and $J$ are ideals in a ring $R$, then we have the related exact sequences:
\begin{equation}
0 \longrightarrow R/I\cap J  \longrightarrow R/I \oplus R/J \longrightarrow R/(I + J) \longrightarrow 0
\end{equation}
and
\begin{equation}
0 \longrightarrow I \cap J  \longrightarrow I \oplus J \longrightarrow  I + J \longrightarrow 0.
\end{equation}
There are well-known generalizations of these sequences to the cases where there are more than two subvarieties, ideals, or quotients, but at this level of generality the resulting Mayer-Vietoris sequence is often not exact.  We will now prove exactness for several cases of interest in this paper.

We start with the case of ideals.  Suppose that $I_1, \ldots, I_n \subset R$ are ideals in a ring $R$.  Define the {\bf Mayer-Vietoris ideal sequence}, $M = \mathbb{M}(I_1, \ldots, I_n)$, to be the
cochain complex
\begin{equation}\label{mvcfirst}
0 \longrightarrow (I_1 \cap \ldots \cap I_n)  \longrightarrow M^0 \longrightarrow M^1 \longrightarrow \ldots \longrightarrow M^{n-1} \longrightarrow 0,
\end{equation}
where
\[ M^p := \bigoplus_{1 \le i_0 < \ldots < i_p \le n}  (I_{i_0} + \ldots + I_{i_p})
\]
and the differential $d^p : M^p \longrightarrow M^{p+1}$ is defined by
\[
(d^p (\phi))_{i_0 \ldots i_{p+1}} = \sum_{j=0}^{p+1} (-1)^j \phi_{i_0 \ldots \hat{i}_j \ldots i_{p+1}}
\]
One checks immediately that $\ker(d^0) = I_1 \cap \ldots \cap I_n$, and that the sequence \eqref{mvcfirst} is in fact a complex. Notice that if $I_i = R$ for all $i$, then $\mathbb{M}(R,\ldots,R)$ is the (reduced) cochain complex of an $(n-1)$-simplex, and so is exact.

The {\bf Mayer-Vietoris quotient complex}, $\mathbb{M}(R/I_1, \ldots, R/I_n)$ is the cochain complex defined in a completely analogous manner:
\begin{equation}\label{mvqc}
0 \longrightarrow R/(I_1 \cap \ldots \cap I_n)  \longrightarrow N^0 \longrightarrow N^1 \longrightarrow \ldots \longrightarrow N^{n-1} \longrightarrow 0,
\end{equation}
where
\[ N^p := \bigoplus_{1 \le i_0 < \ldots < i_p \le n}  R/(I_{i_0} + \ldots + I_{i_p})
\]
and the differential $(d^N)^p : N^p \longrightarrow N^{p+1}$ is defined by the natural quotient maps
\[
((d^N)^p (\phi))_{i_0 \ldots i_{p+1}} = \sum_{j=0}^{p+1} (-1)^j \phi_{i_0 \ldots \hat{i}_j \ldots i_{p+1}}
\]
Again, one checks that \eqref{mvqc} is a complex, and that $\ker (d^N)^0 = R/(I_1 \cap \ldots \cap I_n)$. There is an exact sequence of complexes
\[
0 \longrightarrow \mathbb{M}(I_1, \ldots, I_n) \longrightarrow \mathbb{M}(R, \ldots, R) \longrightarrow \mathbb{M}(R/I_1, \ldots, R/I_n) \longrightarrow 0,
\]
and coupled with the exactness of the middle complex, the complex $\mathbb{M}(I_1, \ldots, I_n)$ is exact precisely when $\mathbb{M}(R/I_1, \ldots, R/I_n)$ is exact.

Finally, given closed subvarieties (or subschemes) $D := D_1 \cup \ldots \cup D_n$ of a variety (or scheme) $X$, we
define the {\bf Mayer-Vietoris sheaf sequence} $\mathbb{M}(D_1, \ldots, D_n)$ in parallel to the above, taking
\[ \mathbb{M}(D_1,\ldots,D_n)^p := \bigoplus_{1 \le i_0 < i_1 < \ldots < i_p \le n} \cO_{D_{i_0} \cap D_{i_1} \cap \ldots \cap D_{i_p}}.\]
Define maps $d^p : \mathbb{M}(D_1, \ldots D_n)^p \longrightarrow \mathbb{M}(D_1, \ldots, D_n)^{p+1}$ in the same manner as for quotients of ideals above.  The resulting complex $\mathbb{M}(D_1, \ldots, D_n)$ has the form

\begin{equation}\label{eq:mv}
0 \longrightarrow \cO_{D}
	\longrightarrow \bigoplus_{i=1}^n \cO_{D_i}
	\longrightarrow \bigoplus_{i<j} \cO_{D_i \cap D_j}
	\longrightarrow \ldots
	\longrightarrow \cO_{D_1 \cap D_2 \cap \ldots \cap D_n}
	\longrightarrow
	0\,.
\end{equation}

\subsection{Distributive lattices of ideals and Mayer-Vietoris complexes of ideals}
The relationship between distributive lattices of ideals and the exactness of Mayer-Vietoris ideal sequences was explained in \cite{2008arXiv0811.3997M}.
In this section, we summarize their results, and then show that several sets of ideals of interest form distributive lattices, so that their Mayer-Vietoris sequences are exact.
This is the key technical fact that will allow us to show in \S\ref{sec:mvexact} that certain Mayer-Vietoris complexes of sheaves are exact.

Fix a (commutative) ring $R$ and ideals $J_1$, \ldots, $J_r$ of $R$. This set of ideals generates a {\it lattice of ideals}: the join of two ideals is their sum,
and the meet of two ideals is their intersection.  The smallest set of ideals that contains $J_1, \ldots, J_r$ and is closed under these two operations is the lattice of ideals generated by $J_1, \ldots, J_r$.

This lattice of ideals is called
\emph{distributive} if  for every three ideals $L_1, L_2, L_3$ in the lattice, one has
 \[ L_1 \cap (L_2 + L_3) = (L_1 \cap L_2) + (L_1 \cap L_3). \]
Equivalently, the lattice is distributive if and only if for
every three ideals $L_1, L_2, L_3$ in the lattice, one has
 \[ L_1 + (L_2 \cap L_3) = (L_1 + L_2) \cap (L_1 + L_3). \]

The importance of this notion is the following useful characterization due to Maszczyk.

\begin{theorem}[Maszczyk \cite{2008arXiv0811.3997M}]
For a set of ideals $J_1, \ldots, J_n$ of a ring $R$, the following are equivalent:

(a) The lattice of ideals generated by $J_1, \ldots, J_n$ is distributive.

(b) $\mathbb{M}(J_1, \ldots, J_n)$ is exact, and hence so is $\mathbb{M}(R/J_1, \ldots, R/J_n)$.
\end{theorem}

The following is a typical distributive lattice of ideals:
\begin{example}
Let $S = \C[x_1, \ldots, x_n]$ be a polynomial ring over a field $\mathbb{F}$.  Then the lattice of ideals generated by $\{(x_1), \ldots, (x_n)\}$ is the set of square-free monomial ideals of $S$.
This is a distributive lattice of ideals.
\end{example}

A square-free monomial ideal $M \subset \mathbb{F}[x_1, \ldots, x_n]$ has a unique description as an irredundant intersection of monomial prime ideals
  \[ M = \bigcap_{\alpha \in \Lambda} \langle x_{i} \mid i \in \alpha \rangle,\]
where the intersection is over a uniquely defined set of subsets of $[1..n]$.  The ideals in the intersection are the minimal primes of $M$.

A key example is a generalization of this to the case where $F = \{ f_1, \ldots, f_n \} \subset R$ is a regular sequence.  We will only consider the case where $R$ is a local ring, or a graded ring, and in these cases any permutation of the $f_i$ remains a regular sequence.

Given $F = \{f_1, \ldots, f_n\} \subset R$, define a map $\phi : \Z[x_1, \ldots, x_n] \longrightarrow R$, where $\phi(x_i) = f_i$.  In this section, we call an ideal $I \subset R$ an \emph{$F$-square-free
ideal} if there is a square-free monomial ideal $M$ such that $I$ is generated by $\phi(M)$, i.e., $I$ is generated by square-free products of the $f_i$.  We will say that $I$ is an \emph{$F$-square-free monomial ideal} if it is $F$-square-free, and
$I$ can be written as the intersection
  \[ I = \bigcap_{\alpha \in \Lambda} \langle f_{i} \mid i \in \alpha \rangle,\]
where
  \[ M = \bigcap_{\alpha \in \Lambda} \langle x_{i} \mid i \in \alpha \rangle.\]

\begin{theorem}~\label{thm:distributive}
Let $F = \{f_1, \ldots, f_n\}$ be a regular sequence in the maximal ideal of a local ring $R$.

(a)  The lattice of ideals generated by $(f_1), \ldots, (f_n)$ is exactly the set of $F$-square-free ideals.

(b) This lattice of ideals is distributive.
\end{theorem}

This follows from the following more precise lemma, which we prove by induction on $n$.

\begin{lemma}~\label{lemmaA}
Let $F = \{f_1, \ldots, f_{n}\}$ be a regular sequence.  Let $F' := \{ f_1, \ldots, f_{n-1}\}$. The following statements hold.

(a) If $J$ is an $F'$-square-free ideal, then $(J : f_n) = J$.

(b) If $I_i, J_i$ are $F'$-square-free ideals, then $L_1 = I_1 f_n + I_2$ and $L_2 = J_1 f_n + J_2$ obey
\[ L_1 \cap L_2 = (I_1 \cap J_1 + I_1 \cap J_2 + I_2 \cap J_1)f_n + I_2 \cap J_2. \]

(c) If $I$ and $J$ are $F'$-square-free ideals, then
\[ ((f_n) + I) \cap ((f_n) + J) = ((f_n) + (I \cap J)). \]

(d)  If $I$ and $J$ are $F'$-square-free ideals, then
\[ (If_n + J) : f_n = I+J. \]

(e) Every $F$-square-free ideal is an $F$-square-free monomial ideal.

(f) If $L_1$ and $L_2$ are $F$-square-free ideals, then
$L_1 \cap L_2$ is also $F$-square-free.

(g) If $L_1, L_2, L_3$ are $F$-square-free ideals, then
\[ L_1 \cap (L_2 + L_3) = L_1 \cap L_2 + L_1 \cap L_3. \]
\end{lemma}

\medskip
\noindent{\bf Proof.}
We prove this by induction.  The case $n=2$ is immediate.  Suppose the statements are true for a number of elements of $F$ less than $n$.

(a) By part (e) in the induction hypothesis,
  \[ J = \bigcap_{\alpha \in \Lambda} \,\langle f_{i} \mid i \in \alpha \rangle,\]
  where all the $f_i$ that appear have $i < n$.  Then $J : f_n$ is the intersection of
   \[ \langle f_{i} \mid i \in \alpha \rangle : f_n = \langle f_{i} \mid i \in \alpha \rangle, \]
whose intersection is again $J$.

(b) The right hand side is clearly contained in the left hand side, so let $\gamma = \alpha_1 f_n + \alpha_2 = \beta_1 f_n + \beta_2 \in L_1 \cap L_2$. One shows that $\gamma$ is contained in the right
hand side (where $\alpha_1 \in I_1, \alpha_2 \in I_2, \beta_1 \in J_1, \beta_2 \in J_2$).  Thus, $(\alpha_1 - \beta_1)f_n = \beta_2 - \alpha_2 \in I_2 + J_2$.  Therefore, by part (a), since $I_2 + J_2$ is $F'$-square-free, $\alpha_1 - \beta_1 \in I_2 + J_2$.  Therefore, there exists $ \delta_1 \in I_2, \delta_2 \in J_2$ such that $\alpha_1 - \beta_1 = \delta_1 - \delta_2$. Now plug this back into the formula for $\gamma$ to get $\delta_1 f_n + \alpha_2 = \delta_2 f_n + \beta_2 \in I_2 \cap J_2$.  Since this is already in the right hand side, we can subtract it from $\gamma$, obtaining an element $\gamma' = (\alpha_1 - \delta_1) f_n = (\beta_1 - \delta_2)f_n$ that is in the right hand side precisely when $\gamma$ is.  Since $f_n$ is a nonzero divisor,
$\alpha_1 - \delta_1 = \beta_1 - \delta_2$, so $\gamma' \in ((I_1 + I_2) \cap (J_1 + J_2)) f_n$.  Combining this with the induction hypothesis of part(g), we have that $(I_1 + I_2) \cap (J_1 + J_2) = I_1 \cap J_1 + I_1 \cap J_2 + I_2 \cap J_1 +  I_2 \cap J_2$, and therefore $\gamma'$, and hence $\gamma$, is contained in the right hand side.

(c)  This follows immediately from (b), by taking $I_1 = J_1 = R$.

(d) This follows immediately from (b): $(If_n + J) \cap (f_n) = (I+J) f_n$, and so $(If_n + J) : f_n = I+J$.

(e) For any ideal $I$, and $f \in R$, such that $I : f^2 = I:f$, we always have $I = (I : f) \cap ((f) + I)$. Let $L = If_n + J$ be an $F$-square-free ideal. If $I = 0$, then the inductive
hypothesis gives that $L = J$ is $F'$-square-free monomial, and is therefore $F$-square-free.  If instead, $f_n \in L$, then the induction hypothesis gives a decomposition for $J$, and part (c)
gives the required decomposition for $L$.  Otherwise, by part (d) we have $L = (I+J) \cap ((f_n)+J)$.  Both of these have decompositions, by (e) (for $n-1$) and (c), and putting them together gives a decomposition for $L$.

(f) This follows immediately from (b) and the inductive hypothesis for (f).

(g) This follows immediately from part (b).
$\square$

\bigskip
\noindent{\bf Proof of Theorem~\ref{thm:distributive}.}
The set of $F$-square-free ideals is plainly closed under addition of ideals.
By Lemma~\ref{lemmaA}(b), the lattice of ideals generated by $(f_1), \ldots, (f_n)$ contains the set of $F$-square-free ideals.  Also by Lemma~\ref{lemmaA}(f), the set of $F$-square-free ideals is closed under intersection, and so the lattice of ideals generated by $(f_1), \ldots, (f_n)$ is the set of $F$-square-free ideals.
Finally, by Lemma~\ref{lemmaA}(g), the lattice is distributive.
$\square$

\subsection{Mayer-Vietoris sequences corresponding to divisors and curves} \label{sec:mvexact}

Recall that $D \subset X$ is called a prime divisor on $X$ if it is irreducible and codimension one in $X$.

\begin{proposition}   \label{MVtoricprop}
If $X$ is a simplicial toric variety, and $D_1, \ldots, D_r$ are prime torus-invariant Weil divisors on $X$, then the Mayer-Vietoris sheaf sequence $\mathbb{M}(D_1, \ldots, D_r)$
is an exact sequence of $\cO_X$-modules.
\end{proposition}

\noindent {\bf Proof.} Let $S = \C[x_1, \ldots, x_N]$ be the Cox ring of $X$, where $D_i$ corresponds to the ring variable $x_i$, for $1 \le i \le r$.
Consider the Mayer-Vietoris ideal sequence $\mathbb{M}(x_1, \ldots, x_r)$.  This is exact, as the $x_i$ generate a distributive lattice, and so $\mathbb{M}(S/x_1, \ldots, S/x_r)$ is also exact.  Sheafification of
this exact sequence of graded $S$-modules remains exact, but this sheafification is precisely $\mathbb{M}(D_1, \ldots, D_r)$. $\square$

\begin{proposition} \label{MVprop}
  If $X$ is a smooth variety, and $D_1, \ldots, D_r$ are effective divisors, such that the intersection of
  each set of $n$ of these is either empty, or has codimension $n$ in $X$,
then the Mayer-Vietoris sheaf sequence $\mathbb{M}(D_1, \ldots, D_r)$
is an exact sequence of $\cO_X$-modules.

\end{proposition}

\noindent {\bf Proof.}  We prove exactness locally.  Let $p \in X$ be a point.  Suppose that the set of $D_i$ that pass through $p$
is $\{D_1, \ldots, D_n\}$. Localizing the Mayer-Vietoris sheaf complex at $p$ results in
the complex $\mathbb{M}(R/f_1, \ldots, R/f_n)$, where $R = \cO_{X,p}$, and the Cartier divisor $D_i$
is defined locally in this ring by $f_i$.  Note that, by hypothesis, the elements $f_1, \ldots, f_n$ form a regular sequence in the maximal ideal of $R$.
Therefore, Theorem~\ref{thm:distributive} shows that $\mathbb{M}(f_1, \ldots, f_n)$ is exact.  This implies that $\mathbb{M}(R/f_1, \ldots, R/f_n)$ is also exact,
which implies that the original complex is exact at each $p \in X$, proving the proposition. $\square$
\bigskip

\noindent
Notice that this proof can be generalized to the case when $X$ is equidimensional, not necessarily smooth, and the $D_i$ are Cartier divisors with the given intersection properties.
\bigskip

\noindent
There also exists a corresponding exact sequence in the case of curves simply using the corresponding normalization without necessitating the above technology.

\begin{lemma}\label{MVlemmacurve}
Suppose $X$ is a smooth, projective variety and assume that $C_{1},\ldots,C_{r}$ are smooth, irreducible curves on $X$, and that the corresponding closed subscheme $C \coloneqq C_{1} \cup \ldots \cup C_{r}$ is a nodal, reducible curve with components $C_{i}$. Then the Mayer-Vietoris sequence $\mathbb{M}(C_{1},\ldots,C_{r})$ is an exact sequence of $\mathcal{O}_{X}$-modules.
\end{lemma}
\noindent {\bf Proof.} By hypothesis, the normalization $f \colon \tilde{C} \rightarrow C$ is the scheme corresponding to the disjoint union of the smooth, irreducible components $C_{i}$. Thus, the normalization induces the following short exact sequence
\begin{equation}
\mathcal{O}_{C}  \xhookrightarrow{} f_{*}\bigoplus_{i}\mathcal{O}_{C_{i}} \twoheadrightarrow f_{*}\bigoplus_{i}\mathcal{O}_{C_{i}}/\mathcal{O}_{C}\,,
\end{equation}
where the quotient sheaf is a torsion sheaf with support precisely on the nodes of $C$. By considering the sequence locally at the nodes, we thus obtain the following exact sequence of $\mathcal{O}_{C}$-modules
\begin{equation}
\mathcal{O}_{C} \xhookrightarrow{} f_{*}\bigoplus_{i}\mathcal{O}_{C_{i}} \twoheadrightarrow g_{*}\bigoplus_{j}\mathcal{O}_{p_{j}}\,,
\end{equation}
where $g$ is the natural closed immersion and $p_{j}$ correspond to the nodes of $C$. Higher direct images of closed immersions vanish and hence we obtain an exact sequence of $\mathcal{O}_{X}$-modules yielding exactness of $\mathbb{M}(C_{1},...,C_{r})$.
$\square$
\subsection{Background on spectral sequences}

We will now highlight the essential aspects of hypercohomology spectral sequences needed to follow the
computation of $h^{\bullet}(\mathcal{O}_{D})$.
We begin by describing the basic usage of the two spectral sequences corresponding
to a double complex.  Instead of going into detail about filtrations,
we assume that
the entries of the second and subsequent pages of the spectral sequence are finite-dimensional vector spaces.
We are describing ``first quadrant'' spectral sequences, as that is what we need for hypercohomology.

We start with a bounded double complex of vector spaces, $C = \{ C^{p,q}, d_h, d_v \}$, i.e.~a collection of vector spaces $C^{p,q}$, where $C^{p,q} = 0$ outside of the box $0 \le p \le M$,
$0 \le q \le M$, for some $M$, horizontal differentials
$d_h^{p,q} : C^{p,q} \longrightarrow C^{p+1,q}$, and vertical differentials
$d_v^{p,q} : C^{p,q} \longrightarrow C^{p,q+1}$ satisfying $d_h^2 = d_v^2 = 0$,
and the compatibility condition that they anticommute:
$d_h d_v = - d_v d_h$.
More generally, we could allow $C^{p,q}$ to be modules, or objects in some Abelian category,
and the maps would then be morphisms in this category.

There are two spectral sequences corresponding to $C$, $'E^{p,q}_r$ and
$''E^{p,q}_r$.  We will describe the first, from which the second is easily obtained.
A spectral sequence is a set of pages
$E^{p,q}_r$, for $r=0, 1, 2, \ldots$.  Each page is a two-dimensional array of vector
spaces $E^{p,q}_r$, for $0 \le p, q \le M$, together with a map $d_r$.
The zeroth page, $E_0$, starts with $C^{p,q}$ in the $(p,q)$
spot, and the map is the
vertical differential $d_0^{p,q} := d_v^{p,q} : C^{p,q} \longrightarrow C^{p,q+1}$.
The differential $d_r$ at each step satisfies the equation $d_r^2 = 0$, and maps
\[ d_r^{p,q} : E^{p,q}_r \longrightarrow E^{p+r,q-r+1}_r. \]
The first page is obtained from the zeroth page by setting
\[ E^{p,q}_1 := \ker d_0^{p,q} / \Ima(d_0^{p,q-1}). \]
The differential $d_1$ is simply the
horizontal map
\[ d_1^{p,q} : E^{p,q}_1 \longrightarrow E^{p+1,q}_1 \]
induced by the
horizontal differential on $C$.
In general, given the $r$-th page of the spectral sequence, the $(r+1)$-st page is the array
\[E^{p,q}_{r+1} := \ker d_r^{p,q} / \Ima(d_r^{p-r,q+r-1})\,.\]
By iterating this procedure, the spectral sequence eventually converges when all terms on the page stabilize. For a first quadrant spectral sequence, the spectral sequence converges to the cohomology of the total complex
\[E^{p,q}_{2} \implies H^{p+q}({\rm{Tot}}(C)) \]
where the total complex is a complex with the $n$-th term defined by \[{\rm{Tot}}(C)^{n} = \bigoplus_{p+q = n}C^{p,q}\] with the natural differential defined by $d = d_{v} + d_{h}$.

\subsection{The hypercohomology spectral sequence}

We now explain
the utility of the spectral sequence discussed above in computing sheaf cohomology.
Let us fix a smooth complex projective variety $X$ and closed subvarieties $D_{1},\ldots,D_{n}$.
We assume that the intersection of each set of $k$ of the $D_i$ is either empty, or has codimension $k$ in $X$, so that by Proposition \ref{MVprop} the corresponding Mayer-Vietoris sheaf sequence $\mathbb{M}(D_{1},\ldots,D_{n})$ is an exact sequence.

Let $\{\mathcal{O}_{D}\}$ denote a complex localized in degree 0 with the term $\mathcal{O}_{D}$ and let
\newline $\{\bigoplus_{i=1}^{n}\mathcal{O}_{D_{i}} \longrightarrow \ldots \longrightarrow \mathcal{O}_{D_{1} \cap \ldots \cap D_{n}}\}$ denote the corresponding complex beginning in degree 0. These complexes are quasi-isomorphic and hence, as objects in the bounded derived category $D^{b}(X)$,
they
are isomorphic. We have the natural global sections functor $\Gamma \colon {\rm{Qcoh}}(X) \rightarrow {\rm{Vec}}(\mathbb{C})$, and by deriving on the right and restricting to the full subcategory of bounded complexes of quasicoherent sheaves with coherent cohomology, we obtain the induced derived functor $R\Gamma \colon D^{b}(X) \rightarrow D^{b}({\rm{Vec}}(\mathbb{C}))$. For a given complex of coherent sheaves $\mathcal{F}^{\bullet}$, we then compute the hypercohomology groups, or higher derived functors, by taking cohomology, namely $H^{i}(X,\mathcal{F}^{\bullet}) = H^{i}R\Gamma(\mathcal{F}^{\bullet})$.

In order to compute the hypercohomology groups of a complex $\mathcal{F}^{\bullet}$, we take a quasi-isomorphic complex of injective objects given by $f \colon \mathcal{F}^{\bullet} \rightarrow \mathcal{I}^{\bullet}$ and compute the cohomology $H^{i}(\Gamma(\mathcal{I}^{\bullet}))$. Such a complex is often constructed as the total complex of the Cartan-Eilenberg resolution $\mathcal{J}^{\bullet,\bullet}$ of $\mathcal{F}^{\bullet}$ where existence is proved by applying the horseshoe lemma. In particular, the hypercohomology $\mathbb{H}(\mathcal{F}^{\bullet})$ is independent of the resolution. However, in practice, injective resolutions are usually hard to find explicitly, and so one often resorts to taking acyclic resolutions.

We wish to compute
\begin{equation}\label{eq:divcohom}
H^{i}(D,\mathcal{O}_{D}) = \mathbb{H}^{i}\Bigl(X, \bigl\{\bigoplus_{i=1}^{n}\mathcal{O}_{D_{i}} \longrightarrow \ldots \longrightarrow \mathcal{O}_{D_{1} \cap \ldots \cap D_{n}}\bigr\}\Bigr)
\end{equation}
In the general situation, given a scheme $(X,\mathcal{O}_{X})$, and given a bounded-below complex $\mathcal{F}^{\bullet}$ of $\mathcal{O}_{X}$-modules, we have the following computational tool:
\begin{lemma}
There exists a spectral sequence with
\begin{equation*}
E_{2}^{p,q} = H^{p}\Bigl(H^{q}(X, \mathcal{F}^{\bullet})\Bigr)
\end{equation*}
converging to $\mathbb{H}^{p+q}(X, \mathcal{F}^{\bullet})$.
\end{lemma}
\newpage

\section{Stratification of Toric Hypersurfaces and Hodge Numbers of Strata}  \label{sec:stratification}

In this appendix we will review a number of results about the stratification of toric varieties and the associated stratification of hypersurfaces.
In particular, we will review how the algorithm of \cite{DK} can be used to conveniently read off the Hodge numbers of toric strata. Although all of these techniques are in principle well-known, a succinct exposition, especially in the physics literature, is lacking. We refer the reader to
\cite{cox2011toric,Danilov,Kreuzer:2006ax} for an introduction to toric geometry. The original reference for how to compute Hodge numbers of toric strata for hypersurfaces is \cite{DK}; see \cite{Batyrev,Klemm:1996ts,Braun:2016igl} for a number of applications close to physics.

\subsection{Toric varieties and stratification}\label{sec:toricstrat}

By definition, a toric variety $V$ is characterized as containing an open dense algebraic torus, $(\C^*)^n$, the action of which (on itself) extends to the whole variety. One can think of an $n$-dimensional toric variety as a (partial) compactification of $(\C^*)^n$ in which various lower-dimensional algebraic tori are added. These can be thought of as `limits' of the original $(\C^*)^n$, so that the action of $(\C^*)^n$ on itself extends naturally to the whole variety.

This structure can be summarized by a combinatorial object called a fan. A fan $\Sigma$ is a collection of strongly convex rational polyhedral cones such that the face of every cone counts as a cone in the fan and two cones only intersect along a face of each. Here, strong convexity demands that no subspace (except the origin) of the ambient vector space is contained in any cone. One can think of each cone as being spanned by a finite number of rays with primitive generators $v_i$ that are elements of a lattice $N$. Finally, we (mostly) make the further assumption that every cone is simplicial, i.e. any $d$-dimensional cone is spanned by $d$ rays. In this case the associated toric variety has at most orbifold singularities and is factorial.

Let us label the (generators of the) rays of $\Sigma$ by $v_i$, with the index $i$ running from $1$ to $r$. The toric variety $\mathbb{P}_\Sigma$ associated with the fan $\Sigma$ can then be described in analogy to complex projective space by associating a homogeneous coordinate $z_i$ to each ray of the fan and forming the quotient
\begin{equation}\label{eq:toricasquotient}
\mathbb{P}_\Sigma = \frac{ \C^r - SR }{G} \, .
\end{equation}
This data is encoded in the fan as follows. First, the exceptional set or Stanley Reisner ideal, $SR$ corresponds to all collections $\{v_i | i \in I\}$ for which the cones $v_i$ do not share a common higher-dimensional cone. Wheneve this is the case, the set $SR$ contains the subspace of $\C^r$ in which the associated collection of coordinates $\{z_i | i \in I\}$ vanishes simultaneously. The Abelian group $G$ is given as the kernel of the homomorphism
\begin{equation}\label{eq:toruscoords}
z_i \rightarrow t_k = \prod_j z_j^{\langle e_k , v_j \rangle}
\end{equation}
where the $e_k$ form a basis of the lattice $M$. Note that this implies that whenever there is a linear relation
\begin{equation}
 \sum_i a_i v_i = 0\,,
\end{equation}
there is an associated one-parameter subgroup $\C^* \in G$ which acts as
\begin{equation}
 (z_1,\cdots,z_r) \sim (\lambda^{a_1}z_1,\cdots,\lambda^{a_r}z_r) \,
\end{equation}
for $\lambda \in \C^*$.
In most cases of interest, $G$ can be completely described through relations like this. In general, however, $G$ may contain discrete factors.
Note that the above discussion implies that the (complex) dimension of $V$ equals the (real) dimension of $N$. In the following, we also use the notation $\mathbb{P}_\Sigma$ to denote the toric variety determined by the fan $\Sigma$.

For all $z_i \neq 0$, the coordinates $t_k$ defined in \eqref{eq:toruscoords} parameterize the open dense $(\C^*)^n$ giving the toric variety its name. The strata that are added to turn $(\C^*)^n$ into a nontrivial toric variety are encoded in the fan as follows. We first associate the unique zero-dimensional cone $\sigma^{[0]}$, i.e. the origin, of $\Sigma$ with $(\C^*)^n$. Each $d$-dimensional cone $\sigma^{[d]}_i$ is naturally associated with the homogeneous coordinates $z_i$ corresponding to its generating rays. Choosing the basis $e_k$ in \eqref{eq:toruscoords} from lattice vectors contained in the dual cone
\begin{equation}\label{eq:dualcones}
\langle \check{\sigma},\sigma \rangle  \geq 0 \, ,
\end{equation}
which sits in $\check{\sigma} \in M\otimes \R$, where $M$ is the dual lattice to $N$, we may take a limit of \eqref{eq:toruscoords} where we set the $z_i = 0$ for all $v_i$ generating $\sigma$. As a $d$-dimensional cone $\sigma^{[d]}$ has $d$ generators, this lands us on an $n-d$-dimensional algebraic torus
$T_{\sigma^{[d]}} \simeq (\C^*)^{n-d}$ and the definition of a fan ensures that all of these strata are consistently sewn together.

Denoting the set of $d$-dimensional cones by
\begin{equation}
 \Sigma(d) = \{\sigma^{[d]}_k\} \, ,
\end{equation}
we can hence stratify any toric variety according to the data of our fan as
\begin{equation}\label{eq:toricstrata}
 \mathbb{P}_\Sigma = \coprod_d \coprod\limits_{\sigma^{[d]}_k \in \Sigma(d)} T_{\sigma^{[d]}_k}\quad  = \quad\coprod_d \coprod\limits_{\sigma^{[d]}_k \in \Sigma(d)} (\mathbbm{C}^{*})^{n-d}\, .
\end{equation}
In terms of the homogenous coordinates $z_i$, the stratum of each cone $\sigma$ is described by setting
\begin{equation}
\{z_i = 0 \quad \forall v_i \in \sigma \} \hspace{.5cm} \mbox{and} \hspace{.5cm} \{z_i \neq 0 \quad \forall v_i \notin \sigma \}\, .
\end{equation}

Let us discuss some examples. The fan corresponding to $\mathbb{P}^1$ lives in $\R$ and is composed of three cones: the origin, a ray generated by $1$ and a ray generated by $-1$. We recover the standard presentation from \eqref{eq:toricasquotient} as
\begin{equation}
 \mathbb{P}^1 = \frac{ (z_1,z_2)  - \{(0,0)\}}{(z_1,z_2) \sim (\lambda z_1, \lambda z_2)} \, .
\end{equation}
The open dense $\C^*$ can be described by the coordinate $z_1/z_2$, or equivalently by $z_2/z_1$, and the strata corresponding to the two one-dimensional cones are simply given by $z_1=0$ and $z_2=0$, respectively. We hence recover the description of $\mathbb{P}$ as the Riemann sphere, i.e. adding the point at infinity to $\C$.

\begin{figure}
 \begin{center}
  \includegraphics[height=8cm]{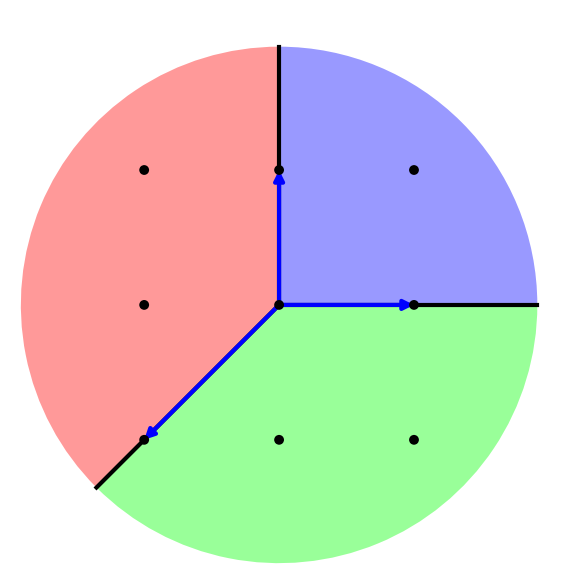}\caption{\label{fig:p2fan}The fan of $\mathbb{P}^2$ contains a zero-dimensional cone, three rays (one-dimensional cones) and three two-dimensional cones. }
 \end{center}
\end{figure}

The fan corresponding to $\mathbb{P}^2$ is shown in figure \ref{fig:p2fan}. Again, \eqref{eq:toricasquotient} gives the standard presentation as
\begin{equation}
 \mathbb{P}^2 = \frac{ (z_1,z_2,z_3)  - \{(0,0,0)\}}{(z_1,z_2,z_3) \sim (\lambda z_1, \lambda z_2, \lambda z_3)} \, .
\end{equation}
From the fan, we can directly read off the stratification data \eqref{eq:toricstrata}, which becomes
\begin{equation}\label{eq:stratp2}
 \mathbb{P}^2 = (\C^*)^2 \coprod_{i=1..3} \C^* \coprod_{k=1..3} pt.
\end{equation}

In the bulk of this paper we are not interested in the geometry of toric varieties per se, but rather in the geometry of algebraic subvarieties. In the simplest setting, such an algebraic subvariety $Z$ is given as the vanishing locus of a section of a line bundle $\mathcal{L}$ on $V$. For such an algebraic subvariety $Z$, one easily obtains a stratification of $Z$ from that of $V$ if all toric strata meet $Z$ transversely. In this case we can define
an $n-d-1$-dimensional stratum
\begin{equation}
 Z_{\sigma^{[d]}} = Z \cap T_{\sigma^{[d]}}
\end{equation}
for every $d$-dimensional cone $\sigma^{[d]} \in \Sigma$. The stratification of the hypersurface $Z$ is then simply
\begin{equation}\label{eq:stratHS}
 Z =  \coprod_d \coprod\limits_{\sigma^{[d]}_k \in \Sigma(d)}Z_{\sigma^{[d]}_k} \, .
\end{equation}

Let us illustrate this in the simple example of hypersurface of degree $k$ in $\mathbb{P}^2$, i.e. we consider the vanishing locus of a homogeneous polynomial $P_k(z_1,z_2,z_3)$ of degree $k$ in the homogeneous coordinates of $\mathbb{P}^2$. Call the resulting curve $C_k$.
By adjunction one finds that such a hypersurface is a Riemann surface of genus
\begin{equation}
g(k) = \frac{(k-1)(k-2)}{2}  \, .
\end{equation}
Let us now see how $C_k$ is stratified using \eqref{eq:stratHS}. First consider the lowest dimensional strata of $\mathbb{P}^2$, which correspond to the highest-dimensional cones. Each of the three two-dimensional cones of the fan of $\mathbb{P}^2$ corresponds to a point on $\mathbb{P}^2$
that is defined by setting two of the three homogeneous coordinates to zero. For a sufficiently generic polynomial $P_{k}$, such points will never lie on $Z$, so that these strata do not contribute to $Z$ (indeed they are supposed to be $2-2-1=-1$-dimensional). For each of the three one-dimensional cones in the fan of $\mathbb{P}^2$, there is a stratum $\C^*$ obtained by setting one of the three homogeneous coordinates to zero while forbidding the remaining two from vanishing. Intersecting this with $P_{k} = 0$ we find that $Z_{\sigma^{[1]}}$ consists of $k$ points for every one-dimensional cone $\sigma^{[1]}$. Finally, the complex one-dimensional stratum $Z_{\sigma^{[0]}}$ is a Riemann surface of genus $g(k)$ with $3 \cdot k$ points excised. Below, we will show how to reproduce these numbers combinatorially by using the Newton polytope of $P_{k}$.

\subsection{Toric varieties and divisors from polytopes}\label{sect:normalfans}

In order to describe how to determine the geometry of strata $ Z_{\sigma^{[d]}} $ combinatorially, we need to present the situation of interest from a slightly different perspective. Note first that the data defining the topology of $Z$ consists of the toric variety $V$ (or, equivalently, the fan $\Sigma$) and a line bundle $\mathcal{L}$ on $V$. We can write the divisor class $D$ of $\mathcal{L}$ in terms of the toric divisors $D_i$ corresponding to rays of $\Sigma$ as
\begin{equation}\label{eq:divisorandNF}
D = c_1(\mathcal{L}) = \sum_i a_i D_i \, .
\end{equation}
The group of holomorphic sections of $\mathcal{L}$ is then given by a polytope $\Delta$, known as the Newton polytope, defined by
\begin{equation}
P_{D} = \{ m \in M \, |\, \langle m, v_i \rangle \geq -a_i \, \forall \, v_i \}\, .
\end{equation}
More explicitly, we may use the monomials
\begin{equation}\label{eq:sectlbundle}
p(m) = \prod_i z_i^{\langle m,v_i \rangle + a_i}
\end{equation}
as a basis for the group of sections. This provides a convenient way to find the zeroth
cohomology group (which is in fact the only non-vanishing one) of a divisor $D$ (line bundle $\mathcal{L}$), as this counts global holomorphic sections of
$\mathcal{L}$:
\begin{equation}
h^{0}(\mathbb{P}_{\Sigma},\mathcal{O}_{\mathbb{P}_{\Sigma}}(D)) = |P_{D} \cap M| \, .
\end{equation}
Interestingly, $\Delta$ determines both the line bundle $\mathcal{L}$ and (a blowdown of) the toric variety $V$. To any polytope $\Delta$ in the $M$-lattice, we can associate its normal fan $\Sigma_n(\Delta)$, which then gives rise to a toric variety $\mathbb{P}_{\Sigma_n(\Delta)}$ along with a divisor $D$. This works as follows. To every face $\Theta^{[k]}$ of the polytope $\Delta$, we associate a cone
\begin{equation}
\check{\sigma}_n(\Theta^{[k]}) = \bigcup_{r \geq 0} r \cdot (p_\Delta - p_{\Theta^{[k]}})
\end{equation}
where $p_\Delta$ is any point lying inside $\Delta$ and $p_\Theta^{[k]}$
is any point lying inside $\Theta^{[k]}$. The dual cones $\sigma_n(\Theta^{[k]})$ (defined as in \eqref{eq:dualcones}) form a complete fan that is called the normal fan $\Sigma_n(\Delta)$ of $\Delta$. Here, $k$-dimensional faces $\Theta^{[k]}$ of $\Delta$ are associated with $d=n-k$-dimensional cones $\sigma_n(\Theta^{[k]})$. In particular, the cones of $\Sigma(\Delta)$ of highest dimension are associated with the vertices of $\Delta$.

On the toric variety $\mathbb{P}_{\Sigma_n(\Delta)}$, the polytope $\Delta$ then determines a line bundle (Cartier divisor) via a strongly convex support function $\psi_\Delta$.  For each cone of maximal dimension of $\Sigma_n(\Delta)$, $\Psi_\Delta$ can be described by
using its dual vertex $m_i$ and setting
\begin{equation}
\left.\Psi_\Delta\right|_{\sigma_n(m_i)}  (p) =  \langle m_i , p \rangle
\end{equation}
for each point $p$ in $\sigma_n(m_i)$. This also determines $\Psi_\Delta$ for all cones of lower dimension.
The divisor
\begin{equation}
D_\Delta = \sum_{\nu_i \in  \Sigma_n(1)}  a_i D_i
\end{equation}
can then be determined from
\begin{equation}
\left.\Psi_\Delta\right|_{\sigma_n(m_i)}  (\nu_i) = - a_i \,\,\, \forall \nu_i \in \sigma_n(m_i)\,.
\end{equation}
With the numbers $a_i$ we can recover a basis for the sections of the line bundle $\mathcal{L}$ via \eqref{eq:sectlbundle}.

The above relations can be used to associate a cone-wise linear support function $\Psi_D$ to any divisor $D$. If the line bundle
associated to $D$ is ample, $\Psi_D$ is strongly convex (i.e. it is convex and different for each cone of maximal dimension). A toric variety is projective
iff its fan is the normal fan of a lattice polytope \cite{Fulton}.

Let us come back to the example of $\mathbb{P}^2$ and a hypersurface $C_k$ determined by $P_k = 0$. We may write
\begin{equation}
 [P_k] =  k D_3 \, .
\end{equation}
in terms of toric divisors, so that $a_1 = a_2=0$ and $a_3 = k$. As shown in figure \ref{fig:p2fan}, the one-dimensional cones of $\Sigma$ are generated by the vectors $v_1 = (1,0),\, v_2=(0,1),\, v_3=(-1,-1)$. The Newton polytope $\Delta_k$ corresponding to $P_k = 0$ has vertices
\begin{equation}\label{eq:vertsdegreerinp2}
(k,0), (0,k), (0,0)
\end{equation}
and its integral points correspond to all monomials of homogeneity degree $k$ by using \eqref{eq:sectlbundle}. In particular
\begin{equation}
p((k,0)) = z_1^{\langle (k,0),(1,0) \rangle} z_2^{\langle (k,0),(0,1) \rangle } z_3^{\langle (k,0),(-1,-1) \rangle +k } = z_1^k\,, \hspace{.5cm} p((0,k)) = z_2^k \,, \hspace{.5cm} p((0,0)) = z_3^k
\end{equation}

\subsection{Resolution of singularities}\label{sect:res}

As there is a one-to-one correspondence between faces of $\Delta$ and cones of $\Sigma_n(\Delta)$, we may write the stratification \eqref{eq:stratHS} in terms of faces of $\Delta$ instead of cones of $\Sigma_n(\Delta)$:
\begin{equation}\label{eq:stratHSfromdelta}
 Z =  \coprod_{k}\coprod_{\Theta^{[k]}_i} Z_{\Theta^{[k]}_i} \, .
\end{equation}
As faces of dimension $k$ correspond to cones of dimension $d = n-k$, the dimension of these strata is $n-d-1 = k-1$. Note that $\Delta$ counts as a face of itself, which corresponds to the zero-dimensional cone of $\Sigma_n(\Delta)$. We will discuss in Appendix \ref{sect:dkalgorithm} how to determine the topology of the strata $ Z_{\Theta^{[k]}_i}$ appearing above from the Newton polytope.

What we have ignored so far, however, is that the normal fan $\Sigma_n(\Delta)$ in most cases (contrary to our simple example) does not give rise to a smooth toric variety. In fact, the fan $\Sigma_n(\Delta)$ does not even need to be simplicial.\footnote{A fan is simplicial if all of its $d$-dimensional cones are generated by $d$ rays. Fans that are not simplicial give rise to toric varieties with singularities that are more general than orbifold singularities. In particular, not every Weil divisor is $\mathbb{Q}$-Cartier for such varieties.}
We are hence interested in a refinement $\Sigma$ of the fan $\Sigma_n(\Delta)$ in order to resolve the hypersurface $Z$. Fortunately, this process results in only minor modifications of the stratification \eqref{eq:stratHSfromdelta} above. In a slight abuse of notation, we will use the same letter $Z$ also for the resolved hypersurface. Under a refinement $\pi: \Sigma \rightarrow \Sigma(\Delta)$, the stratification associated with faces of $\Delta$ becomes:
\begin{equation}\label{eq:stratZresolved}
 Z = Z_{\Delta} \coprod_i Z_{\Theta^{[n-1]}_i} \coprod_{k\geq 2} \coprod_l E_{\Theta^{[n-k]}_l} \times Z_{\Theta^{[n-k]}_l} \, .
\end{equation}
Here $E_{\Theta^{[n-k]}}$ is the exceptional set of the refinement of the cone in $\Sigma(\Delta)$ associated with the face $\Theta^{[n-k]}$,
\begin{equation}
 E_{\Theta^{[n-k]}} = \coprod_{i=0}^{n-k-1} (\C^*)^{i} \, .
\end{equation}
For every $l$-dimensional cone in $\pi^{-1}(\sigma)$, where $\sigma \in \Sigma(\Delta)$ is the cone in the normal fan associated with
the face $\Theta^{[n-k]}$, there is a corresponding stratum $(\C^*)^{k-l}$ in $E_{\Theta^{[n-k]}}$.

\subsection{Computing Hodge-Deligne numbers of strata}\label{sect:dkalgorithm}

In order to compute the Hodge numbers of toric hypersurfaces and, more generally, toric strata in such hypersurfaces, we need to introduce a further piece of machinery. The strata appearing in the stratification \eqref{eq:stratZresolved} naturally carry a mixed Hodge structure. In very simple terms, this means that the Hodge-Deligne numbers
\begin{equation}
 h^{p,q}(H^k(X,\mathbb{C}))
\end{equation}
can be nonzero even when $p+q \neq k$, see e.g. \cite{voisin2002hodge,voisin2003hodge} for a proper introduction.

These data can be packed into the numbers (which we will also call Hodge-Deligne numbers in the following)
\begin{equation}
 e^{p,q}(X) = \sum_k (-1)^k  h^{p,q}(H^k(X,\mathbb{C})) \, ,
\end{equation}
which are convenient for a number of reasons. First of all, in case these is a pure Hodge structure, they agree (up to a sign) with the usual Hodge numbers. Secondly, they behave in the same way as the topological Euler characteristic under unions and products of spaces
\begin{align}
e^{p,q}(X_1 \amalg X_2) & = \, e^{p,q}(X_1) + e^{p,q}(X_2) \\
e^{p,q}(X_1 \times X_2) & = \sum_{\substack{p_1+p_2 = p \\ q_1+q_2=q}} e^{p_1,q_1}(X_1)\, \cdot \,e^{p_2,q_2}(X_2) \, .
\end{align}

Hence knowing the numbers $e^{p,q}(Z_{\Theta^{[k]}})$ (and of $(\C^*)^n)$ is sufficient to compute the Hodge numbers of a toric hypersurface. This information has been supplied (in the form of an algorithm) in the work of \cite{DK}. Before reviewing their algorithm, let us first discuss toric varieties themselves to illustrate the method. We have that
\begin{equation}
e^{p,q}((\C^*)^n) = \delta_{p,q} (-1)^{n+p} \binom{n}{p} \, .
\end{equation}
For a smooth toric variety $V$, we hence see that $h^{p,0} = 0$ for $p>0$. A direct consequence of this formula combined with the stratification of a toric variety read off from its fan is the standard formula \cite{Fulton} for the nontrivial Hodge numbers of a smooth $n$-dimensional toric variety.
Letting $|\Sigma(l)|$ denote the number of $l$-dimensional cones in $\Sigma$ we can write
\begin{equation}
 h^{p,p}(\mathbb{P}_\Sigma) = \sum_{k=0}^n (-1)^{p+k} \binom{k}{p}|\Sigma(n-k)|\, .
\end{equation}

For the smooth hypersurface strata $Z_{\Theta^{[k]}}$, the following relations, shown in \cite{DK}, allow to compute the Hodge-Deligne numbers. First we have that for $p>0$
\begin{align}\label{eq:ep0}
 e^{p,0}(Z_{\Theta^{[k]}}) = (-1)^{k-1} \sum_{\Theta^{[p+1]} \leq \Theta^{[k]}} \ell^*(\Theta^{[p+1]})
\end{align}
where $\ell^*(\Theta^{[p+1]})$ counts the number of lattice points in the relative interior of $\Theta^{[p+1]}$ and the sum runs over all faces
$\Theta^{[p+1]}$ of dimension $p+1$ contained in the face $\Theta^{[k]}$. Note that this sum only has one term for $p+1=k$ which corresponds to the face
$\Theta^{[k]}$ itself.

The remaining Hodge-Deligne numbers satisfy the `sum rule'
\begin{equation}\label{eq:DKsumrule}
 (-1)^{k-1} \sum_q e^{p,q}(Z_{\Theta^{[k]}}) = (-1)^p \binom{k}{p+1} + \varphi_{k-p}(\Theta^{[k]}) \, ,
\end{equation}
where the function $\varphi_n(\Theta^{[k]})$ is defined by
\begin{equation}
\varphi_n(\Theta^{[k]}) := \sum_{j = 1}^n (-1)^{n+j} \binom{k+1}{n-j}\ell^*(j\Theta) \, .
\label{eq:def-varphi}
\end{equation}
Here $j \Theta$ denotes the polytope that is found by scaling the face $\Theta$ by $j$.

For a face of dimension $k \geq 4$ there is the simple formula
\begin{equation}\label{eq:e31}
 e^{k-2,1}(Z_\Theta^{[k]}) = (-1)^{k-1}\left(\varphi_2(\Theta^{[k]}) -  \sum_{\Theta^{[k-1]} \leq \Theta^{[k]}} \varphi_1(\Theta^{[k-1]}) \right) .
\end{equation}
Finally, for higher Hodge numbers, $ p+q\geq n$, one has
\begin{equation}
 e^{p,q}(Z_{\Theta^{[n]}}) =  \delta_{p,q}(-1)^{n+p+1}\binom{n}{p+1} \, .
\end{equation}

By subsequent application of these formulae, one can derive combinatorial formulae for the Hodge-Deligne numbers of strata $Z_{\Theta^{[k]}}$ for arbitrarily high $k$. It is convenient to use $\ell^n(\Theta)$ to denote the number of points on the $n$-skeleton of $\Theta$. Let us derive the Hodge-Deligne numbers $e^{p,q}$ of strata $Z_{\Theta^{[k]}}$ for $k \leq 3$. First,
\begin{equation}\label{eq:e00Z}
 e^{0,0}(Z_{\Theta^{[k]}}) = (-1)^{k-1}\left( \ell^1(\Theta^{[k]}) - 1 \right) \, ,
\end{equation}
so that $e^{0,0}(Z_{\Theta^{[1]}}) = \ell^*(\Theta^{[1]})+1$, i.e. the stratum $Z_{\Theta^{[1]}}$ consists of $\ell^*(\Theta^{[1]})+1$ points.
Hence
\begin{equation}
e^{p,q}(Z_{\Theta^{[1]}}) =
\begin{array}{|cc}
  0 & 0 \\
 \ell^*(\Theta^{[1]})+1 & 0 \\
 \hline
\end{array}
\end{equation}

For $Z_{\Theta^{[2]}}$, we immediately find $e^{1,1}(Z_{\Theta^{[2]}}) =1 $. Furthermore, $e^{1,0}(Z_{\Theta^{[2]}}) = - \ell^*(\Theta^{[2]})$, and we can write
\begin{equation}
 e^{p,q}(Z_{\Theta^{[2]}}) =
 \begin{array}{|cc}
 - \ell^*(\Theta^{[2]}) & 1 \\
1 - \ell^1(\Theta^{[2]}) & - \ell^*(\Theta^{[2]}) \\
 \hline
 \end{array}
\end{equation}

Similarly, we find for $Z_{\Theta^{[3]}}$ that
\begin{equation}\label{eq:hijofZT3}
 \begin{array}{|ccc}
  \ell^*(\Theta^{[3]}) & 0 & 1 \\
\ell^2(\Theta^{[3]}) - \ell^1(\Theta^{[3]}) \quad\quad  & -3 + \ell^*(2\Theta^{[3]})-4\ell^*({\Theta^{[3]}}) - \ell^2(\Theta^{[3]}) + \ell^1(\Theta^{[3]})& 0 \\
\ell^1(\Theta^{[3]}) - 1 & \ell^2(\Theta^{[3]}) - \ell^1(\Theta^{[3]})  &   \ell^*(\Theta^{[3]}) \\
\hline
 \end{array}
\end{equation}
and for $Z_{\Theta^{[4]}}$
\begin{equation}
\begin{aligned}
 e^{0,0}(Z_{\Theta^{[4]}}) & = 1 - \ell^1(\Theta^{[4]}) \\
 e^{1,0}(Z_{\Theta^{[4]}}) & =  -\ell^2(\Theta^{[4]}) + \ell^1(\Theta^{[4]}) \\
 e^{2,0}(Z_{\Theta^{[4]}}) & = - \ell^3(\Theta^{[4]}) + \ell^2(\Theta^{[4]}) \\
 e^{2,1}(Z_{\Theta^{[4]}}) & = \ell^3(\Theta^{[4]}) - \ell^2(\Theta^{[4]}) - \varphi_2(\Theta^{[4]}) \\
 e^{2,2}(Z_{\Theta^{[4]}}) & = -4  \, .
\end{aligned}
\end{equation}

Let us return to the example of a degree $k$ hypersurface in $\mathbb{P}^2$. The stratification of $\mathbb{P}^2$ has already been discussed in \S\ref{sec:toricstrat}, see \eqref{eq:stratp2}, and its Newton polytope $\Delta_k$ is given in \eqref{eq:vertsdegreerinp2}.
The open dense torus $(\C^*)^2$ in $\mathbb{P}^2$ gives rise to an open dense subset of $C_k$, i.e. a Riemann surface with a number of points excised. Its Hodge-Deligne numbers are given by
 \begin{equation}
 \begin{aligned}
 e^{p,q}(Z_{\Delta_k})  & =
 \begin{array}{|cc}
 - \ell^*(\Delta_k) & 1 \\
1 - \ell^1(\Delta_k) & - \ell^*(\Delta_k) \\
 \hline
 \end{array} \\
& =
 \begin{array}{|cc}
 - (k-1)(k-2)/2 & 1 \\
1 - 3k & - (k-1)(k-2)/2 \\
 \hline
 \end{array}
 \end{aligned}
\end{equation}
The stratum associated with each of the three 1-faces $\Theta^{[1]}$ of $\Delta_k$ consists of
\begin{equation}
1 + \ell^*(\Theta^{[1]}) =  k
\end{equation}
points. We hence recover the results derived in \S\ref{sec:toricstrat} for this example.

\subsection{Calabi-Yau hypersurfaces and reflexive polytopes}

Let us now come back to our prime interest, which is in Calabi-Yau manifolds. We assume that the dimension of the vector space $N\otimes \mathbb{R}$ containing the fan $\Sigma$ is $n$, so that we get an n-dimensional toric variety $\mathbb{P}_\Sigma$ in which we want to embedd a Calabi-Yau hypersurface of dimension $n-1$.

In order for a hypersurface to be Calabi-Yau, its defining polynomial must be a section of the anticanonical bundle $-K_{\mathbb{P}_\Sigma}$. The corresponding divisor is hence given by
\begin{equation}
D_{-K} = c_1(\mathbb{P}_\Sigma) = \sum_{\nu_i \in \Sigma(1)} D_i  \, .
\end{equation}
In the notation of \eqref{sect:normalfans}, we hence have $a_i = 1$ for all $i$ and we can identify sections of $-K_{\mathbb{P}_\Sigma}$ with the set of lattice points on the polytope
\begin{equation}
\Delta \equiv \{m \in M | \langle m,\nu \rangle \geq -1 \,\, \forall \nu \in \Sigma(1)\} \, \subset M \otimes \mathbb{R} .
\end{equation}
A general section of $-K_{\mathbb{P}_\Sigma}$, the zero locus of which defines a Calabi-Yau hypersurface, is then given by
\begin{equation}\label{eq:cypolynomial_reflexive_pair}
\sum_m \alpha_m\, p(m) = \sum_{m \in \Delta} \prod_{\nu_i \in \Sigma(1)}\alpha_m z_i^{\langle m,\nu_i \rangle + 1}
\end{equation}
for complex constants $\alpha_m$.

In general, the vertices of the polytope $\Delta$ are not lattice points of $M$ and we are not guaranteed that a generic section of $-K_{\mathbb{P}_\Sigma}$ defines a smooth (or even irreducible) Calabi-Yau hypersurface. If all of the vertices of $\Delta$ are contained in $M$ (in which case we will call $\Delta$ a lattice polytope), it follows that the vertices $\nu_i$ are all sitting on a lattice polytope $\Delta^\circ$, defined by
\begin{equation}
 \langle \Delta, \Delta^\circ \rangle \geq -1 \, .
\end{equation}
$\Delta^\circ$ is called the polar dual of $\Delta$, and $\Delta$ and $\Delta^\circ$ are called a reflexive pair \cite{Batyrev} if they are both lattice polytopes. Any lattice polytope whose polar dual is also a lattice polytope is called reflexive. A necessary condition for reflexivity is that the origin is the unique interior point of the polytope in question.

Repeating the construction of \S\ref{sect:normalfans}, Calabi-Yau hypersurfaces in toric varieties are naturally constructed from reflexive pairs of lattice polytopes $\Delta,  \Delta^\circ$. Starting from a lattice polytope $\Delta$, we may construct its normal fan $\Sigma_n(\Delta)$. In the case of a reflexive pair, this is equal to the fan over the faces of $\Delta^\circ$. Of course, such a fan does not in general define a smooth toric variety. In fact, the cones need not even be simplicial. However, there is a natural maximal projective crepant partial (MPCP) desingularization that can be found as follows. Using \eqref{eq:cypolynomial_reflexive_pair}, it follows that any refinement of the fan $\Sigma_n(\Delta)$ that only introduces rays generated by lattice points $\nu$ on $\Delta^\circ$ is crepant, i.e.~preserves the Calabi-Yau property. We may hence find a MPCP desingularization by a fan refinement
$\Sigma \rightarrow \Sigma_n(\Delta)$ for which all lattice points $\nu_i$ on $\Delta^\circ$ are employed and for which $\mathbb{P}_\Sigma$ is a projective toric variety. This is equivalent to finding a fine regular star triangulation $T$ of $\Delta$. Here, fine means that all lattice points of $\Delta^\circ$ are used, and star means that every simplex contains the origin as a vertex.\footnote{The simplices of the triangulations are hence the cones of the fan $\Sigma$, cut off at the surface of $\Delta^\circ$. Note that one may start from any triangulation $T$ of $\Delta^\circ$, restrict it to a triangulation $T_{\partial \Delta^\circ}$ of the faces of $\Delta^\circ$ and then simply construct the cones over $T_{\partial \Delta^\circ}$.}
Projectivity of a toric variety is equivalent to its fan being the normal fan of a lattice polytope. While the toric variety $\mathbb{P}_{\Sigma_n(\Delta)}$ is projective by construction, this is not necessarily true for a refinement $\Sigma \rightarrow \Sigma_n(\Delta)$. Triangulations $T$ for which the associated fan $\Sigma(T)$ is the normal fan of a polytope are called regular (or projective, coherent) in the literature, see \cite{rambau} for more details. Finally, a fine triangulation is in general not sufficient to completely resolve\footnote{Of course, we can always find non-crepant resolutions by introducing rays generated by lattice points outside of $\Delta^\circ$.} all singularities of $\mathbb{P}_{\Sigma_n(\Delta)}$ meeting a generic Calabi-Yau hypersurface of dimension $\geq 4$ \cite{Batyrev}. The reason for this is that having a fine triangulation of a 2-face of a reflexive polytope implies that the corresponding cones do not lead to any singularities. The first dimension in which singularities can persist even for fine triangulations of a reflexive polytope is
$n=5$, i.e.~Calabi-Yau fourfolds. For Calabi-Yau threefolds we are considering four-dimensional polytopes. Here, simplices in 3-faces can lead to pointlike singularities of the ambient toric variety even for fine triangulations, but these do not meet a generic Calabi-Yau hypersurface. In contrast, the four-dimensional cones associated with three-simplices for five-dimensional polytopes lead to singularities along curves which may meet a generic Calabi-Yau fourfold hypersurface.

For a pair of $n$-dimensional reflexive polytopes, there is a one-to-one correspondence between the faces $\Theta^{[k]}$ of $\Delta$ and the faces $\Theta^{\circ [n-k-1]}$ of $\Delta^\circ$ defined by
\begin{equation}
 \langle \Theta^{[k]} , \Theta^{\circ [n-k-1]} \rangle = -1 \, .
\end{equation}
Under the resolution induced by the refinement $\Sigma \rightarrow \Sigma_n(\Delta)$, the stratum $Z_\Theta^{[k]}$ of a Calabi-Yau hypersurface, which corresponds to a $(k-1)$-dimensional subvariety of $Z$ before resolution, is changed according to the new simplices introduced in the dual face $\Theta^{\circ [n-k-1]}$. A simplex of dimension $l$ (a cone of dimension $l+1$) corresponds to a subvariety of $Z$ of dimension $n-l-2$. Hence a simplex of dimension $l$ that is contained in the interior of a face $\Theta^{\circ [n-k-1]}$ corresponds to a subvariety of the form
\begin{equation}
 Z_{\Theta^{[k]}} \times (\C^*)^{n-l-k-1} \, .
\end{equation}
Note that vertices $\Theta^{[0]}$ of $\Delta$, which are dual to the faces $\Theta^{\circ[n-1]}$ of maximal dimension, correspond to $-1$-dimensional varieties, i.e. they do not contribute in the stratification of $Z$. This persists after the resolution $\Sigma \rightarrow \Sigma_n(\Delta)$.  A simple intersection argument shows that none of the divisors corresponding to points $\nu$ interior to a face of maximal dimension (and hence none of the other strata corresponding to simplices interior to a face of maximal dimension) intersect a smooth Calabi-Yau hypersurface. For any face of maximal dimension (also called a facet) $\Theta^{\circ [n-1]}$ we can find a normal vector $n$ (this is the dual vertex $\Theta^{[0]}$) such that $\langle n,\Theta^{\circ [n-1]}\rangle\,= -1$. This means that there is a linear relation of the form
\begin{equation}
 \sum_{\nu_i \in \Theta^{\circ [n-1]}} D_i + \sum_{\nu_j\notin \Theta^{\circ [n-1]}} a_j D_j = 0 \, ,
\end{equation}
for some integers $a_j$. Let us now assume we have refined $\Sigma$ such that there is a point $\nu_p$ interior to the facet $\Theta^{\circ [n-1]}$. The associated divisor $D_p$ can only have a nonzero intersection with divisors $D_k$ for which $\nu_k$ also lies in $\Theta^{\circ [n-1]}$, as all others necessarily
lie in different cones of the fan $\Sigma$. This means that the above relation implies
\begin{equation}
D_p \cdot \sum_{\nu_i \in \Theta^{\circ [n-1]}} D_i = 0 \, ,
\end{equation}
where we sum over all toric divisors coming from points on $\Theta^{\circ [n-1]}$. The Calabi-Yau hypersurface is given as the zero locus of
a section of $-K_{\mathbb{P}^n_\Sigma} = \sum_j D_j$, where we sum over all toric divisors. We now find
\begin{equation}
D_p \cdot \sum_j D_j = D_p \cdot \sum_{\nu_i \in \Theta^{\circ [n-1]}} D_i = 0 \, ,
\end{equation}
by using the same argument again. Hence $D_p$ does not meet a generic Calabi-Yau hypersurface. Correspondingly, a refinement of $\Sigma$ introducing $\nu_p$ does not have any influence on $Z$. As the strata corresponding to simplices of dimension $\geq 1$ interior to $\Theta^{\circ [n-1]}$ can be thought
of as (an open subset of) intersections of divisors, at least one of which corresponds to an interior point of $\Theta^{\circ [n-1]}$, none of the simplices interior to a face of maximal dimension gives rise to any subvariety of $\mathbb{P}_\Sigma$ meeting $Z$. Correspondingly, such strata do not appear in the stratification of $Z$. The fact that all simplices contained in faces of maximal dimension of $\Delta^\circ$ do not contribute to Calabi-Yau hypersurfaces means that we can ignore such faces when constructing a triangulation of $\Delta^\circ$.

Using the methods explained above, one can derive combinatorial formulas for the Hodge numbers of toric Calabi-Yau hypersurfaces \cite{Batyrev} that do not depend on the specific triangulation chosen. For a Calabi-Yau hypersurface of dimension $n-1$ which is embedded in a toric variety of dimension $n$ we have to consider a pair of reflexive polytopes of dimension $n$ and stratification gives
\begin{align}
h^{1,1}(Z)   & = \ell(\Delta^\circ) - (n+1) -
 \sum_{\Theta^{\circ[n-1]}} \ell^*(\Theta^{^\circ [n-1]})
+ \sum_{(\Theta^{\circ[n-2]},\Theta^{[1]})}
     \ell^*(\Theta^{\circ[n-2]})\ell^*(\Theta^{[1]})
 \label{eq:bath11} \\
h^{n-2,1}(Z) & = \ell(\Delta) - (n+1)
  - \sum_{\Theta^{[n-1]}} \ell^*(\Theta^{[n-1]})
  + \sum_{(\Theta^{[n-2]},\Theta^{\circ[1]})}
      \ell^*(\Theta^{[n-2]})\ell^*(\Theta^{\circ[1]})
 \label{eq:bath31} \\
h^{m,1}(Z) & = \sum_{(\Theta^{\circ[n-m-1]}, \Theta^{[m]})}
 \ell^*(\Theta^{\circ[n-m-1]}) \ell^*(\Theta^{[m]})\quad
\mbox{for} \quad n-2 > m > 1. \label{eq:bath21}
\end{align}
Note that these numbers only make sense for a smooth Calabi-Yau hypersurface, which is only guaranteed without further investigation for Calabi-Yau hypersurfaces of dimension $\leq 3$.

Although the above formulas for $h^{1,1}(Z)$ and $h^{n-2,1}(Z)$ are derived using the stratification technique of \cite{DK}, they have a straightforward explanation. In particular, the formula for $h^{n-2,1}(Z)$ counts the number of complex structure deformations by counting the number of monomial deformations appearing in the defining equation and subtracting the dimension of the automorphism group of $\mathbb{P}_\Sigma$. Finally, the last term in \eqref{eq:bath31} corrects for the fact that not all deformations are realized as polynomial deformations. Similarly, the formula for $h^{1,1}(Z)$ counts the number of inequivalent divisors of $\mathbb{P}_\Sigma$ that meet $Z$, with a correction term taking into account that some divisors of $\mathbb{P}_\Sigma$ become reducible on $Z$.

As is apparent from the above formulae, exchanging the roles of $\Delta$ and $\Delta^\circ$ exchanges $h^{1,1}(Z)\, \leftrightarrow \, h^{2,1}(Z)$. This is how mirror symmetry is realized for toric Calabi-Yau hypersurfaces.

\subsection{Topology of subvarieties of Calabi-Yau threefolds}\label{sect:topcy3divs}

In this section we describe the topology of subvarieties of Calabi-Yau hypersurfaces in toric varieties that are obtained by restricting toric subvarieties of the ambient space. For ease of notation we restrict to the case $n=4$, i.e.~Calabi-Yau threefolds, but a similar analysis may be carried out in higher dimensions.  As we have already explained, we only need to consider simplices on the 2-skeleton of $\Delta^\circ$, as strata of $\mathbb{P}_\Sigma$ associated with simplices interior to 3-faces of $\Delta^\circ$ do not meet a smooth Calabi-Yau hypersurface. Each $l$-simplex of a triangulation $T$ of $\Delta^\circ$ corresponds to a $l+1$-dimensional cone in the fan $\Sigma$ and hence to an open stratum $(\C^*)^{4-(l+1)}$ in $\mathbb{P}_\Sigma$. Depending on the location of the simplex on $\Delta^\circ$, the defining equation of the Calabi-Yau hypersurface will only constrain some of the $\C^*$ factors, while others will lie entirely in the Calabi-Yau hypersurface. The reason for this is in the resolution process $\Sigma \, \rightarrow \Sigma_n(\Delta)$. If we work with the singular varieties determined by $\Sigma_n(\Delta)$, every $k$-dimensional
face $\Theta^{\circ [k]}$ of $\Delta^\circ$ gives rise to a stratum $Z_{\Theta^{[4-k-1]}}$ of dimension $2-k$. This stratum is given as a hypersurface in $(\C^*)^{3-k}$. The resolution process $\Sigma \, \rightarrow \Sigma_n(\Delta)$, described in Appendix \ref{sect:res}, yields
\begin{equation}
Z_{\Theta^{[4-k-1]}}\, \rightarrow Z_{\Theta^{[4-k-1]}} \times E_{\Theta^{[4-k-1]}} \, .
\end{equation}
The factor $E_{\Theta^{[4-k-1]}}$ is determined by the simplices contained in the relative interior of the face $\Theta^{\circ [k]}$. Every $l$-simplex contained in the relative interior of a $k$-dimensional face $\Theta^{\circ [k]}$ contributes a $(\C^*)^{k-l}$ to $E_{\Theta^{[4-k-1]}}$, and hence it contributes a stratum
\begin{equation}
 Z_{\Theta^{[4-k-1]}} \times (\C^*)^{k-l}
\end{equation}
to $Z$. Note that the factor $Z_{\Theta^{[3-k]}}$ is common to all of the strata originating from simplices contained in a chosen face. For a Calabi-Yau threefold, this correspondence is such that $Z_{\Theta}$ is a two-dimensional variety for vertices, a curve for strata interior to 2-faces and a collection of points for strata interior to two-dimensional faces.

The closed subvarieties of $\mathbb{P}_\Sigma$, and hence of $Z$, associated with a simplex $t$ are found by collecting all lower-dimensional simplices $u$ attached to $t$ (i.e. $t\subset \partial u$) and taking the disjoint union of the associated strata. As the Hodge-Deligne numbers $e^{p,q}$ are additive, this provides an efficient way to find the Hodge numbers of the associated subvariety. Again, we may neglect all simplices that are contained in the relative interior of faces of maximal dimension (three in this case) as these do not contribute any strata to a Calabi-Yau hypersurface. In the following, we will explicitly write down the resulting stratifications of various subvarieties and compute their Hodge numbers.

At this point, we will adopt a different notation than in the rest of this appendix. As in the main text of the paper, we focus on threefolds, and so only need to distinguish vertices, edges, 2-faces, and 3-faces (facets) of $\Delta^\circ$, which we denote by $v$, $e$, $f$ and $c$, respectively. These are dual to 3-faces, 2-faces, edges and vertices on the $M$-lattice polytope $\Delta$, which are consequently denoted by $v^\circ$, $e^\circ$, $f^\circ$ and $c^\circ$. We hope this does not confuse the reader.

\subsubsection{Vertices}

Let us consider a divisor $D_i$ for which the associated lattice point $\nu_i = v$ is a vertex. The vertex has a dual face
$v^\circ$ on $\Delta$ that contributes an open two-dimensional stratum $Z_{v^\circ}$. Furthermore, there will be 1-simplices contained in
edges $e$ (dual to faces $e^\circ$) ending on $v$ contributing $Z_{e^\circ}$ as well as 1-simplices on 2-faces $f$ (dual to 1-faces $f^\circ$) contributing $Z_{f^\circ}\times \C^*$. Finally there are 2-simplices on faces $f$ (dual to 1-faces $f^\circ$) contributing $Z_{f^\circ}$. Hence such divisors contain the irreducible hypersurface $Z_{v^\circ}$ as an open dense set that is compactified by the other strata. Note that $Z_{f^\circ}$ is just a collection of $\ell^*(f^\circ)+1$ points.

Collecting all of these strata we find the stratification of $D_i$ to be
\begin{equation}\label{eq:stratYvertex}
D_i = Z_{v^\circ}  \amalg_{e \geq v} Z_{e^\circ}\times \left( pt \right)
\amalg_{f \geq v}  Z_f^{\circ}\times \left(\sum \C^* \amalg \sum pt \right) \, .
\end{equation}

With the stratification \eqref{eq:stratYvertex} at hand, we can start computing the Hodge numbers. For $h^{1,0}(D_i)$, only the
first two strata contribute and we find
\begin{equation}
\begin{aligned}
h^{1,0}(D_i) & = -e^{1,0}(D_i) \\
 & = -\left(e^{1,0}(Z_{v^\circ}) + \sum_{e \supset v} e^{1,0}(Z_{e^\circ}) \right) \\
 & = -\left(\sum_{e \supset v} \ell^*(e^\circ) - \sum_{e \supset v} \ell^*(e^\circ) \right)
 & = 0
\end{aligned}
\end{equation}
For $h^{2,0}(D_i)$, only the first stratum in  \eqref{eq:stratYvertex} contributes and we find
\begin{equation}
 h^{2,0}(D_i) = e^{2,0}(Z_{v^\circ})  = \ell^*(Z_{v^\circ}) \, .
\end{equation}

Finally, we can compute $h^{1,1}(D_i)$. Here, we find a contribution from $Z_{v^\circ}$, computed in \eqref{eq:hijofZT3}, as well as a contribution
\begin{equation}
\begin{aligned}
& \,\, \sum_{e \supset v} e^{1,1}(Z_{e^\circ}) = &\,\,\sum_{e \supset v}  1
 \end{aligned}
\end{equation}
from any edge $e$ emanating from the vertex $v$. Furthermore, 1-simplices interior to any 2-face $f$ (dual to $f^\circ$) connected to the vertex $v$ contribute
\begin{equation}
\begin{aligned}
  & \sum_{f \supset e}  e^{0,0}(Z_{f^\circ}) \times \sum_{t_1} e^{1,1}(\C^*) \\
= & \sum_{f \supset e}  (\ell^*(f^\circ)+1)  \sum_{t_1 \supset \nu_i} 1
\end{aligned}
\end{equation}
Summing it all up, the result is
\begin{equation}
\begin{aligned}
 h^{1,1}(D_i) = & -3 + \ell^*(2 v^\circ)-4\ell^*(v^\circ) - \ell^2(v^\circ) + \ell^1(v^\circ) \\
& + \sum_{e \supset v}  1 + \sum_{f \supset v}  (\ell^*(f^\circ )+1)\sum_{t_1 \supset \nu_i} 1
\end{aligned}
\end{equation}

\subsubsection{Simplices interior to 1-faces of $\Delta^\circ$}
Let us first consider divisors originating from points $\nu_i$ interior to edges $e$ of $\Delta^\circ$ dual to
2-faces $e^\circ$ of $\Delta$. The 0-simplex corresponding to $\nu_i$ is of the form $Z_{e^\circ} \times \C^*$, and the two
1-simplices on the edge $e$ containing $\nu_i$ correspond to $Z_{e^\circ}$. One-simplices containing $\nu_i$ that are
interior to 2-faces $f$ (dual to $f^\circ$) contribute $\ell^*(f^\circ) + 1$ points times $\C^*$ and 2-simplices contribute
$\ell^*(f^\circ) + 1$ points.

\begin{figure}[!h]
 \begin{center}
 \scalebox{.5}{ 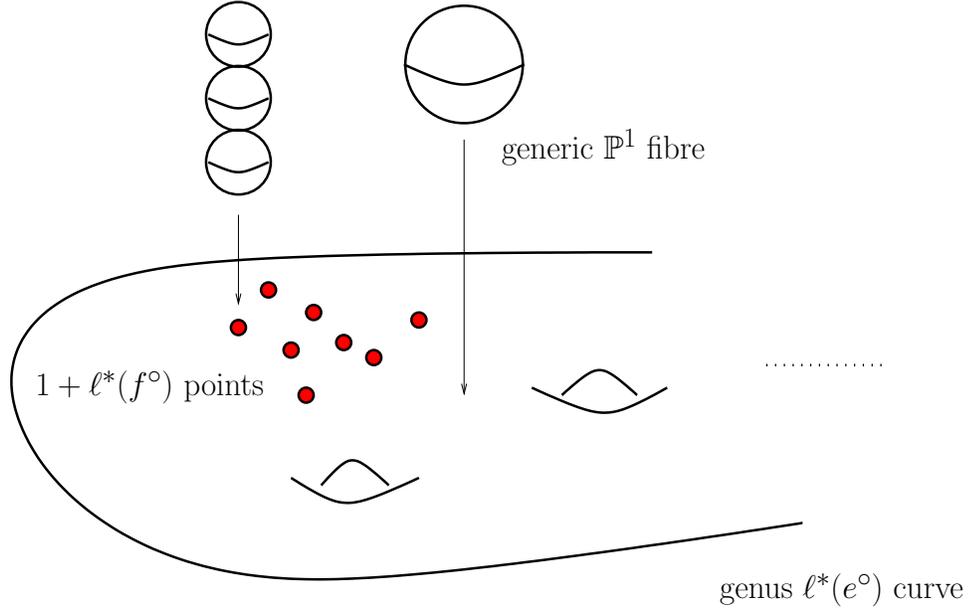 }
 \caption{The fibration structure of a divisor $D_i$ of $Z$ associated with a lattice point $\nu_i$ interior to an edge $e$ of a four-dimensional polytope. The base is a genus $g = \ell^*(e^\circ)$ curve and the generic fiber is a $\mathbb{P}^1$.
 For each neighboring 2-face $f \supset e$, there are $1+\ell^*(f^\circ)$ points over which the fiber degenerates into as many $\mathbb{P}^1$s as there are 1-simplices $t_1$ on $f$ that contain $\nu_i$. \label{fig:div_in_1D_face}}
 \end{center}
 \end{figure}

The open and dense stratum of such divisors (which originates from the simplex $\nu_i$) is simply the product of a curve of genus $\ell^*(e^\circ)$ (with $\ell^1(e^\circ)$ points excised) times a $\C^*$, which gets compactified by the remaining strata. We may think of these as (the open dense subsets of) the intersection of $D_i$ with `neighboring' divisors. The two 1-simplices along the edge $e$ partially compactify $Z_{e^\circ} \times \C^*$ to $Z_{e^\circ} \times \mathbb{P}^1$. The remaining $\C^*$'s and points sit over the $\ell^1(e^\circ)$ points excised in the open curve $Z_{e^\circ}$. We may hence think of such divisors as follows: they are flat fibrations of a $\mathbb{P}^1$ over a curve of genus $\ell^*(e^\circ)$. Over $\ell^1(e^\circ)$ points, the fiber $\mathbb{P}^1$ degenerates into a chain of $\mathbb{P}^1$'s, as determined by number of 1-simplices (and 2-simplices) attached to $\nu_i$ lying on neighboring 2-faces $f \supset v$. To see the details of how this works first note that
\begin{equation}
\ell^1(e^\circ) = \sum_{f^\circ \subset e^\circ} \ell^*(f^\circ) + 1 \, .
\end{equation}
Over each of the $\ell^*(f^\circ)+1$ points that are excised due to the face $f^\circ \subset e^\circ$, we find the strata corresponding to the one- (and two-) simplices interior to the dual face $f \supset e$. Hence over $\ell^*(f^\circ)+1$ points, where $f^\circ,f$ are dual faces, the generic fiber  $\mathbb{P}^1$ of $D_i$ degenerates into a number of $\mathbb{P}$'s equal to the number of 1-simplices which attached to $\nu_i$ and interior to $f$. A cartoon of this is shown in figure \ref{fig:div_in_1D_face}.

From this analysis of the fibration structure, we expect that
\begin{equation}
 h^{1,1}(D_i) =  2 + \sum_{f \supset e} \left(\ell^*(f^\circ)+1\right) \cdot (-1 + \sum_{t_1\supset \nu_i} 1)\, ,
\end{equation}
which will be confirmed by a direct computation using the stratification below.

As explained above, the stratification of $D_i$ is
\begin{equation}\label{eq:stratYedge}
D_i = Z_{e^\circ} \times \left(\C^* + 2pts \right)
\amalg_{f \supset e} Z_{f^\circ}\left(\sum \C^* + \sum pt \right) \, .
\end{equation}
Here, the $\C^*$ multiplying $Z_{e^\circ}$ is due to $\nu_i$ ($k=2,l=1$), whereas the 2 points correspond
to the two 1-simplices on $e$ containing $\nu_i$ ($k=2,l=2$). Each $\C^*$ multiplying $Z_{f^\circ}$
corresponds to a 1-simplex containing $\nu_i$ that is interior to the face $f$ and each $pt$ corresponds
to a 2-simplex containing $\nu_i$ that is interior to $f$.

Again, $h^{0,0}(D_i)=1$ as $D_i$ is irreducible. The computation for $h^{1,0}$ now becomes
\begin{equation}
 \begin{aligned}
h^{1,0}(D_i) & = -e^{1,0}(Z_{e^\circ}\cdot\left(\C^* \amalg 2pts\right))  \\
 & = \ell^*(e^\circ)\cdot (e^{0,0}(\C^*)+2e^{0,0}(pt)) \\
 & = \ell^*(e^\circ) \, .
 \end{aligned}
\end{equation}
We have $h^{2,0}(D_i) = e^{2,0}(D_i) = 0$ as no stratum contributes. Already for the highest stratum
$Z_{e^\circ}$, we have to count interior points to 3-dimensional faces of $e^\circ$, of which there are none. These Hodge numbers fit with the fibration structure discussed above.

Finally, let us compute $h^{1,1}(D_i)$. Here we need
\begin{equation}
\begin{aligned}
h^{1,1}(D_i) & =  e^{1,1}(Z_{e^\circ}) \cdot \left(e^{0,0}(\C^*) + 2 e^{0,0}(pt) \right) +
e^{0,0}(Z_{e^\circ}) \cdot e^{1,1}(\C^*) \\
& + \sum_{f^\circ \subset e^\circ} e^{0,0}(Z_{f^\circ}) \cdot \sum_{t_1\supset \nu_i} 1 \\
& = 1 \cdot (-1+2) + (1-\ell^1(e^\circ)) \cdot 1 +  \sum_{f^\circ \subset e^\circ} (1+\ell^*(f^\circ))\cdot \sum_{t_1\supset \nu_i} 1\\
& = 2 - \left( \sum_{f^\circ\subset e^\circ} \ell^*(f^\circ) + 1\right) +  \sum_{f^\circ \subset e^\circ} (1+\ell^*(f^\circ))\cdot \sum_{t_1\supset \nu_i} 1 \\
& = 2 + \sum_{f^\circ \subset e^\circ} (1+\ell^*(f^\circ))\cdot \left( -1 + \sum_{t_1\supset \nu_i} 1 \right) \, ,
\end{aligned}
\end{equation}
as predicted from the analysis of the fibration of $D_i$ carried out above.

Similarly, one may analyze curves $C$ that correspond to 1-simplices $t_1$ interior to a 1-face $e$. Here, the stratification is
\begin{equation}
 C = Z_{e^\circ} + \sum_{f^\circ\subset e^\circ} Z_f^\circ \times (pt) \, .
\end{equation}
The stratum $Z_{e^\circ}$ is a curve of genus $ \ell^*(e^\circ)$ with a number of points excised. The second term is due to the unique 2-simplex attached to $t_1$ on every face $f \supset e$, which consists of $\ell^*(f) + 1$ points for each two-dimensional face containing $e$. It supplies the points that compactify $Z_{e^\circ}$ to $C$. It follows immediately that the genus of $C$ is
\begin{equation}
h^{1,0}(C) = \ell^*(e^\circ) \, .
\end{equation}
This fits with the fact that the union of all strata corresponding to simplices in the interior of $e$ sits over the curve $C$, so that two neighboring divisors $D_i$ and $D_j$ intersect in the common base of both their fibrations.

\subsubsection{Simplices interior to 2-faces of $\Delta^\circ$}

\begin{figure}[!h]
 \begin{center}
 \scalebox{.6}{ 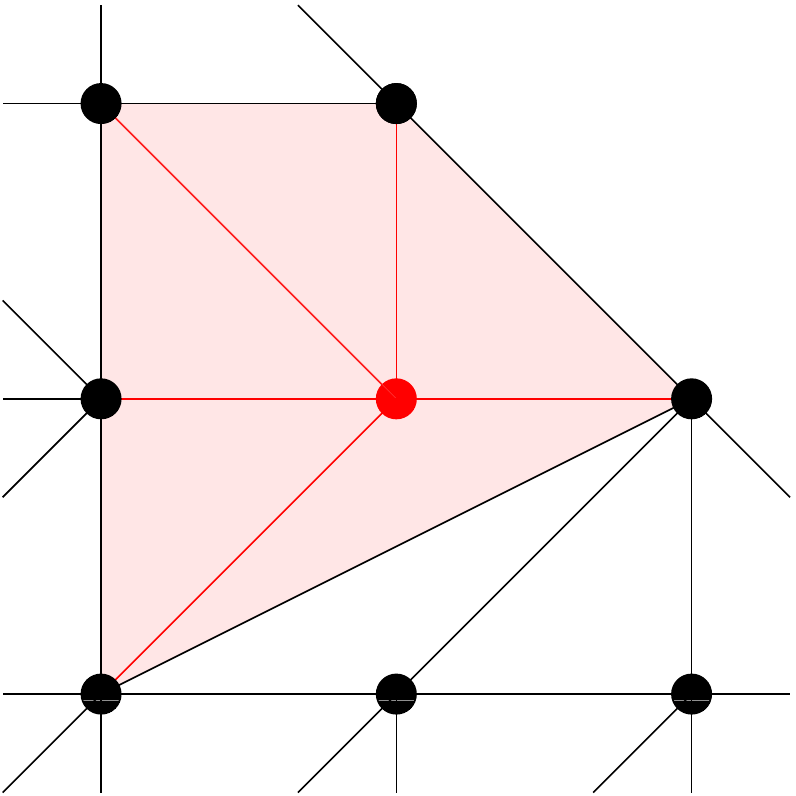 }
 \hspace{1cm}
 \includegraphics[height=5cm]{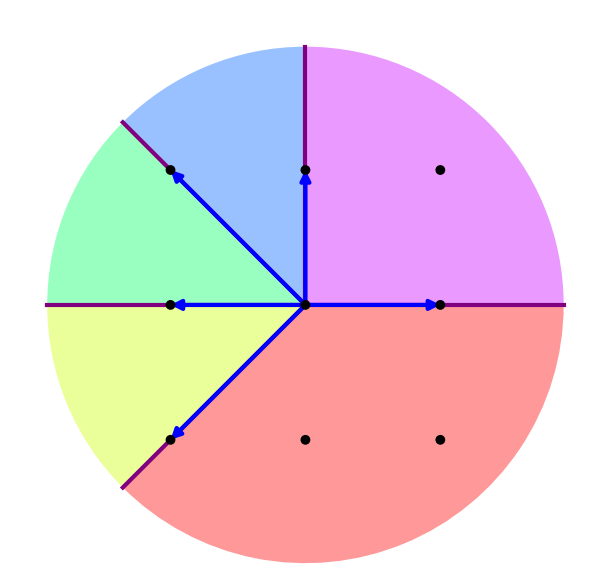}
 \caption{On the left hand side, the neighborhood of a lattice point $\nu_i$ inside a 2-face $f$ with a triangulation. We have colored the simplices containing $\nu_i$ in red. These contribute to the star fan $star(\nu_i)$ shown on the right hand side. \label{fig:starfan}}
 \end{center}
 \end{figure}

Again, let us first consider 0-simplices $\nu_i$ interior to a 2-face $f$ of $\Delta^\circ$. The open dense subset of $D_i$ originating from $\nu_i$ is given by
\begin{equation}
 Z_{f^\circ} \times (\C^*)^2 \, ,
\end{equation}
while 1-simplices (2-simplices) containing $\nu_i$ compactify $D_i$ this by contributing $Z_{f^\circ} \times \C^*$ and  $Z_{f^\circ} \times (pts)$. All in all, the stratification of $D_i$ is
\begin{equation}\label{eq:stratD2dface}
D_i = Z_{f^\circ}\times\left((\C^*)^2 + \sum_{t_1 \supset \nu_i} \C^* + \sum_{t_2 \supset \nu_i} pt \right)  \, .
\end{equation}
where $t_1$ and $t_2$ are simplices interior to $f$. As $Z_{f^\circ}$ is just a collection of $\ell^*(f^\circ) +1 $ points, the divisors considered here are, in general, reducible with each irreducible component being a toric variety $\mathbb{P}_{star(\nu_i)}$ determined by the stratification above. Starting from the triangulation of a 2-face, we may construct the star fan w.r.t. $\nu_i$ to find the fan of the toric variety $\mathbb{P}_{star(\nu_i)}$, see figure \ref{fig:starfan}.

From this, it immediately follows that
\begin{equation}
 h^{1,1}(\mathbb{P}_{star(\nu_i)}) = -2 + \sum_{t1 \subset \nu_i} 1\, .
\end{equation}
The same result is easily recovered from the stratification \eqref{eq:stratD2dface}:
\begin{equation}\label{eq:h11in2dface}
\begin{aligned}
 h^{1,1}(D_i) & = e^{0,0} (Z_{f^\circ}) \times \left(e^{1,1}((\C^*)^2))+ \sum_{t_1 \supset \nu_i} e^{1,1}(\C^*)  \right) \\
& = (\ell^*(f^\circ) + 1)\cdot(-2 + \sum_{t1 \subset \nu_i} 1) \, .
 \end{aligned}
\end{equation}

Similarly, the closed subvariety associated with each 1-simplex interior to $f$ is $\ell^*(f^\circ) + 1$ times a $\mathbb{P}^1$ and the closed subvariety associated with each 2-simplex consists of $\ell^*(f^\circ) + 1$ points. This implies that any two (three) divisors associated with points interior to $f$ with a nonzero intersection will intersect in a collection of $\ell^*(f^\circ) + 1$ disjoint $\mathbb{P}^s$'s (points).

\subsubsection{An example}

Let us consider a (slightly) nontrivial example to see the above machinery at work. Consider a reflexive polytope $\Delta^\circ$ with vertices
\begin{equation}
 \begin{aligned}
v_0&= \left[-1, -3, -9, -14\right] \\
v_1&= \left[0, -2, -6, -9\right] \\
v_2&= \left[0, 0, 0, 1\right] \\
v_3&= \left[0, 0, 1, 0\right] \\
v_4&= \left[0, 1, 0, 0\right] \\
v_5&= \left[1, 0, 0, 0\right]
 \end{aligned}
\end{equation}

The 3-faces of $\Delta^\circ$ as well as the vertices of $\Delta^\circ$ spanning them, their numbers of interior points and the dual vertices on $\Delta$ are
\begin{equation}
 \begin{aligned}
c_0&=<v_0,v_1,v_2,v_3>\hspace{.5cm}&\ell^*(c_0)&=0\hspace{.5cm}&\leftrightarrow& \hspace{.5cm} c^\circ_0=[ 0,  8, -1, -1] \\
c_1&=<v_0,v_1,v_2,v_4>\hspace{.5cm}&\ell^*(c_1)&=0\hspace{.5cm}&\leftrightarrow& \hspace{.5cm}c^\circ_1=[ 0, -1,  2, -1] \\
c_2&=<v_0,v_2,v_3,v_4>\hspace{.5cm}&\ell^*(c_2)&=0\hspace{.5cm}&\leftrightarrow& \hspace{.5cm}c^\circ_2=[27, -1, -1, -1] \\
c_3&=<v_1,v_2,v_4,v_5>\hspace{.5cm}&\ell^*(c_3)&=0\hspace{.5cm}&\leftrightarrow& \hspace{.5cm}c^\circ_3=[-1, -1,  2, -1] \\
c_4&=<v_0,v_1,v_3,v_4,v_5>\hspace{.5cm}&\ell^*(c_4)&=3\hspace{.5cm}&\leftrightarrow& \hspace{.5cm}c^\circ_4=[-1, -1, -1,  1] \\
c_5&=<v_2,v_3,v_4,v_5>\hspace{.5cm}&\ell^*(c_5)&=0\hspace{.5cm}&\leftrightarrow& \hspace{.5cm}c^\circ_5=[-1, -1, -1, -1] \\
c_6&=<v_1,v_2,v_3,v_5>\hspace{.5cm}&\ell^*(c_6)&=0\hspace{.5cm}&\leftrightarrow& \hspace{.5cm}c^\circ_6=[-1,  8, -1, -1]
 \end{aligned}
\end{equation}
The 3-faces of $\Delta$ dual to the vertices of $\Delta^\circ$ are
\begin{equation}
 \begin{aligned}
v_0\hspace{.5cm}&\leftrightarrow \hspace{.5cm} v^\circ_0 = <c^\circ_4, c^\circ_1, c^\circ_2, c^\circ_0> &\hspace{.5cm}\ell^*(v^\circ_0)=1 \\
v_1\hspace{.5cm}&\leftrightarrow \hspace{.5cm}v^\circ_1 = <c^\circ_4, c^\circ_3, c^\circ_0, c^\circ_1, c^\circ_6>&\hspace{.5cm}\ell^*(v^\circ_1)=0 \\
v_2\hspace{.5cm}&\leftrightarrow \hspace{.5cm}v^\circ_2 = <c^\circ_5, c^\circ_2, c^\circ_0, c^\circ_3, c^\circ_1,c^\circ_6>&\hspace{.5cm}\ell^*(v^\circ_2)=54 \\
v_3\hspace{.5cm}&\leftrightarrow \hspace{.5cm}v^\circ_3 = <c^\circ_5, c^\circ_2, c^\circ_4, c^\circ_0,c^\circ_6>&\hspace{.5cm}\ell^*(v^\circ_3)=22 \\
v_4\hspace{.5cm}&\leftrightarrow \hspace{.5cm}v^\circ_4 = <c^\circ_5, c^\circ_2, c^\circ_4, c^\circ_1,c^\circ_3>&\hspace{.5cm}\ell^*(v^\circ_4)=4 \\
v_5\hspace{.5cm}&\leftrightarrow \hspace{.5cm}v^\circ_5 = <c^\circ_5, c^\circ_6, c^\circ_4,c^\circ_3>&\hspace{.5cm}\ell^*(v^\circ_5)=1
 \end{aligned}
\end{equation}
The edges of $\Delta^\circ$ and their dual 2-faces on $\Delta$ are
\begin{equation}\label{eq:edges_vs_faces}
 \begin{aligned}
e_0&=<v_0,v_1>\hspace{.5cm}&\ell^*(e_0)&=0\hspace{.5cm}&\leftrightarrow \hspace{.5cm}&e^\circ_0&=<c^\circ_4, c^\circ_1, c^\circ_0>\hspace{.5cm}&\ell^*(e_0)&=0 \\
e_1&=<v_0,v_2>\hspace{.5cm}&\ell^*(e_1)&=0\hspace{.5cm}&\leftrightarrow \hspace{.5cm}&e^\circ_1&=<c^\circ_1, c^\circ_2, c^\circ_0>\hspace{.5cm}&\ell^*(e_1)&=7 \\
e_2&=<v_1,v_2>\hspace{.5cm}&\ell^*(e_2)&=1\hspace{.5cm}&\leftrightarrow \hspace{.5cm}&e^\circ_2&=<c^\circ_3, c^\circ_0, 1, c^\circ_6>\hspace{.5cm}&\ell^*(e_2)&=0 \\
e_3&=<v_0,v_3>\hspace{.5cm}&\ell^*(e_3)&=0\hspace{.5cm}&\leftrightarrow \hspace{.5cm}&e^\circ_3&=<c^\circ_4, c^\circ_0, c^\circ_2>\hspace{.5cm}&\ell^*(e_3)&=4 \\
e_4&=<v_1,v_3>\hspace{.5cm}&\ell^*(e_4)&=0\hspace{.5cm}&\leftrightarrow \hspace{.5cm}&e^\circ_4&=<c^\circ_4, c^\circ_6, c^\circ_0>\hspace{.5cm}&\ell^*(e_4)&=0 \\
e_5&=<v_2,v_3>\hspace{.5cm}&\ell^*(e_5)&=0\hspace{.5cm}&\leftrightarrow \hspace{.5cm}&e^\circ_5&=<c^\circ_5, c^\circ_2, c^\circ_0, c^\circ_6>\hspace{.5cm}&\ell^*(e_5)&=108 \\
e_6&=<v_0,v_4>\hspace{.5cm}&\ell^*(e_6)&=0\hspace{.5cm}&\leftrightarrow \hspace{.5cm}&e^\circ_6&=<c^\circ_4, c^\circ_1, c^\circ_2>\hspace{.5cm}&\ell^*(e_6)&=1 \\
e_7&=<v_1,v_4>\hspace{.5cm}&\ell^*(e_7)&=2\hspace{.5cm}&\leftrightarrow \hspace{.5cm}&e^\circ_7&=<c^\circ_4, c^\circ_3, c^\circ_1>\hspace{.5cm}&\ell^*(e_7)&=0 \\
e_8&=<v_2,v_4>\hspace{.5cm}&\ell^*(e_8)&=0\hspace{.5cm}&\leftrightarrow \hspace{.5cm}&e^\circ_8&=<c^\circ_5, c^\circ_2, c^\circ_1, c^\circ_3>\hspace{.5cm}&\ell^*(e_8)&=27 \\
e_9&=<v_3,v_4>\hspace{.5cm}&\ell^*(e_9)&=0\hspace{.5cm}&\leftrightarrow \hspace{.5cm}&e^\circ_9&=<c^\circ_5, c^\circ_2, c^\circ_4>\hspace{.5cm}&\ell^*(e_9)&=13 \\
e_{10}&=<v_4,v_5>\hspace{.5cm}&\ell^*(e_{10})&=0\hspace{.5cm}&\leftrightarrow \hspace{.5cm}&e^\circ_{10}&=<c^\circ_5, c^\circ_3, c^\circ_4>\hspace{.5cm}&\ell^*(e_{10})&=1 \\
e_{11}&=<v_1,v_5>\hspace{.5cm}&\ell^*(e_{11})&=0\hspace{.5cm}&\leftrightarrow \hspace{.5cm}&e^\circ_{11}&=<c^\circ_4, c^\circ_3, c^\circ_6>\hspace{.5cm}&\ell^*(e_{11})&=0 \\
e_{12}&=<v_2,v_5>\hspace{.5cm}&\ell^*(e_{12})&=0\hspace{.5cm}&\leftrightarrow \hspace{.5cm}&e^\circ_{12}&=<c^\circ_5, c^\circ_6, c^\circ_3>\hspace{.5cm}&\ell^*(e_{12})&=7 \\
e_{13}&=<v_3,v_5>\hspace{.5cm}&\ell^*(e_{13})&=0\hspace{.5cm}&\leftrightarrow \hspace{.5cm}&e^\circ_{13}&=<c^\circ_5, c^\circ_6, c^\circ_4>\hspace{.5cm}&\ell^*(e_{13})&=4
 \end{aligned}
\end{equation}
and finally the 2-faces on $\Delta^\circ$ and their dual edges on $\Delta$ together with their numbers of interior points are
\begin{equation}
\begin{aligned}
f_0&=<v_0,v_1,v_2>\hspace{.5cm}&\ell^*(f_0)&=0\hspace{.5cm}&\leftrightarrow \hspace{.5cm}&f^\circ_0&=<c^\circ_1, c^\circ_0>\hspace{.5cm}&\ell^*(f_0)&=2\\
f_1&=<v_0,v_1,v_3>\hspace{.5cm}&\ell^*(f_1)&=0\hspace{.5cm}&\leftrightarrow \hspace{.5cm}&f^\circ_1&=<c^\circ_4, c^\circ_0>\hspace{.5cm}&\ell^*(f_1)&=0\\
f_2&=<v_1,v_2,v_3>\hspace{.5cm}&\ell^*(f_2)&=0\hspace{.5cm}&\leftrightarrow \hspace{.5cm}&f^\circ_2&=<c^\circ_6, c^\circ_0>\hspace{.5cm}&\ell^*(f_2)&=0\\
f_3&=<v_0,v_2,v_3>\hspace{.5cm}&\ell^*(f_3)&=0\hspace{.5cm}&\leftrightarrow \hspace{.5cm}&f^\circ_3&=<c^\circ_0, c^\circ_2>\hspace{.5cm}&\ell^*(f_3)&=8\\
f_4&=<v_0,v_1,v_4>\hspace{.5cm}&\ell^*(f_4)&=0\hspace{.5cm}&\leftrightarrow \hspace{.5cm}&f^\circ_4&=<c^\circ_4, c^\circ_1>\hspace{.5cm}&\ell^*(f_4)&=0\\
f_5&=<v_0,v_2,v_4>\hspace{.5cm}&\ell^*(f_5)&=0\hspace{.5cm}&\leftrightarrow \hspace{.5cm}&f^\circ_5&=<c^\circ_1, c^\circ_2>\hspace{.5cm}&\ell^*(f_5)&=2\\
f_6&=<v_1,v_2,v_4>\hspace{.5cm}&\ell^*(f_6)&=1\hspace{.5cm}&\leftrightarrow \hspace{.5cm}&f^\circ_6&=<c^\circ_3, c^\circ_1>\hspace{.5cm}&\ell^*(f_6)&=0\\
f_7&=<v_0,v_3,v_4>\hspace{.5cm}&\ell^*(f_7)&=0\hspace{.5cm}&\leftrightarrow \hspace{.5cm}&f^\circ_7&=<c^\circ_4, c^\circ_2>\hspace{.5cm}&\ell^*(f_7)&=1\\
f_8&=<v_2,v_3,v_4>\hspace{.5cm}&\ell^*(f_8)&=0\hspace{.5cm}&\leftrightarrow \hspace{.5cm}&f^\circ_8&=<c^\circ_5, c^\circ_2>\hspace{.5cm}&\ell^*(f_8)&=27\\
f_9&=<v_1,v_4,v_5>\hspace{.5cm}&\ell^*(f_9)&=0\hspace{.5cm}&\leftrightarrow \hspace{.5cm}&f^\circ_9&=<c^\circ_4, c^\circ_3>\hspace{.5cm}&\ell^*(f_9)&=0\\
f_{10}&=<v_2,v_4,v_5>\hspace{.5cm}&\ell^*(f_{10})&=0\hspace{.5cm}&\leftrightarrow \hspace{.5cm}&f^\circ_{10}&=<c^\circ_5, c^\circ_3>\hspace{.5cm}&\ell^*(f_{10})&=2\\
f_{11}&=<v_1,v_2,v_5>\hspace{.5cm}&\ell^*(f_{11})&=0\hspace{.5cm}&\leftrightarrow \hspace{.5cm}&f^\circ_{11}&=<c^\circ_3, c^\circ_6>\hspace{.5cm}&\ell^*(f_{11})&=2\\
f_{12}&=<v_3,v_4,v_5>\hspace{.5cm}&\ell^*(f_{12})&=0\hspace{.5cm}&\leftrightarrow \hspace{.5cm}&f^\circ_{12}&=<c^\circ_5, c^\circ_4>\hspace{.5cm}&\ell^*(f_{12})&=1\\
f_{13}&=<v_1,v_3,v_5>\hspace{.5cm}&\ell^*(f_{13})&=0\hspace{.5cm}&\leftrightarrow \hspace{.5cm}&f^\circ_{13}&=<c^\circ_4, c^\circ_6>\hspace{.5cm}&\ell^*(f_{13})&=0\\
f_{14}&=<v_2,v_3,v_5>\hspace{.5cm}&\ell^*(f_{14})&=0\hspace{.5cm}&\leftrightarrow \hspace{.5cm}&f^\circ_{14}&=<c^\circ_5, c^\circ_6>\hspace{.5cm}&\ell^*(f_{14})&=8
\end{aligned}
\end{equation}

The Hodge numbers of the corresponding mirror pair of Calabi-Yau threefolds $Z$ and $\tilde{Z}$ can be quickly found with these numbers by evaluating \eqref{eq:bath11} and \eqref{eq:bath21}
\begin{equation}
\begin{aligned}
 h^{1,1}(Z) = h^{2,1}(\tilde{Z}) =  6\\
 h^{1,1}(\tilde{Z}) = h^{2,1}(Z) = 228
\end{aligned}
\end{equation}

\begin{figure}[!h]
 \begin{center}
\scalebox{.4}{ 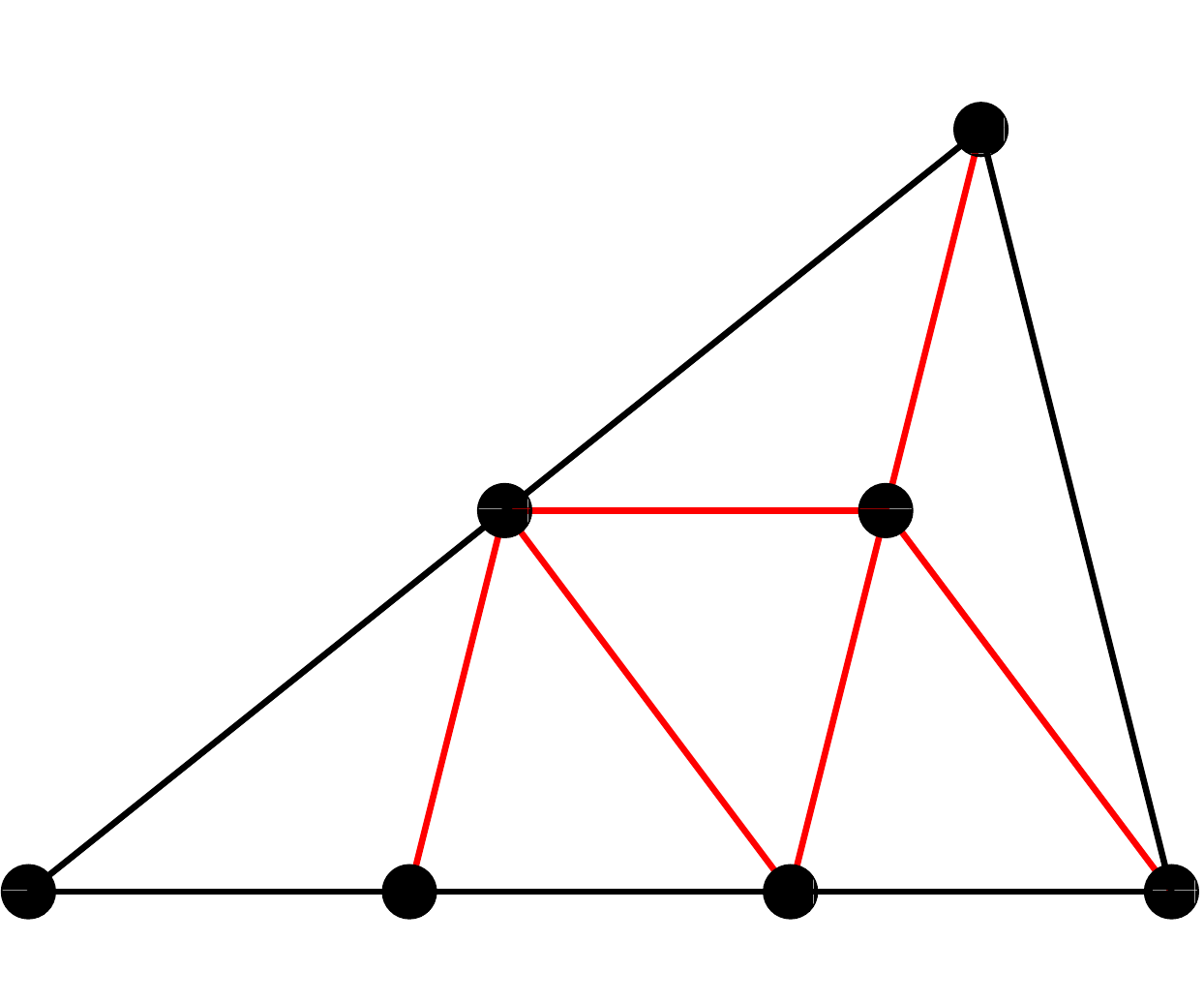  }
\hspace{2cm}
\scalebox{.4}{ 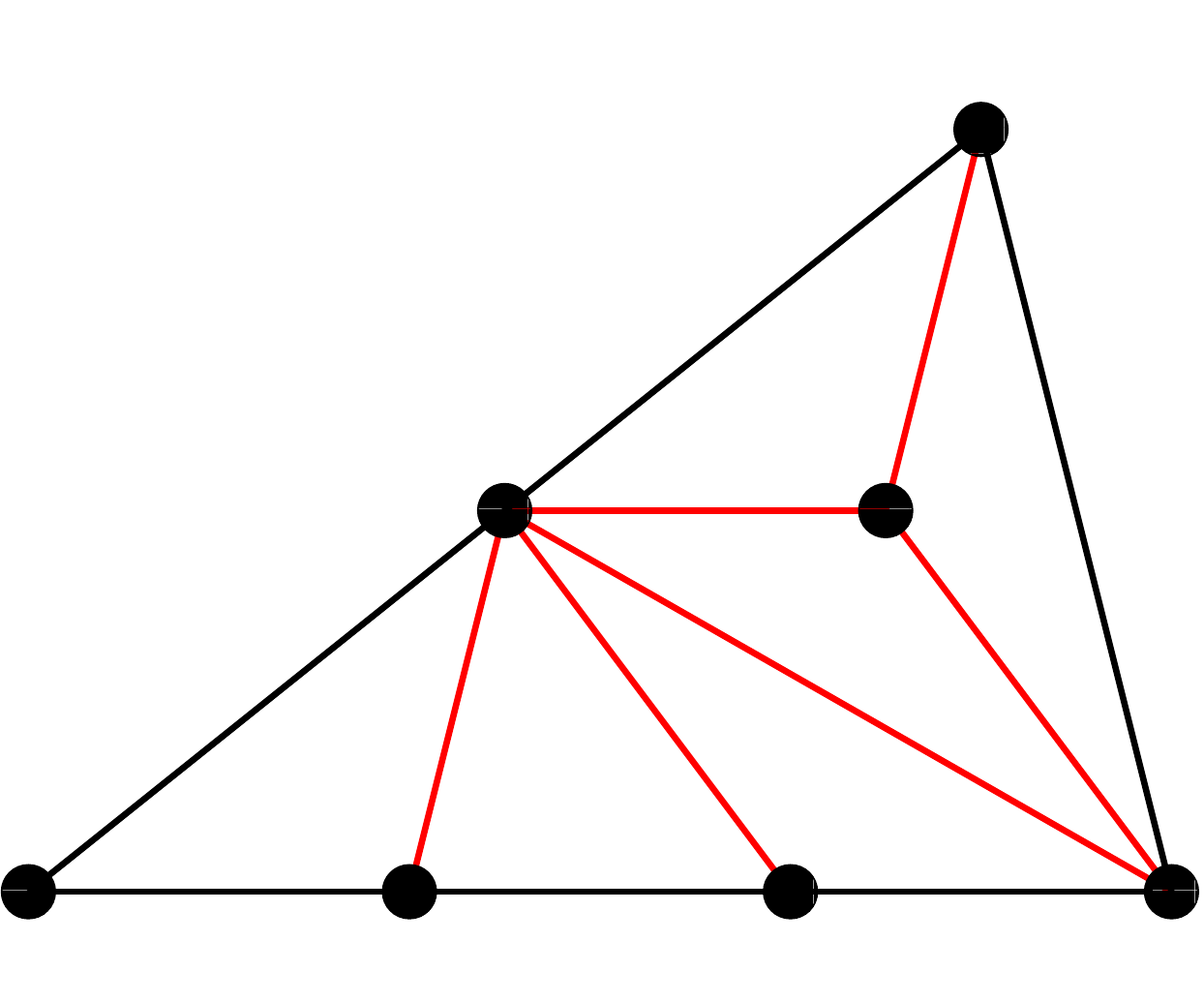  }
\\
\vspace{.5cm}
\scalebox{.4}{ 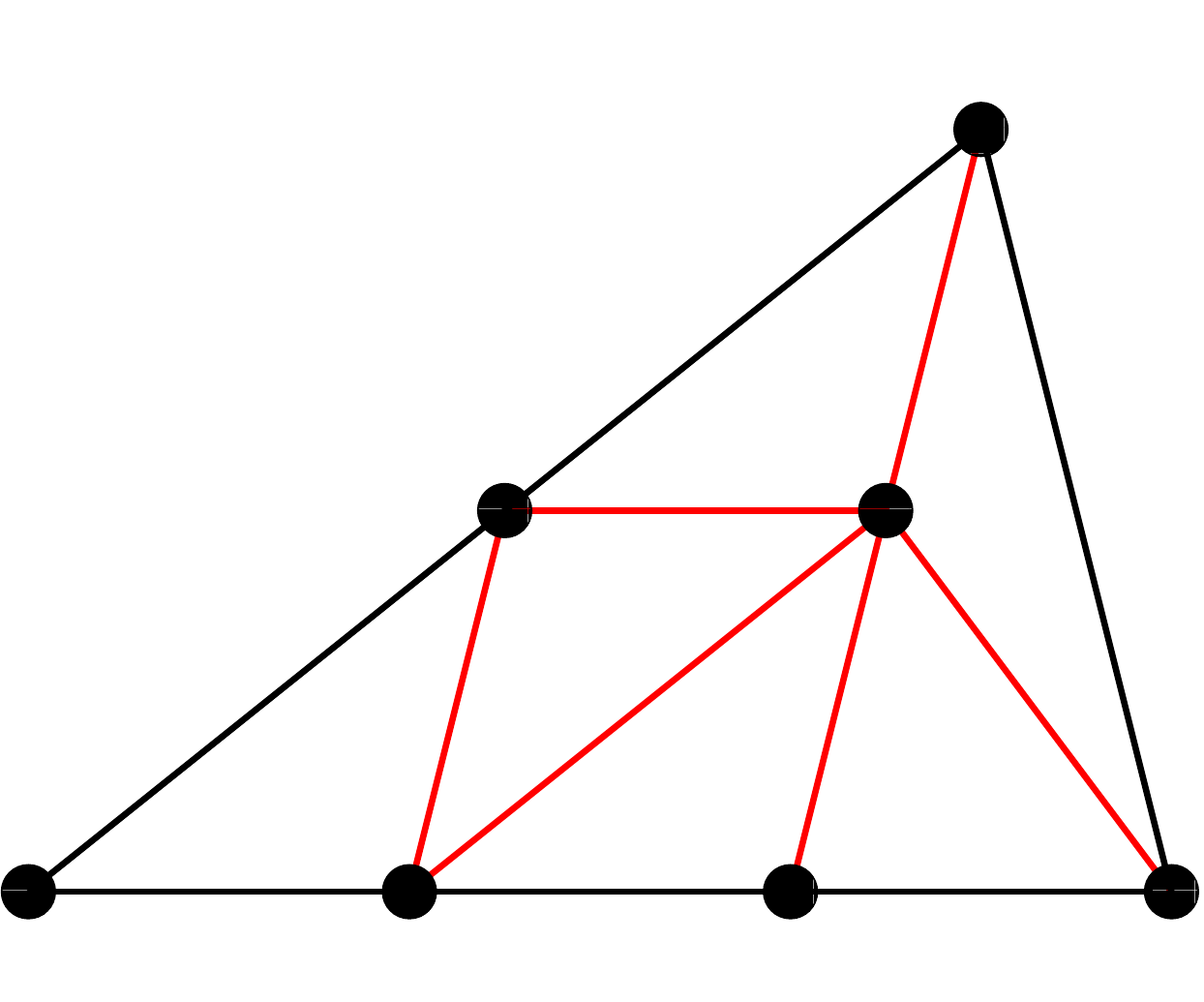  }
\hspace{2cm}
\scalebox{.4}{ 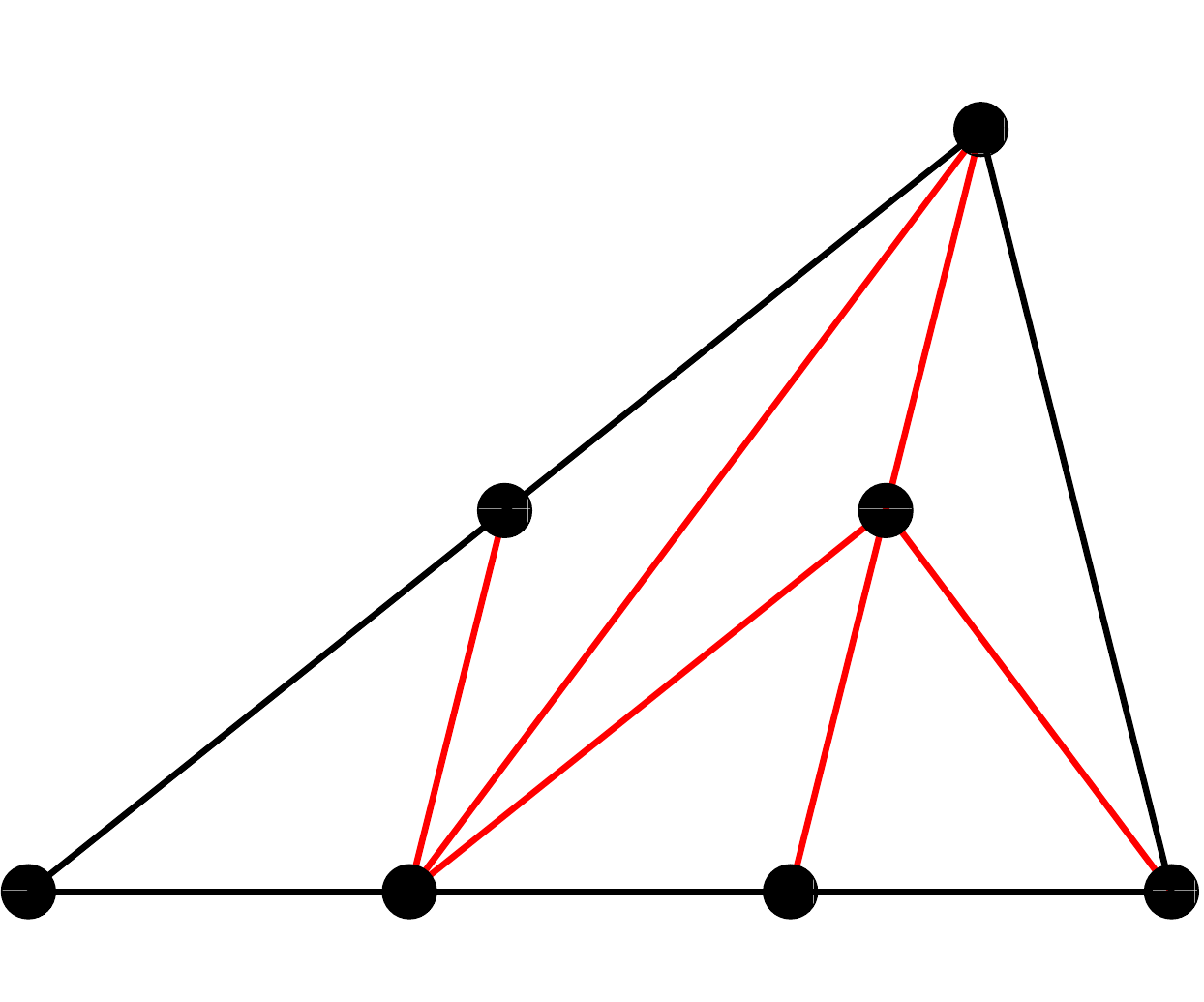  }
\\
\vspace{.5cm}
\scalebox{.4}{ 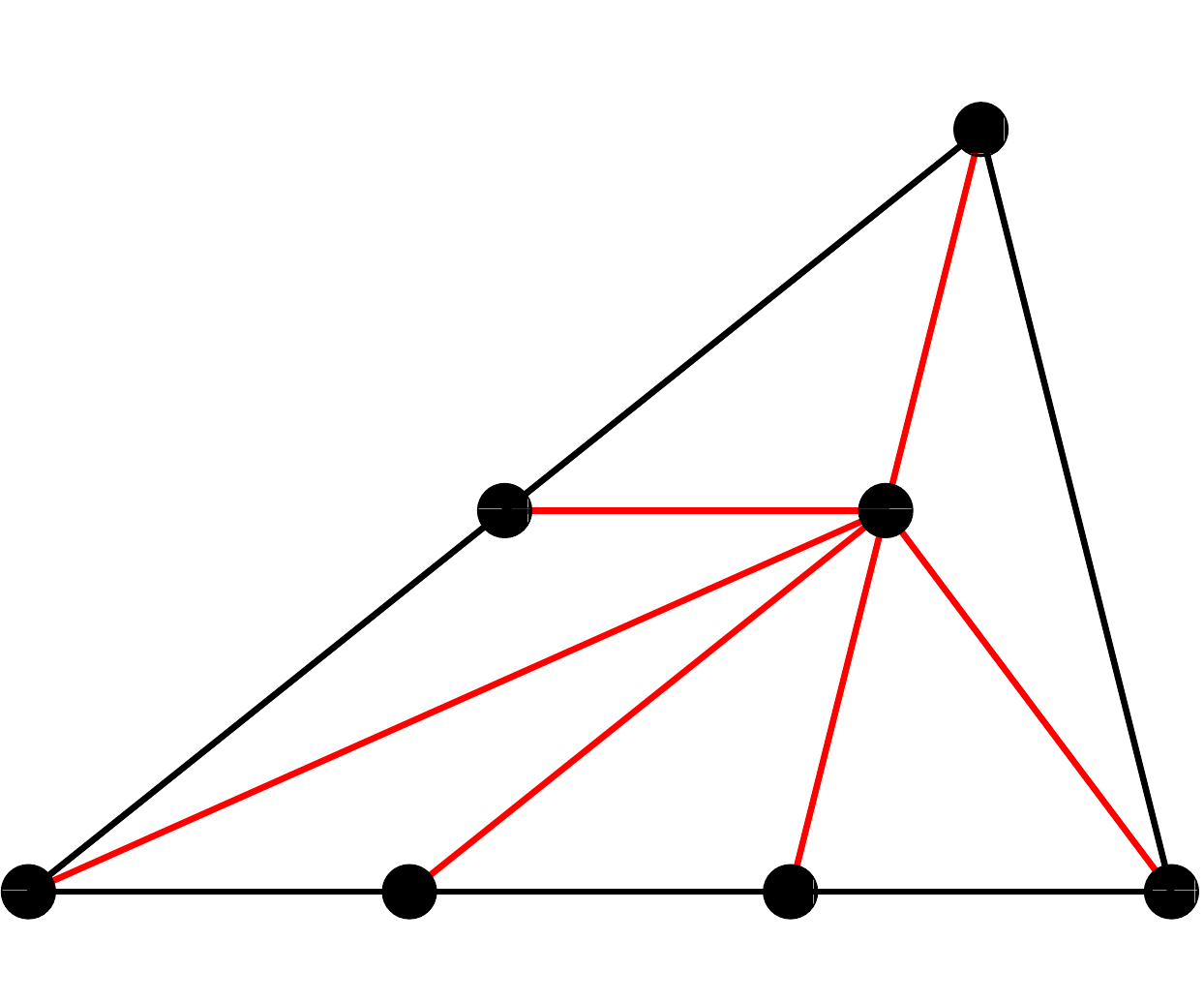  }
\caption{The only nontrivial 2-face $f_6$ of $\Delta^\circ$ along with its five possible triangulations. \label{2dface_ex}}
 \end{center}
 \end{figure}

There is a single 2-face $f_6$ that requires triangulation. This face, with its integral points and its bounding edges, as well as its triangulations, is shown in figure \ref{2dface_ex}.
Let us discuss the topology of the divisors $D_i$ that correspond to the lattice points $\nu_i,\,\, i = 1 \cdots 7$ sitting on this face. They are
\begin{equation}
 \begin{aligned}
\nu_1 &=  [0,-2,-6,-9] &= v_1 \\
\nu_2 &= [0,-2,-4,-6] &   \\
\nu_3 &= [0,0,-2,-3] &  \\
\nu_4 &= [0,1,0,0] &= v_4 \\
\nu_5 &= [0,-1,-3,-4]&   \\
\nu_6 &= [0,0,-1,-1] &  \\
\nu_7 &= [0,0,0,1] &=v_2 \\
 \end{aligned}
\end{equation}
As remarked above, some of the Hodge numbers will depend on the triangulation. Let us choose the triangulation shown in the upper left of figure \ref{2dface_ex}. The divisors $D_1$, $D_4$ and $D_7$ correspond to vertices, so that we conclude $h^{1,0}$ vanishes for all three and
\begin{equation}
\begin{aligned}
h^{2,0}(D_1) & = \ell^*(v^\circ_1) = 0 \\
h^{2,0}(D_4) &= \ell^*(v^\circ_4) = 4 \\
h^{2,0}(D_7) &= \ell^*(v^\circ_2) = 54
\end{aligned}
\end{equation}
no matter which triangulation is chosen. Let us now compute the Hodge numbers $h^{1,1}$ for the triangulation in the upper left of Figure \ref{2dface_ex}, for which we have to evaluate
\begin{equation}
\begin{aligned}
h^{1,1}(D_i) = & -3 + \ell^*(2 v^\circ)-4\ell^*(v^\circ) - \ell^2(v^\circ) + \ell^1(v^\circ) \\
& + \sum_{e \supset v}  1 + \sum_{f \supset v}  (\ell^*(f^\circ )+1)\sum_{t_1 \supset \nu_i} 1
\end{aligned}
\end{equation}
Note that there are no other 1-faces except $e_2$ and $e_7$, both of which are on $f_6$, with interior points and all 2-faces of $\Delta^\circ$ are simplicial. Hence there can be no other 1-simplices except the ones in $f_6$, shown in Figure \ref{2dface_ex}, which contain $\nu_1$, $\nu_4$ or $\nu_7$ and are interior to a 2-face. Furthermore, $\ell^*(f^\circ_6)=0$. The number of edges $e_i$ containing each of the three vertices in questions is found from \eqref{eq:edges_vs_faces}.
\begin{equation}
\begin{aligned}
\#e \supset v_1 & = 5 \\
\#e \supset v_4 & = 5 \\
\#e \supset v_2 & = 5
\end{aligned}
\end{equation}
We finally find that
\begin{equation}
\begin{aligned}
h^{1,1}(D_1) & = 1 + 5 + 0 = 6 \\
h^{1,1}(D_4) & = 77 + 5 + 1 = 83 \\
h^{1,1}(D_7) & = 398 + 5 + 1 = 404
\end{aligned}
\end{equation}
Only the last number depends on the triangulation chosen.

Let us now investigate the points $\nu_i$ interior to 1-faces $e_i$. Here $h^{2,0}(D_i) = 0$ for all cases. We start with $v_5$ which is contained in $e_2$. We hence learn from \eqref{eq:edges_vs_faces} that $h^{1,0}(D_5) = 0$. For $h^{1,1}$, we have to evaluate
\begin{equation}
2 + \sum_{f^\circ \subset e^\circ} (1+\ell^*(f^\circ))\cdot \left( -1 + \sum_{t_1\supset \nu_i} 1 \right)
\end{equation}
In the triangulation we are considering, there are three 1-simplices contained in $f_6$ which each contribute $1$ (as $\ell^*(f^\circ)=0$) to $h^{1,1}$. For each 2-face apart form $f_6$, there can only be a single 1-simplex containing $\nu_5$, so that we conclude
\begin{equation}
 h^{1,1}(D_5) = 4 \, .
\end{equation}
As described in general above, this means we should think of $D_5$ as a fibration of a $\mathbb{P}^1$ over another $\mathbb{P}^1$ for which the fiber degenerates into a union of three $\mathbb{P}^1$s over a single point in the base.

The other two points $\nu_2$ and $\nu_3$ interior to edges are contained in the same edge, $e_7$, so that
\begin{equation}
h^{1,0}(D_2) =  h^{1,0}(D_2) = \ell^*(e^\circ_7) = 0 \, .
\end{equation}
For the triangulation chosen, $\nu_2$ only connects to a single vertex inside $f_6$, whereas $\nu_3$ connects to two, hence
\begin{equation}
h^{1,1}(D_2) = 2 \hspace{1cm} h^{1,1}(D_3) = 3\, .
\end{equation}

Finally, there is $\nu_6$. As it is interior to a 2-face, it is $n =\ell^*(f_6)+1 = 1$ copies of a toric variety. This toric variety can be directly read off from the star fan to be the Hirzebruch surface $\mathbb{F}_1$ for the triangulation chosen. Hence
\begin{equation}
 h^{1,0}(D_6) =  h^{2,0}(D_6) = 0 \hspace{1cm} h^{1,1}(D_6) = 2 \, .
\end{equation}

A similar discussion can now easily be made for other triangulations. We can e.g. consider a flop taking us from the triangulation on the upper left to the one on the upper right. This will decrease the $h^{1,1}$ of $D_3$ and $D_6$ by one, whereas the $h^{1,1}$ of $D_5$ and $D_4$ are increased by one.

\subsection{Hodge numbers $h^{i,0}$ of toric divisors of Calabi-Yau $n$-folds} \label{sec:anydim}

For Calabi-Yau manifolds of higher dimension than $3$, the same technique as used above can be used to find topological data of toric divisors restricted to a Calabi-Yau hypersurface. Whereas Hodge numbers such as $h^{1,1}(D_i)$ will depend on the triangulation, one can derive a remarkably simple formula for the Hodge numbers $h^{i,0}$. For a smooth Calabi-Yau $n-1$-fold associated with a pair of $n$-dimensional reflexive polytopes $\Delta^\circ, \Delta$ and a lattice point $\nu$ in the relative interior of a face $\Theta^{\circ[n-d]}$ of dimension $n-d$, the associated divisor $\widehat{D}$ is such that
\begin{equation}\label{eq:masterhi0formula}
h^{i,0}(\widehat{D}) = \delta_{i,d-2} \,\, \ell^*(\Theta^{[d-1]}) \, ,
\end{equation}
where $d>2$. Here $\Theta^{[d-1]}$ is the face of $\Delta$ dual to the face $\Theta^{\circ[n-d]}$ containing $\nu_D$. Furthermore, $h^{0,0}(\widehat{D}) = 1$ holds for $d>2$, as all such divisors are connected. Formula \eqref{eq:masterhi0formula}, which we will prove in the following, is the central result of this section. Note that \eqref{eq:masterhi0formula} reduces to the corresponding relations derived above for Calabi-Yau threefolds (where $n=4$). In the threefold case, divisors associated with vertices ($d=4$) only have a non-vanishing $h^{2,0}$ and divisors associated with points interior to edges ($d=3$) only have a non-vanishing $h^{1,0}$.

Let us assume that we are given a pair of reflexive polytopes $\Delta^\circ,\Delta$ and a triangulation giving rise to a smooth projective toric variety\footnote{More generally, it is enough to for the singularities of $\mathbb{P}_\Sigma$ to miss a generic Calabi-Yau hypersurface, i.e. the only cones of $\Sigma$ of lattice volume $\geq 1$ are sitting inside faces of maximal dimension of $\Delta^\circ$.} $\mathbb{P}_\Sigma$. Let $D$ be the toric divisor associated with a lattice point $\nu_D$ contained in the relative interior of a face $\Theta^{\circ [n-d]}$ of dimension $n-d$. We are interested in the Hodge numbers of a divisor $\widehat{D} = D \cap Z$. Any toric divisor $D$ is composed of the strata associated with all cones that contain the ray over $\nu_D$. As before, these descend to a subset of the strata of $Z$ and we can sum their Hodge-Deligne numbers $e^{p,q}$ to find the Hodge numbers of $\widehat{D}$.

We first note that the cases $d=1$ and $d=2$ are trivial. In the first case, $d=1$, $D$ does not give rise to any divisor on $Z$, while in the second case, $d=2$, the face $\Theta^{\circ [n-2]}$ is dual to a face of dimension $n-(n-2)-1= 1$, denoted by $\Theta^{[1]}$, and it enjoys a stratification of the form
\begin{equation}
\widehat{D} = Z_{\Theta^{[1]}} \times \left[ \mbox{strata of the form} (\C^*)^i \right]
\end{equation}
so that such divisors have $h^{0,0} = \ell^*(\Theta^{[1]}) + 1$ disconnected components which are all smooth toric varieties. Hence the only nontrivial Hodge numbers of such divisors are $h^{i,i}(\widehat{D})$.

We hence assume $d > 2$ in the following. Let us start by writing down the stratification of an arbitrary divisor of $Z$ descending from a toric divisor. It is given by
\begin{equation}\label{eq:generalstratofdivisor}
\widehat{D} = \amalg_{k, \Theta^{\circ [n-d+k]}\supset \nu_D} Z_{\Theta^{[d-1-k]}}\times \sum \left[ pt. + \cdots + (\C^*)^{n-d+k} \right] \, .
\end{equation}
As usual, $(\Theta^{\circ [n-d+k]},\Theta^{[d-1-k]})$ are a pair of dual faces. For a divisor inside a face $\Theta^{\circ[n-d]}$, only strata on neighboring faces containing $\Theta^{\circ[n-d]}$ contribute. On $\Delta$, this can be expressed by saying that only faces $\Theta^{[d-1-k]}$ of $\Theta^{[d-1]}$ contribute. For each such face, the toric strata $(\C^*)^l$ in each term originate from various simplices of the triangulation on the faces $\Theta^{\circ [n-d+k]}$ dual to $\Theta^{[d-1-k]}$ that contain the point $\nu_D$.
In particular, a $(\C^*)^{n-d+k}$ originates from the point $\nu_D$ itself, a $(\C^*)^{n-d+k-1}$ originates from every 1-simplex in the interior of $\Theta^{\circ [n-d+k]}$ containing $\nu_D$, and so on. Finally, the highest-dimensional simplices, which are $n-d+k$-dimensional, give rise to points in the above expression.

Our main tool in deriving \eqref{eq:masterhi0formula} will be \eqref{eq:ep0}. Only the strata $Z_{\Theta^{[d-1-k]}}$ can potentially contribute to $h^{i,0}$, as $e^{i,0}((\C^*)^k) = 0$ for $i>0$. Hence we will only need to evaluate $e^{i,0}(\Theta^{[d-1-k]})$ and $e^{0,0}$ of the sum over simplices on the right-hand side of \eqref{eq:generalstratofdivisor}.

The first conclusion that can be drawn directly from \eqref{eq:ep0} is that $h^{i,0}(\widehat{D})=0$ whenever $i > d-2$. In this case, none of the strata
$Z_{\Theta^{[d-1-k]}}$ can contribute, as we would need to count points in faces of dimension $i+1$ in the face $Z_{\Theta^{[d-1-k]}}$, but even for $k=0$, there are no such faces. The geometric reason for this is that any divisor associated with a point $\nu_D$ inside a face of dimension $n-d$ should be thought of as an exceptional divisor originating from the resolution discussed in \S\ref{sect:res}. Correspondingly, each such divisor is a fibration of a toric variety of dimension $n-d$ (which degenerates over various subloci) over an irreducible manifold of dimension $d-2$. Hence the highest possible $i$ for which $h^{i,0}$ is nonzero is $d-2$, as already established.

Indeed, we do get a nonzero contribution whenever $ i = d-2$. In this case, only the stratum $Z_{\Theta^{[d-1]}}$ (i.e. $k=0$) contributes and we find
\begin{equation}
e^{i,0} = (-1)^{d-2} \ell^*(\Theta^{[d-1]}) \times e^{0,0}\left(\sum (\C^*)^l \mbox{strata}\right)
\end{equation}
The sum on the right hand side runs over all of the simplices on $\Theta^{\circ [n-d]}$ that contain $\nu_D$, including $\nu_D$ itself, and the sign alternates according to the dimension of the simplex in question as $e^{0,0}((\C^*)^l) = (-1)^l$.

If we neglect $\nu_D$, the remaining simplices are arranged such that they form an $n-d-1$-dimensional polyhedron. It can be found by intersecting the various simplices with an $n-d-1$-sphere on $\Theta^{\circ [n-d]}$ centered at $\nu_D$. Here, 1-simplices in the alternating sum, which contribute $(-1)^{n-d-1}$ above, correspond to vertices of the polyhedon.  As this polyhedron is topologically a sphere we can write
\begin{equation}
e^{0,0}\left(\sum (\C^*)^l \mbox{strata}\right) = (-1)^{n-d-1} \chi(S^{n-d-1}) + (-1)^{n-d} = 1
\end{equation}
where $\nu_D$ contributes the second term. As $\widehat{D}$ is a smooth compact manifold we can use $h^{0,i}(\widehat{D}) = (-1)^i e^{i,0}(\widehat{D})$, so that we have shown the case $i = d-2$ of \eqref{eq:masterhi0formula}.

We now proceed to show that $h^{0,i} = 0$ for all $0 < i < d-2$. Depending on $i$, a number of strata $Z_{\Theta^{[k]}}$ from \eqref{eq:generalstratofdivisor} contribute. Starting again from \eqref{eq:generalstratofdivisor}, we can write
\begin{equation}
e^{i,0}(\widehat{D}) =  \sum_{k, \Theta^{\circ [n-d+k]}\supseteq \nu_D} e^{i,0}\left( Z_{\Theta^{[d-1-k]}} \right) \times e^{0,0} \left( \sum \left[ pt. + \cdots + (\C^*)^{n-d+k} \right] \right) \, .
\end{equation}
For every term in the above sum over $k$, we have to find the alternating sum of all of the simplices containing $\nu_D$ on the face $\Theta^{\circ [n-d+k]}$ dual to $\Theta^{[d-1-k]}$ to evaluate the various $e^{0,0}$. For $k=0$, we have already found that this sum simply gives $1$ by relating it to the Euler characteristic of a sphere. For higher values of $k$, we can essentially use a similar argument. In this case, the point $\nu_D$ sits on the codimension-$k$ hyperplane of $\mathbb{R}^{n-d+k}$ defined by the face $\Theta^{\circ [n-d]}$. Furthermore, the face $\Theta^{\circ [n-d+k]}$ will be bounded by other hyperplanes of dimension greater than or equal to $n-d$, so that the set of all simplices on $\Theta^{\circ [n-d+k]}$ connecting to $\nu_D$ will correspond to a triangulation of an open subset of a sphere of dimension $n-d+k-1$. This has Euler characteristic $1$ in even and $-1$ in odd dimensions. To fix the sign, note that points on this sphere correspond again to 1-simplices, which in turn have a factor of $(-1)^{n-d+k-1}$ in the sum. As such points contribute $1$ in the computation of the Euler characteristic, the contribution of the sum over simplices to $e^{0,0}$ is always equal to $1$.

\noindent
Using \eqref{eq:ep0}, we are hence led to
\begin{align}\label{eq:h0iofanydivisor}
e^{i,0}(\widehat{D}) & =  \sum_{k, \Theta^{\circ [n-d+k]}\supset \nu_D} e^{i,0}\left( Z_{\Theta^{[d-1-k]}} \right) \nonumber \\
& = \sum_{k, \Theta^{\circ [n-d+k]}\supseteq \nu_D} (-1)^{d-1-k} \sum_{\Theta^{[i+1]} \subseteq \Theta^{[d-1-k]} } \ell^*(\Theta^{[i+1]})\,.
\end{align}
Note that each face containing $\nu_D$ appears multiple times and with alternating signs in the above expression. In particular, a single face
$\Theta^{[i+1]}$ can appear multiple times in a single term in the sum over $k$. Let us consider a single such face $\Theta^{[i+1]}$ and find how often it appears with which signs. First, note that we may equally well phrase the problem in terms of faces of $\Delta^\circ$. Given the face $\Theta^{\circ [n-d]}$ containing $\nu_D$ and the face $\Theta^{\circ [n-i-2]}$ dual to $\Theta^{[i+1]}$, the factor multiplying $\ell^*(\Theta^{[i+1]})$ for a fixed face of dimension $[i-1]$ in the above sum is then simply
\begin{equation}\label{eq:altsumfacesh0i}
 \sum_{k, \Theta^{\circ [n-d]} \subseteq \Theta^{\circ [n-d+k]} \subseteq \Theta^{\circ [n-i-2]}} (-1)^{d-1-k} \, .
\end{equation}
The contribution proportional to $\ell^*(\Theta^{[i+1]})$, the dual face of which is $\Theta^{\circ [n-i-2]}$, is hence given by counting all $n-d+k$-dimensional faces containing $\Theta^{\circ [n-d]}$ and contained in $\Theta^{\circ [n-i-2]}$. To compute this quantity, we again interpret this as an Euler characteristic of a topological space as follows. Consider a sphere of dimension $n-i-2-(n-d)-1 = d-i-3$ centered at $\nu_D$ and orthogonal to the face $\Theta^{\circ [n-d]}$. All the faces contributing to the sum above, except $\Theta^{\circ [n-d]}$ itself, will give rise to a decomposition of one closed half of this sphere, which has Euler characteristic $+1$ in any dimension. The contribution of the highest-dimensional stratum on this half-sphere has $k=d-i-2$, so that it contributes $(-1)^{i-3}$ to the alternating sum in \eqref{eq:altsumfacesh0i}. As its dimension is $d-i-3$, it contributes $(-1)^{d-i-3}$ to the Euler characteristic, so that the sum in \eqref{eq:altsumfacesh0i}, still neglecting the face $\Theta^{\circ [n-d]}$, is $(-1)^{d}$. The face $\Theta^{\circ [n-d]}$ contributes $(-1)^{d-1}$, so that these two terms always cancel and the sum \eqref{eq:altsumfacesh0i} vanishes for any pair of faces $\Theta^{\circ [n-d]}$ and $\Theta^{\circ [n-i-2]}$. Hence the sum \eqref{eq:h0iofanydivisor} vanishes except when $d=i-2$, when only one term in \eqref{eq:altsumfacesh0i} contributes. This completes the proof of \eqref{eq:masterhi0formula}.

\newpage

\section{Computation of $h^2$ for a Divisor on a 2-Face}\label{s:theproof}

\noindent In this appendix we give an alternative computation of $h^2(\mathcal{O}_D)$ in a special case (defined below).
We will compute $h^2(\mathcal{O}_D)$ directly in terms of a counting of lattice points in $\Delta$, arriving at a result that coincides with Theorem~\ref{MasterTheorem} in this subcase.  This computation provides an alternative perspective to that of the spectral sequence in \S\ref{sec:spectral}.

\subsection{Preliminaries}
We begin by assembling some elementary results about divisors and Calabi-Yau hypersurfaces in toric varieties.  Let $V$ be a four-dimensional simplicial toric variety, with $X$ a Calabi-Yau hypersurface in $V$, and let $\widehat{D}$ denote a divisor in $V$.  We write $D = \widehat{D}\, \cap \,X$.
\begin{proposition}
Serre duality gives
\begin{equation} \label{SerreV}
h^{i}(\mathcal{O}_V(-\widehat{D}))=h^{4-i}(\mathcal{O}_V(\widehat{D}-X))
\end{equation}
on $V$, and
\begin{equation}  \label{SerreX}
h^i(\mathcal{O}_X(-D)) = h^{3-i}(\mathcal{O}_X(D))
\end{equation}
on $X$.
\end{proposition}
Let us now assume $\widehat{D}$ is effective. We then have $h^0(\mathcal{O}_{V}(-\widehat{D})) = 0$, and so $h^4(\mathcal{O}_{V}(\widehat{D}-X)) = 0$.
Using also that $X = \sum \widehat{D}_i$, where the $\widehat{D}_i$ are effective, we have $h^0(\mathcal{O}_{V}(\widehat{D}-X)) = 0$, and so
$h^{4}(\mathcal{O}_V(-\widehat{D}))=0$.  In addition because $D$ is effective we have $h^0(\mathcal{O}_X(-D))=0$ and so $h^{3}(\mathcal{O}_X(D))=0$.

Because $V$ is a toric fourfold, we have the relation
\begin{equation} \label{toricbullet}
h^{\bullet}(V,\mathcal{O}_V) = (1,0,0,0,0)\,.
\end{equation}
Using Serre duality (as in \cite{cox2011toric}) in the long exact sequence in cohomology induced by
the Koszul sequence
\begin{equation} \label{koszulseqVX}
0  \rightarrow \mathcal{O}_{V}(-X)\rightarrow \mathcal{O}_{V}\rightarrow \mathcal{O}_{X}\rightarrow 0\,,
\end{equation}
one immediately finds that
the Hodge numbers $h^{\bullet}(X,\mathcal{O}_X)$ obey
\begin{equation} \label{Xbullet}
h^{\bullet}(X,\mathcal{O}_X) = (1,0,0,1)\,.
\end{equation}
Similarly, we can establish:
\begin{proposition} \label{hatDprop}
For a space $S$, define $\tilde{h}^{i}(S):=h^i(S)$ for $i>0$, and $\tilde{h}^{0}(S):=h^0(S)-1$.
Then the following relations hold, for $0 \le i \le 3$:
\begin{equation} \label{hatDpropeq}
\tilde{h}^i(\widehat{D},\mathcal{O}_{\widehat{D}}) = h^{i+1}(\mathcal{O}_V(-\widehat{D}))=h^{3-i}(\mathcal{O}_V(\widehat{D}-X))\,.
\end{equation}
\end{proposition}
We consider the Koszul sequence for $\widehat{D} \subset V$, which reads
\begin{equation} \label{koszulseqV}
0  \rightarrow \mathcal{O}_{V}(-\widehat{D})\rightarrow \mathcal{O}_{V}\rightarrow \mathcal{O}_{\widehat{D}}\rightarrow 0\,.
\end{equation}
This induces the long exact sequence in cohomology
\begin{center}
\begin{tikzpicture}[descr/.style={fill=white,inner sep=1.5pt}]
        \matrix (m) [
            matrix of math nodes,
            row sep=1em,
            column sep=2.5em,
            text height=1.5ex, text depth=0.25ex
        ]
        { 0 & H^0(\mathcal{O}_{V}(-\widehat{D})) & H^0(\mathcal{O}_{V}) & H^0(\mathcal{O}_{\widehat{D}}) \\
            & H^1(\mathcal{O}_{V}(-\widehat{D})) & H^1(\mathcal{O}_{V}) & H^1(\mathcal{O}_{\widehat{D}}) \\
            & H^2(\mathcal{O}_{V}(-\widehat{D})) & H^2(\mathcal{O}_{V}) & H^2(\mathcal{O}_{\widehat{D}}) \\
            & H^3(\mathcal{O}_{V}(-\widehat{D})) & H^3(\mathcal{O}_{V}) & H^3(\mathcal{O}_{\widehat{D}})\\
            & H^4(\mathcal{O}_{V}(-\widehat{D}))&H^4(\mathcal{O}_{V}) & 0\\
        };

        \path[overlay,->, font=\scriptsize,>=latex]
        (m-1-1) edge (m-1-2)
        (m-1-2) edge (m-1-3)
        (m-1-3) edge (m-1-4)
        (m-1-4) edge[out=355,in=175](m-2-2)
        (m-2-2) edge (m-2-3)
        (m-2-3) edge (m-2-4)
        (m-2-4) edge[out=355,in=175] (m-3-2)
        (m-3-2) edge (m-3-3)
        (m-3-3) edge (m-3-4)
        (m-3-4) edge[out=355,in=175] (m-4-2)
        (m-4-2) edge (m-4-3)
        (m-4-3) edge (m-4-4)
        (m-4-4) edge[out=355,in=175] (m-5-2)
        (m-5-2) edge (m-5-3)
        (m-5-3) edge (m-5-4);\,.
\end{tikzpicture}
\end{center}
Applying (\ref{toricbullet}) leads to the first equality in (\ref{hatDpropeq}).
The second equality then follows from (\ref{SerreV}). $\square$
\begin{corollary} \label{h3dhat}
If $\widehat{D} \neq X$ we have
\begin{equation}
h^3(\widehat{D},\mathcal{O}_{\widehat{D}}) = h^{0}(\mathcal{O}_V(\widehat{D}-X)) =0\,,
\end{equation}
while if $\widehat{D}=X$ we have
\begin{equation}
h^3(\widehat{D},\mathcal{O}_{\widehat{D}}) = h^{0}(\mathcal{O}_V(\widehat{D}-X)) =1\, .
\end{equation}
\end{corollary}
In close parallel to Proposition \ref{hatDprop}, we can show the following:
\begin{proposition} \label{Dprop}
The following relations hold, for $0 \le i \le 2$:
\begin{equation} \label{Dpropeq}
\tilde{h}^i(D,\mathcal{O}_{D})
=\tilde{h}^{2-i}(X,\mathcal{O}_X(D))\,.
\end{equation}
\end{proposition}
We consider the Koszul sequence for $D \subset X$, which reads
\begin{equation} \label{koszulseqX}
0  \rightarrow \mathcal{O}_{X}(-D)\rightarrow \mathcal{O}_{X}\rightarrow \mathcal{O}_{D}\rightarrow 0\,.
\end{equation}
Applying (\ref{Xbullet}) and (\ref{SerreX}) to the long exact sequence in cohomology induced by (\ref{koszulseqX}) yields (\ref{Dpropeq}).
$\square$

In particular, we have
\begin{corollary} \label{previouslylemma7}
\begin{equation}
h^{2}(D,\mathcal{O}_D) = h^{0}(X,\mathcal{O}_X(D)) - 1\,.
\end{equation}
\end{corollary}

\subsection{Relating $h^2$ to toric data}

We now state the condition that defines the special case treated in this appendix.
\begin{definition} \label{def2faced}
Let $\widehat{D}$ be a square-free divisor in $V$ corresponding to a collection of lattice points $\{u_I\} \subset \Delta^{\circ}$.  We call $\widehat{D}$ {\bf{face-limited}}
if the $u_I$ are all contained in a single 2-face $f$ of $\Delta^{\circ}$.
We call a divisor $D$ in $X$ face-limited if $D:= \widehat{D} \cap X$ with $\widehat{D}$ face-limited.
\end{definition}
Notice that $\RV_D$ contains all the layers of the ravioli complex $\RV$ over $f$.  Thus, $D$ corresponds not just to $\RV_D \subset \RV$, but also to a subcomplex ${\mathcal{T}}_D \subset {\mathcal{T}}$. 

We now examine the simplicial complex ${\mathcal{T}}_D$
associated to $D$.
\begin{lemma} \label{h2gh2hatd}
Let $\widehat{D}$ be a face-limited divisor in $V$, let $D = \widehat{D} \cap X$, and let ${\mathcal{T}}_D$ be the associated simplicial complex.
Then
\begin{equation}\label{h2gh2hatdeq}
h^i (\widehat{D},\mathcal{O}_{\widehat{D}})= h^i({\mathcal{T}}_D)\,.
\end{equation}
\end{lemma}
This follows from the spectral sequence associated to the generalized Mayer-Vietoris sequence on $V$ given in Proposition \ref{MVtoricprop} of Appendix \ref{sec:mvc}.  $\square$
\begin{corollary} \label{2faceh2hat}
Let $\widehat{D}$ be a face-limited divisor in $V$,
and let $D = \widehat{D} \cap X$ be the corresponding face-limited divisor in $X$.
Then
\begin{equation} \label{2faceh2eq}
h^2 (\widehat{D},\mathcal{O}_{\widehat{D}})= 0\,.
\end{equation}
\end{corollary}
This follows from (\ref{h2gh2hatdeq}), because $h^2({\mathcal{T}}_D)=0$ for a divisor on a single 2-face. $\square$

We can now relate sections on $X$ to sections on $V$:
\begin{lemma} \label{l:h0v}
Let $\widehat{D}$ be a face-limited divisor in $V$,
and let $D = \widehat{D} \cap X$ be the corresponding face-limited divisor in $X$.
Then $h^0 (X,\mathcal{O}_X (D))= h^0 (V,\mathcal{O}_V (\widehat{D}))$.
\end{lemma}
We tensor the Koszul sequence from $V$ to $X$ with $\mathcal{O}(\widehat{D})$, which reads
\begin{align*}
0 & \rightarrow \mathcal{O}_{V}(\widehat{D}-X)\rightarrow \mathcal{O}_{V}(\widehat{D})\rightarrow \mathcal{O}_{X}(D)\rightarrow 0\, .
\end{align*}

A general square-free divisor $\widehat{D}$ on $V$ is written as $\widehat{D} = \sum a_i \widehat{D}_i$, where the $\widehat{D}_i$ are the toric divisors and
$a_i \in \{ 0, 1 \}$.  Because $X = \sum \widehat{D}_i$ and the $\widehat{D}_i$ are effective,
we have that $h^0(\mathcal{O}_{V}(\widehat{D}-X)) = 0$.  Therefore, to show that $h^0(X,\mathcal{O}_X(D))=h^0(V,\mathcal{O}_V(\widehat{D}))$, we need to show that $h^1(\mathcal{O}_{V}(\widehat{D}-X)) = 0$. By Serre duality, equation (\ref{SerreV}), we can equivalently show that $h^3(\mathcal{O}_{V}(-\widehat{D}))=0$.
Consider the Koszul sequence from $V$ to $\widehat{D}$,
\begin{equation} \label{kosvhatd}
0  \rightarrow \mathcal{O}_{V}(-\widehat{D})\rightarrow \mathcal{O}_{V}\rightarrow \mathcal{O}_{\widehat{D}}\rightarrow 0\,.
\end{equation}
Using (\ref{toricbullet}) in the long exact sequence induced by (\ref{kosvhatd}), we find that
\begin{equation}
h^2(\widehat{D},\mathcal{O}_{\widehat{D}}) = 0 \Rightarrow h^3(\mathcal{O}_{V}(-\widehat{D}))=0\,.
\end{equation}
The lemma follows upon using (\ref{2faceh2eq}). $\square$

We have thus proved:
\begin{corollary} \label{2faceh2}
Let $\widehat{D}$ be a face-limited divisor in $V$,
and let $D = \widehat{D} \cap X$ be the corresponding face-limited divisor in $X$.
Then
\begin{equation} \label{2faceh2dhv}
h^{2}(D,\mathcal{O}_D) = h^{0}(V,\mathcal{O}_V(\widehat{D})) - 1\,.
\end{equation}
\end{corollary}

\subsection{Computation of $h^2$ for a face-limited divisor} \label{s:ch2}

We are now equipped to calculate $h^2\left(D,\mathcal{O}_D\right)$ for an arbitrary
face-limited divisor $D=\sum\limits_i a_i D_i$. We first establish how to compute $h^0(V, \mathcal{O}_V(D))$.
\begin{proposition}\label{prop:coxtoricsections}
Let $V$ be a toric variety corresponding to a fan $\Sigma$, let $\widehat{D}_i$ be the toric divisors on $V$, and for $a_i \in \mathbb{Z}$,
define $\widehat{D} := \sum_{i}a_{i}\widehat{D}_{i}$.  Define the polyhedra $P_{\widehat{D}} = \{m\in M_{\mathbb{R}} | \langle m, u_{i} \rangle \geq -a_{i} \text{ for all }u_i \in \Sigma(1) \}$. Then $h^{0}(V,\mathcal{O}_{V}(\widehat{D})) = |P_{\widehat{D}} \cap M|$.
\end{proposition}
The proof is given in Proposition 4.3.2 of~\cite{cox2011toric}.
\begin{lemma} \label{twofacelV}
Let $\widehat{D}$ be a face-limited divisor in $V$, let $D = \widehat{D} \cap X$ be the corresponding face-limited divisor in $X$,
and let $v$ be the vertices, $e$ the complete edges, and $f$ the complete faces included in ${\mathcal{T}}_D$.\footnote{Because ${\mathcal{T}}_D$ is by assumption contained in a single 2-face $f$, the last sum in (\ref{eq:h2twofaceV}) will only have one term, but we find it useful to write (\ref{eq:h2twofaceV}) in a form that anticipates our result for a completely general square-free divisor.}  Then
\begin{equation}\label{eq:h2twofaceV}
h^0 (V,\mathcal{O}_V(\widehat{D})) = 1+\sum\limits_{v} g(v) + \sum\limits_{e} g(e) +\sum\limits_{f} g(f) \, .
\end{equation}
\end{lemma}
\smallskip
\noindent{\bf Proof.}
Sections of $\mathcal{O}_V (\widehat{D})$ are counted by lattice points $m$ such that $\langle m, u_i \rangle \geq -a_i$, and so
$h^0 (V,\mathcal{O}_V(\widehat{D}))$ can be computed by counting suitable lattice points.  We consider a divisor $\widehat{D} = \sum\limits_{i} \widehat{D}_i$ specified by a set of points $u_i \in f\, , \, i \in 1,\dots, N$, where $f$ is a 2-face.
We label the points not in the set $\{u_i\}$ as $u_a$.
First, note that for any effective divisor the origin $m= \vec{0}$ corresponds to a global section.
As $\widehat{D}$ is by assumption also square-free, the additional sections of $\mathcal{O}_V (\widehat{D})$ are counted by lattice points $m$ such that $\langle m, u_i \rangle \geq -1$, and $\langle m, u_a \rangle \geq 0$. We will count these sections by including points in the set $\{u_i\}$ one by one, and checking how the number of sections changes. In other words, let $\widehat{D} = \sum\limits_{i = 1}^j \widehat{D}_{i}$, where $j \leq N-1$, and let $\widehat{D}^{'} = \sum\limits_{i = 1}^{j + 1} \widehat{D}_{i} = \widehat{D} + \widehat{D}_{j+1}$.
 
The divisor $\widehat{D}_{j+1}$ corresponds to a lattice point $u_{j+1}$.
Then $h^0(V,\mathcal{O}_V(\widehat{D}^{'}))$
equals the number of lattice points $m$ such that
\begin{align}
&\langle m, u_i \rangle \geq -1,\\
&\langle m, u_{j+1} \rangle \geq -1, \\
&\langle m, u_a \rangle \geq 0\, .
\end{align}
On the other hand, $h^0(V,\mathcal{O}_V(\widehat{D}))$
equals the number of lattice points $m$ such that
\begin{align}
&\langle m, u_i \rangle \geq -1 \, , \\
&\langle m, u_a \rangle \geq 0\, .
\end{align}
Thus, $h^0(V,\mathcal{O}_V(\widehat{D}^{'}))-h^0(V,\mathcal{O}_V(\widehat{D}))$ is
the number of lattice points $m$ such that
\begin{align}
&\langle m, u_i \rangle \geq -1,  \label{eqn:cons1}\\
&\langle m, u_{j+1} \rangle = -1, \label{eqn:dualfacet} \\
&\langle m, u_a \rangle \geq 0\, . \label{eqn:cons2}
\end{align}
The points $m$ obeying (\ref{eqn:dualfacet}) are by definition the lattice points in the face $u_{j+1}^{\circ}$ of $\Delta$, dual to the point $u_{j+1}$.
We need to count points in $u_{j+1}^{\circ}$ that also satisfy (\ref{eqn:cons1}) and (\ref{eqn:cons2}).
There are three types of $u_i$: vertices, points interior to edges, and points interior to $f$. We will include them in the set $\{ u_i\}$ in that particular order. First consider the case where $u_{j+1}$ is a vertex of $f$. Then 
we need to solve
\begin{align}
&\langle m, u_{j+1} \rangle = -1\, \\
&\langle m, u_a \rangle \geq 0\, .\label{eqn:greater}
\end{align}
Equation~(\ref{eqn:dualfacet}) defines the facet $u_{j+1}^\circ \in \Delta(3)$. However, any point on the boundary of $u_{j+1}^\circ$
has a dual in $\Delta^\circ$, defined by $\langle m \, , \cdot \rangle = -1$, that violates (\ref{eqn:greater}). The only points that do not violate (\ref{eqn:greater}) are those in the interior of $u_{j+1}^\circ$, as they are dual only to $u_{j+1}$ itself, and therefore $h^0(V,\mathcal{O}_V (\widehat{D}_0) ) = 1+ g(u_{j+1})$.

Next let $u_{j+1}$ correspond to a point internal to an edge $e \in \Delta^\circ(1)$. The condition (\ref{eqn:greater}) is violated unless the entire edge, including the vertices bounding it, is included, since $\langle m, u_{j+1} \rangle = -1$ implies that $\langle m, u_\alpha \rangle = -1$ for any $u_\alpha \subset e$. Therefore, a divisor $D$ corresponding to a complete edge $e$ with vertices $u_a$ and $u_b$ has
\begin{equation}
h^0 (V,\mathcal{O}_V(\widehat{D}) ) =1 +g(v_u) + g(v_u) + g(e)\,.
\end{equation}
In a similar manner we find that including points $u_\alpha$ internal to $f$ in the set $\{ u_i \}$ can only contribute to $h^2$ if every point in the face $f$ is included in $\{ u_i \}$.
$\square$
 
From Corollary \ref{2faceh2} and Lemma \ref{twofacelV} we deduce:
\begin{corollary} \label{twofacel}
Let $D$ be a face-limited square-free divisor in $X$.
Then
\begin{equation}\label{eq:h2twoface}
h^2(D,\mathcal{O}_D ) = \sum\limits_{v} g (v^\circ) + \sum\limits_{e} g(e^\circ) +\sum\limits_{f} g (f^\circ) \, .
\end{equation}
\end{corollary}

\bigbreak

\bibliographystyle{JHEP}
\bibliography{refs}
\end{document}